%% file: main_arxiv.tex
\documentclass{article}

\usepackage[preprint, nonatbib]{neurips_2023}

\usepackage[utf8]{inputenc} 
\usepackage[T1]{fontenc}    
\usepackage[hidelinks]{hyperref}       
\usepackage{url}            
\usepackage{booktabs}       
\usepackage{amsfonts}       
\usepackage{nicefrac}       
\usepackage{microtype}      
\usepackage{xcolor}         
\usepackage{float}

\usepackage[
    sorting=none,
    backend=biber,
    maxcitenames=2,
    mincitenames=2,
    maxbibnames=20
]{biblatex}
\usepackage{amsmath}
\usepackage{bm}
\usepackage{amssymb}
\usepackage{multirow}
\usepackage{graphicx}
\usepackage{subcaption}
\usepackage{colortbl}
\usepackage{array}

\usepackage{siunitx}
\definecolor{cb-green-sea}  {RGB}{  0, 146, 146}
\definecolor{cb-burgundy}   {RGB}{146,   0,   0}
\definecolor{cb-blue}       {RGB}{ 0, 109, 219}

\definecolor{custom-green}{HTML}{00BA38}
\definecolor{custom-red}{HTML}{F8766D}

\newcommand{\eqendp}{\,\text{.}} 
\newcommand{\eqendc}{\,\text{,}} 

\makeatletter
\newcommand{\printfnsymbol}[1]{%
  \textsuperscript{\@fnsymbol{#1}}%
}
\makeatother

\title{Post-Training Network Compression\\for 3D Medical Image Segmentation: Reducing Computational Efforts via Tucker Decomposition}

\author{
Tobias Weber\thanks{These authors contributed equally to this work.}\phantom{\footnotesize *}\textsuperscript{1, 2, 3}\quad Jakob Dexl{\footnotesize\printfnsymbol{1}}\textsuperscript{1, 2}\quad David Rügamer\textsuperscript{2, 3}\quad Michael Ingrisch\textsuperscript{1, 3}\\
\textsuperscript{1} Department of Radiology, University Hospital, LMU Munich\\ 
\textsuperscript{2} Department of Statistics, LMU Munich \\
\textsuperscript{3} Munich Center for Machine Learning (MCML) \\
{\tt\small tobias.weber@stat.uni-muenchen.de} \\ {\tt\small jakob.dexl@med.uni-muenchen.de}
}

\begin{document}

\maketitle

\begin{abstract}

\textbf{Purpose}
We address the computational barrier of deploying advanced deep learning segmentation models in clinical settings by studying the efficacy of network compression through tensor decomposition.
We propose a post-training Tucker factorization that enables the decomposition of pre-existing models to reduce computational requirements without impeding segmentation accuracy.

\textbf{Materials and Methods}
We applied Tucker decomposition to the convolutional kernels of the TotalSegmentator (TS) model, an nnU-Net model trained on a comprehensive dataset for automatic segmentation of 117 anatomical structures.
Our approach reduced the floating-point operations (FLOPs) and memory required during inference, offering an adjustable trade-off between computational efficiency and segmentation quality.
This study utilized the publicly available TS dataset, employing various downsampling factors to explore the relationship between model size, inference speed, and segmentation performance.

\textbf{Results}
The application of Tucker decomposition to the TS model substantially reduced the model parameters and FLOPs across various compression rates, with limited loss in segmentation accuracy.
We removed up to 88\% of the model's parameters with no significant performance changes in the majority of classes after fine-tuning.
Practical benefits varied across different graphics processing unit (GPU) architectures, with more distinct speed-ups on less powerful hardware. 

\textbf{Conclusion}
Post-hoc network compression via Tucker decomposition presents a viable strategy for reducing the computational demand of medical image segmentation models without substantially sacrificing accuracy.
This approach enables the broader adoption of advanced deep learning technologies in clinical practice, offering a way to navigate the constraints of hardware capabilities.

\end{abstract}

\section{Introduction}
Computer-assisted segmentation is one of the cornerstones of medical image analysis. It assists physicians in accurately identifying, measuring, and visualizing anatomical structures and pathological conditions.
The progress in segmentation techniques is largely enabled by deep learning (DL) methods, allowing the automation or simplification of tasks that are otherwise labor- and time-intensive.
Over the years, medical segmentation challenges like BraTS \cite{menze2014multimodal} or the Medical Segmentation Decathlon \cite{antonelli2022medical} have been dominated by DL-based approaches, in most cases based on U-net \cite{ronneberger2015u} and V-net \cite{milletari2016v}.
Until recently, the majority of proposed segmentation approaches shared a common characteristic: training models for a single segmentation task, catering exclusively to specific use cases.

With the maturation of segmentation research, attention has shifted towards general-purpose models capable of labeling a myriad of anatomical structures directly \cite{wasserthal2023totalsegmentator}.
These models fall under the category of \textit{foundation models} \cite{bommasani2021opportunities}.
This trend is particularly driven by the transformer-based \textit{Segment Anything} model \cite{kirillov2023sam}, which has been investigated and adapted for its use in medical contexts \cite{mazurowski2023segment, gu2024segmentanybone, bui2023sam3d}.
An important contribution is the release of the \textit{TotalSegmentator}(TS; \cite{wasserthal2023totalsegmentator}), which is an nnU-Net model \cite{isensee2021nnu} trained from scratch on a massive dataset and offers automatic segmentation of 117 anatomical structures.
TS is deployed with an easy-to-use API including a configured container environment that aids the integration of DL models into clinical practice.

Models like TS require a high-end graphics processing unit (GPU) to be of practical use, yet servers in clinics typically lack advanced hardware.
Even with more modern hardware setups, the inference of a single volume can take several minutes.
Models with reduced computational efforts would provide institutions with limited resources the possibility to use advanced DL segmentation technologies and effectively democratize cutting-edge medical image analysis.

The central component of the TS model involves computationally intensive 3D convolutions.
To reduce the number of floating-point operations (FLOPs), these can be decomposed into multiple simpler operations using Tucker decomposition \cite{tucker1966some, yong2016tucker}.
This method can be applied post-training and enables the factorization of already existing models.
Network compression with Tucker decomposition offers an adjustable trade-off between accuracy and computational efficiency, allowing users to optimize for either faster inference or higher segmentation performance based on their clinical needs.
In this study, we investigate whether the computational effort of 3D multi-organ segmentation with TS can be reduced via Tucker decomposition-based network compression.

\section{Materials and Methods}

%
\begin{figure}[t]
    \centering
    \includegraphics[width=0.85\textwidth]{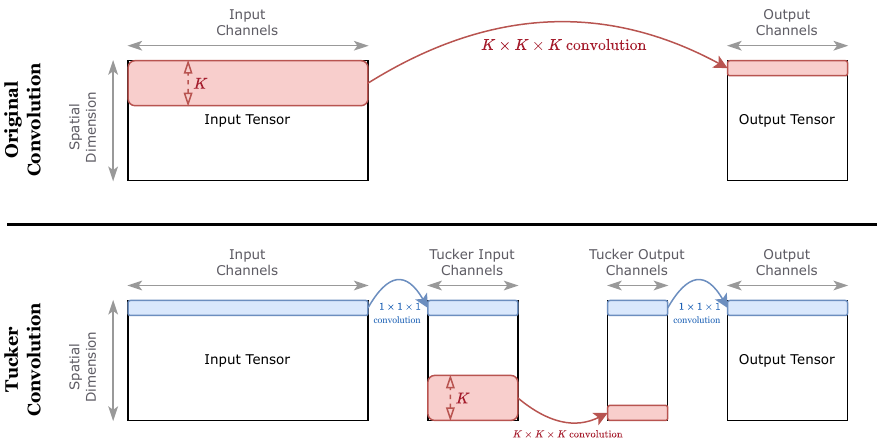}
    \caption{
    Schematic overview of the Tucker-decomposed convolution operation.
    The \textbf{top row} shows the original convolution with a $K \times K \times K$ kernel.
    The Tucker-decomposed convolution (\textbf{bottom row}) achieves its efficiency by first projecting each voxel of the input tensor into a space with a substantially smaller amount of channels using a $1 \times 1 \times 1$ kernel convolution and then performing the (otherwise) costly spatial convolution in this reduced representation space.
    Subsequently, the tensor is projected back into the original output channel domain.
    Note that the spatial dimensions $H \times W \times D$ are represented by a single dimension for visual purposes. 
    }
    \label{fig:tucker-high}
\end{figure}
\subsection{Tucker Decomposition for 3D Convolutional Kernels}

The Tucker decomposition \cite{tucker1966some} is a form of matrix decomposition that decomposes a multi-dimensional tensor into a set of factor matrices connected by a so-called core tensor.
Different amounts of compression can be achieved by choosing the dimensions of this core tensor.
\cite{yong2016tucker} suggested to utilize the Tucker decomposition to compress 2D convolutional neural networks (CNNs) for low-power applications. 
We extended their approach by deriving a Tucker decomposition that can be applied to 3D convolutions.
We briefly describe the concept in the following and refer the interested reader to Supplementary Material~\ref{app:tucker-meth} for a detailed mathematical derivation.

\paragraph{Tucker Decomposition for 3D Convolutions.}

In order to reduce the computational costs, the Tucker decomposition is applied to the convolutional weight kernel to derive a series of three consecutive simple convolution operations instead of performing a single costly convolution (see Figure~\ref{fig:tucker-high}).
Within a CNN, the dimensions of intermediate data representations can be divided into spatial dimensions $H$ (height), $W$ (width), $D$ (depth), and $C$ (channels/filters).
The latter is typically high-dimensional, to increase predictive power. This is, however, also a major factor for its computational burden.
To reduce the amount of computation and FLOPs in general, a Tucker-decomposed layer: (i) projects the high-dimensional $C$ axis of the data tensor into a dimension $\ll C$, (ii) convolutes the data with a $K \times K \times K$ core kernel, where $K$ corresponds to the original convolution's kernel spatial dimensions, and (iii) projects the low-rank channels to the desired output channels.
It is important to emphasize that this approach can be applied to a pre-trained network --- in our use case the \textit{TotalSegmentator} (TS).
In this post-hoc procedure, the existing TS weights were decomposed and the conventional convolutional layers were swapped with the Tucker convolution layers.

\paragraph{Complexity Reduction.}

A 3D convolutional kernel with $I$ input and $O$ output channels has $OIK^3$ parameters.
A Tucker convolution with internal dimensions $T_I \leq I$ (input) and $T_O \leq O$ (output) leads to $T_II$ parameters in the first convolution, $T_O T_IK^3$ parameters in the second convolution, and $OT_O$ parameters in the last convolution.
The advantage of using Tucker convolution compared to standard convolution becomes more pronounced when the dimensionality of the convolution kernel increases, i.e. from 1D (linear) through 2D (square) to 3D (cubic), as the Tucker decomposition leverages a fixed-cost dimensionality reduction in its initial and final layers, which does not depend on the size of the kernel.
If $K$ is fixed, the reduction in parameters is upper bounded by $\frac{OI}{T_O T_I}$ \cite{yong2016tucker}, implying that the Tucker rank should be chosen as low as the target network allows without collapsing.
The computational complexity, in terms of FLOPs, shifts from $OIK^3H'W'D'$ in the original convolution to a structure that benefits from dimensionality reduction in the Tucker convolution with $T_IIHWD + T_O T_IK^3 H'W'D' + OT_O H'W'D'$, where $H'$, $W'$ and $D'$ correspond to the spatial output dimensions.
These terms yield an upper bound equivalent to the complexity of the parameters.

\subsection{Dataset}
 In our study, we used the publicly available TS dataset, comprising 1204 CT images covering 104 classes of important anatomical structures. Our study focuses on the latest release of the TS Dataset (DOI: 10.5281/zenodo.10047292), which includes slightly more samples (1228) and covers a broader range of classes (117).
For evaluation, we use the officially recommended test subset of 89 CTs. Initially, the data was randomly sampled from the University Hospital Basel picture archiving and communication system (PACS) and annotated iteratively using a pseudo labeling and refinement scheme \cite{wasserthal2023totalsegmentator}.

\subsection{Experimental Setup.}

\setcounter{footnote}{0} 
We conducted our experiments using the \texttt{TotalSegmentator} (TS) package, version $2.0.5$, Python version $3.11$, and \texttt{PyTorch} version $2.1$.\footnote{\textbf{Code:} \texttt{https://github.com/ClinicalDataScience/tucker-cnn}}

The TS package provides access to two models: The default model for 1.5mm resolution and a fast version for 3mm resolution.
Although marketed as a single-model interface, the 1.5mm model consists of five separate U-Nets, trained on different label sets, i.e., one separate model for the skeleton, cardiovascular system, gastrointestinal tract, muscles, and other organs.

We conducted our analysis on both the 1.5mm and 3mm models.
We defined a \textit{downsampling factor} (DF) that describes the amount of reduction in the Tucker decomposition of the network.
For example, given a DF of $0.3$, a layer with input channels $I$ and output channels $O$ resulted in internal tucker dimension $T_I = 0.3 \cdot I$ and $T_O = 0.3 \cdot O$.
Our experiments included a variety of DFs, ranging from $0.9$ to $0.05$, with the intent to investigate the compressibility of the network and resulting segmentation performance.
To prevent a collapse of the target network due to limiting the information flow too stringent, we defined a minimum of $8$ channels in the Tucker decomposition. 

We further used two different approaches: A zero-shot approach that only decomposed the network, and a fine-tuned approach, that applied the decomposition and subsequently fine-tuned the model on the TS training set.
For both approaches, we then examined how much of the original model performance was preserved.
For fine-tuning, we used the Adam optimizer with a learning rate of $1\text{e}-5$ and a batch size of $2$ for $25,000$ weight updates.
The model was fine-tuned using the entire public TS training dataset using all available label classes.
Apart from comparisons between the original and compressed model, we also contrasted the performance of our approach with a simple baseline using channel-wise pruning based on weight magnitude.
To achieve this, the L2-norm was calculated for each channel in every weight kernel, and a predetermined fraction of the channels with the smallest values was set to zero.

\subsection{Statistical Analysis}

The segmentation performance of the models was evaluated using the Dice Score and the normalized surface distance metric (NSD) metric.
To compare compression methods, we utilized the compression ratio (CR) defined as the ratio of the number of parameters in the original and the compressed models. 

\paragraph{Quantitative Performance Assessment.}
After running all experiments, we analyzed whether compression has a significant influence on the model performance. 
To this end, we investigated the hypothesis that the difference between the NSD of the original and the Tucker-decomposed model is zero, i.e., that segmentation performance does not differ between the original and compressed model.
Given the complex data situation (repeated observations per patient, possibly interacting effects between DF and the different label classes, and the difference in NSD not following a normal distribution), we employed a logistic mixed model for the NSD difference as the outcome, a random effect for each patient, main effects for DF and the label class, and an interaction effect between the latter two.
Modeling was performed in R $4.3.3$ using the package \texttt{mgcv}\cite{wood2011fast} version $1.9.1$.
We then computed a p-value for the above hypothesis using the estimated mean and standard error for each combination of class label and DF.
To further account for multiple comparisons, we adjusted the resulting p-values using the Holm method.
These p-values allowed us to examine for which classes and compression factors the model performance deteriorates significantly. 

\paragraph{Inference Speed Assessment.}

Aside from assessing the actual segmentation performance, we explored the reduction in parameters and FLOPs.
As the theoretical computational complexity can have different practical implications depending on the amount of parallelization and the size and architecture of modern GPU hardware, we measured the processing time of Tucker-decomposed models in milliseconds on a variety of GPUs with different computational power.
We report results for 32-bit full-precision as well as for 16-bit half-precision inference.
Furthermore, we computed the relative speedup of the Tucker-decomposed network as the ratio of the original versus the compressed model's inference time.
Our experiments included NVIDIA consumer and enterprise GPUs from the following series, ordered by their computational capability:
GTX 1080, GTX 1660, RTX 2070 SUPER, RTX 2080 Ti, RTX 3060, RTX 3090, A6000, A100.

\paragraph{Quality of Approximation Assessment}
We evaluated the retained variance of each layer's weight kernel to quantify the information loss due to the degree of compression.
For this purpose, we reconstructed the kernel from a grid of possible combinations of $T_I$ and $T_O$.
For further information see Appendix~\ref{app:kernel-rec}. 

\section{Results}

\paragraph{Achieved Compression.}
Tucker decomposition resulted in a considerable reduction of model size, expressed in the number of model parameters. 
Table~\ref{tab:flopreduc} shows the reduction in parameters and  GigaFLOPs for the 3mm and all 1.5mm models.
A single one of the five 1.5mm models with $\approx31.19$ million parameters is almost twice as large as the full 3mm model with $16.55$ million parameters, which is also reflected in the number of FLOPS ($480.56$ GigaFLOPs vs.~$369.14$ GigaFLOPs).
Despite their different base architectures, applying the Tucker decomposition on the models had similar effects on the relative compression ratio.
For example, applying a DF of $0.5$, i.e., choosing a Tucker core tensor size equivalent to half of the input and output channels respectively, resulted in a realized parameter reduction of $70.37\%$ (1.5mm) and $70.26\%$ (3mm).
As the Tucker convolution sequence itself has a small overhead, the effective FLOP reduction, which in this case was $68.96\%$ (1.5mm) and $66.14\%$ (3mm), did not fully equal the parameter reduction.
Results show that $90\%$ of FLOP and parameter reduction was achieved by selecting a DF around $0.2$.

\begin{table}[t]
    \centering
    \resizebox{\textwidth}{!}{%
    \begin{tabular}{cc|ccccccccc}
    \toprule
     \multirow{2}{*}{\textbf{Model}} & \multirow{2}{*}{} & \multirow{2}{*}{\textbf{TS}} & \multicolumn{8}{c}{\textbf{Downsampling Factors (DF)}} \\
    \cline{4-11}
    \noalign{\smallskip}
    & & & 0.9  &  0.7 & 0.5 & 0.4 & 0.3 & 0.2 & 0.1 & 0.05 \\
    \midrule
    \multirow{5}{*}{1.5mm} & M-param. & 155.95 & 139.6 & 86.75 & 46.2 & 30.9 & 18.45 & 9.25 & 3.1 & 1.25 \\
    & $\Delta$ & - & 10.48\% & 44.37\%  & 70.37\% & 80.18\% & 88.17\% & 94.06\% & 98.01\%  & 99.20\% \\
    & CR & - & 1.2 & 1.8 & 3.4 & 5.0 & 8.5 & 16.9 & 50.3  & 124.7 \\
    \cmidrule{2-11}
    & G-FLOPs & 2402.8 & 2220.0 & 1407.4 & 745.75 & 515.45 & 325.05 & 189.05 & 128.55 & 120.0\\
    & $\Delta$ & - & 7.6\% & 41.43\%  & 68.96\% & 78.55\% & 86.47\% & 92.13\% & 94.65\%  & 95.01\% \\
        \midrule
    \multirow{5}{*}{3mm} & M-param. & 16.55 & 14.87 & 9.25 & 4.92 & 3.30 & 1.98 & 1.00 & 0.34 & 0.15 \\
    & $\Delta$ & - & 10.11\% & 44.06\%  & 70.26\% & 80.00\% & 88.02\% & 93.95\% & 97.91\%  & 99.08\% \\
    & CR & - & 1.1 & 1.8 & 3.4 & 5.0 & 8.4 & 16.6 & 48.7 & 110.3 \\
    \cmidrule{2-11}
    & G-FLOPs & 369.14 & 359.96 & 230.90 & 124.95 & 87.69 & 56.29 & 33.16 & 21.62 & 19.54 \\
    & $\Delta$ & - & 2.48\% & 37.44\%  & 66.14\% & 76.24\% & 84.75\% & 91.01\% & 94.14\%  & 94.70\% \\
    \bottomrule
    \end{tabular}
    }
    \vspace{2mm}
    \caption{Number of parameters in millions (M-param.), the achieved compression rate (CR), and amount of GigaFLOPs (G-FLOPs) for different downsampling factors (columns) and models (rows) together with the percentual change ($\Delta$) compared the original TS model.}
    \label{tab:flopreduc}
\end{table}
%

\paragraph{Segmentation Quality}

\begin{figure}[p]
    \centering
    \includegraphics[width=\textwidth]{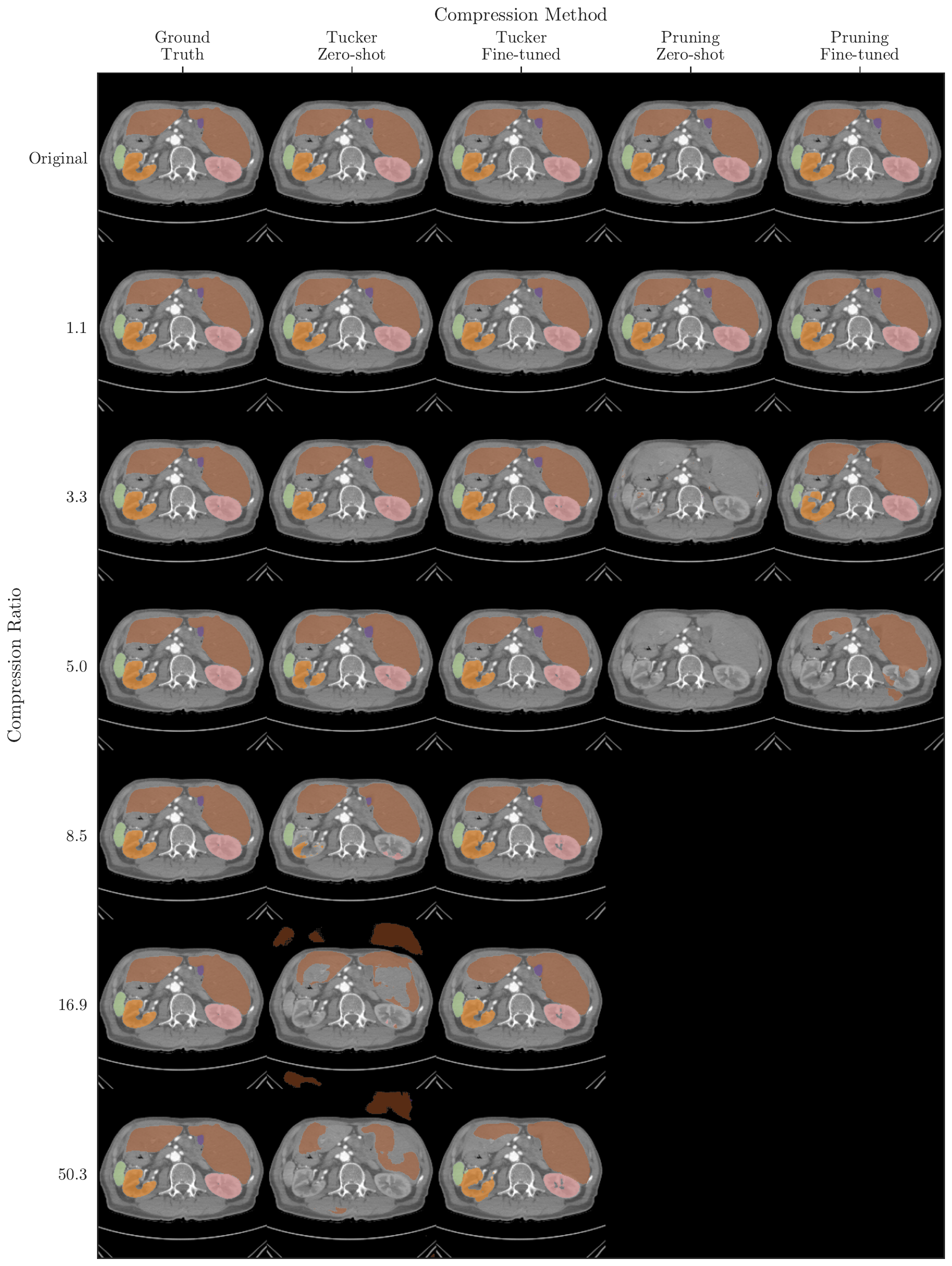}
    \caption{
    Visualization of segmentation performance across different compression methods (\textbf{columns}) applied to an abdominal CT image. The \textbf{rows} show the achieved compression ratios, which were determined by dividing the original model size by that of the compressed model size. 
    The segmented classes include \textit{spleen} (green), \textit{right kidney} (pink), \textit{left kidney} (orange), \textit{gallbladder} (purple), and \textit{liver} (brown).
    The ground truth (column one) remains constant across all evaluated compression ratios and acts as a benchmark for comparison.
    Columns two and three demonstrate that Tucker compression achieved noteworthy segmentation performance even for high CRs.
    Zero-shot Tucker compression introduced artifacts at higher CRs, a limitation that was not observed with fine-tuned Tucker compression.
    In contrast, the segmentation performances of both pruning approaches deteriorated rapidly (columns four and five). 
    }
    \label{fig:abdomen}
\end{figure}

\begin{figure}[p]
    \centering
    \includegraphics[width=0.985\textwidth]{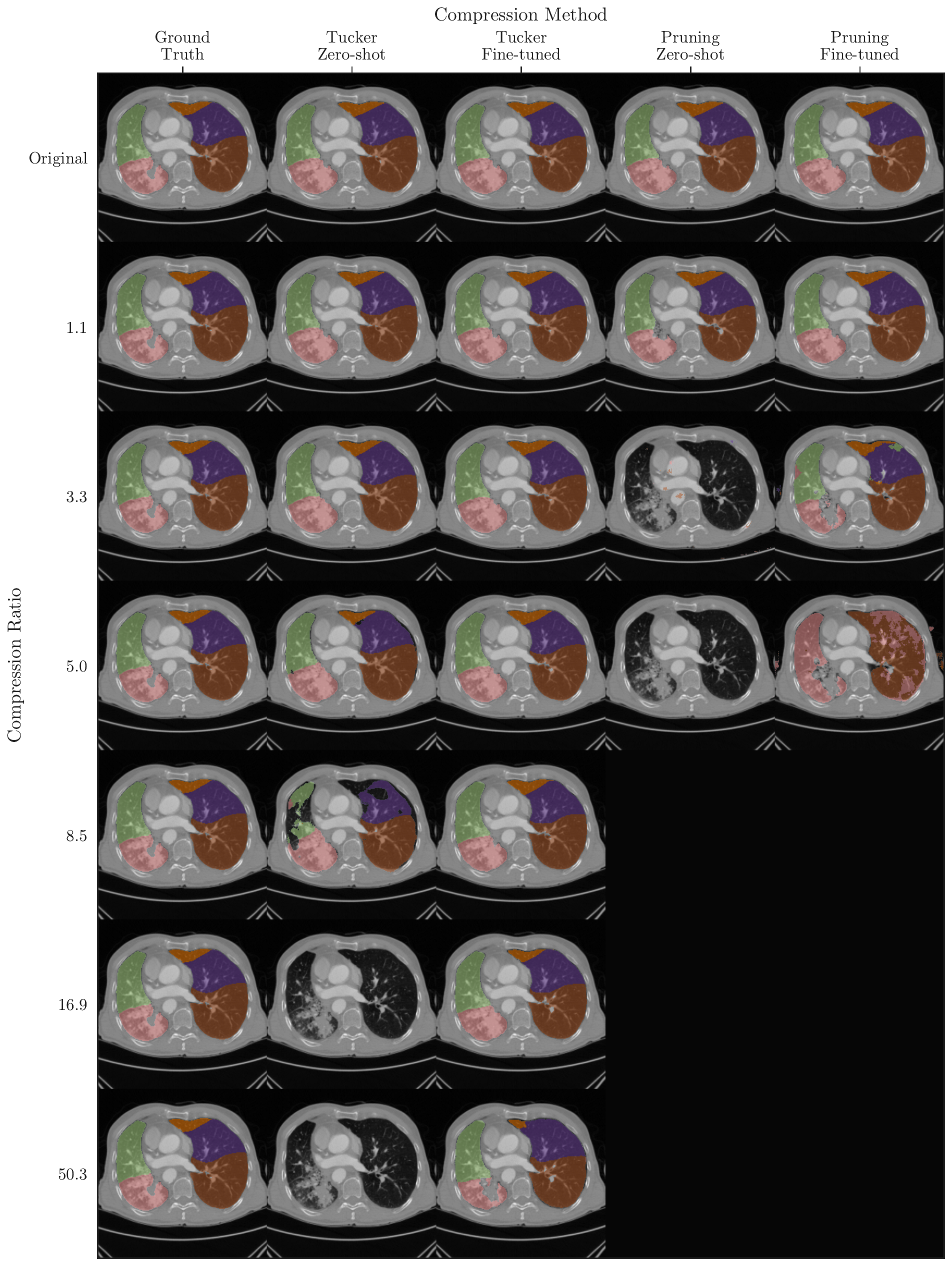}
    \caption{
    Visualization of segmentation accuracy across different compression methods (\textbf{columns}) applied to a thoracic CT image. The \textbf{rows} show the achieved compression ratios, which were determined by dividing the original model size by that of the compressed model size.
    The segmented classes include \textit{lung\_upper\_lobe\_left} (green), \textit{lung\_lower\_lobe\_left} (pink), \textit{lung\_upper\_lobe\_right} (orange), \textit{lung\_middle\_lobe\_right} (purple), \textit{lung\_lower\_lobe\_right} (brown).
    The ground truth (column one) remains constant across all evaluated compression ratios and acts as a benchmark for comparison.
    Columns two and three demonstrate that Tucker compression achieved noteworthy segmentation performance even for high CRs.
    Zero-shot Tucker compression introduced artifacts at higher CRs, a limitation that was not observed with fine-tuned Tucker compression.
    In contrast, the segmentation performances of both pruning approaches deteriorated rapidly (columns four and five).
    Notably, all models failed to segment the pathology in \textit{lung\_lower\_lobe\_left} accurately - this is a general property of the TS package, which is trained to segment normal anatomy.
    }
    \label{fig:thorax}
\end{figure}

We showcase the effect of compression on segmentation accuracy on CT images of the abdomen (Figure~\ref{fig:abdomen}) and thorax (Figure~\ref{fig:thorax}) for a sample of the test set.
Reaching a CR of 3.3, the zero-shot pruning method experienced a full collapse.
Fine-tuning mitigated this degeneration slightly, however, at this level of compression, the pruned model did not identify essential anatomies such as the gallbladder, in contrast to Tucker-decomposed models.
While the zero-shot Tucker-decomposed model began to degenerate at CRs between $5$ and $8.5$, the fine-tuned version achieved satisfactory segmentation performance at CRs up to $16.9$.
For a CR of $50$, the latter segmented most of the spleen and kidneys in the abdominal CT, although it neglected parts of the liver and gallbladder.
Similar behavior was evident in the thorax CT, with visible degeneration in the upper right lung lobe.

\paragraph{Segmentation Performance.}
\begin{figure}[b]
    \centering
    \begin{subfigure}[b]{0.48\textwidth}
        \includegraphics[width=\textwidth]{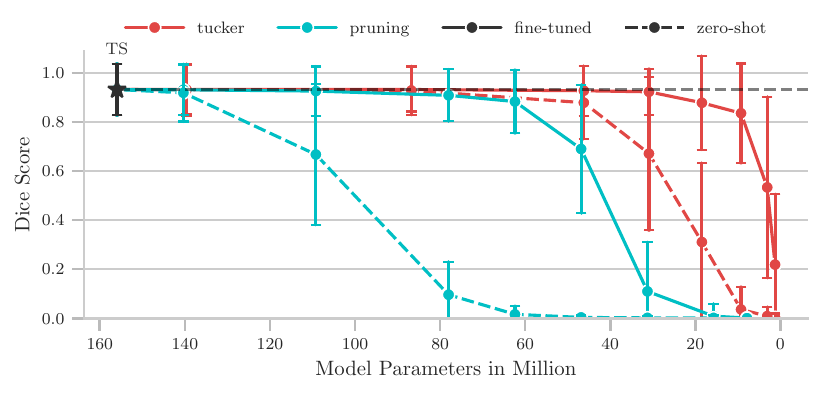}
        \caption{1.5mm model.}
        \label{fig:dice-overview-slow}
    \end{subfigure}
    \begin{subfigure}[b]{0.48\textwidth}
        \includegraphics[width=\textwidth]{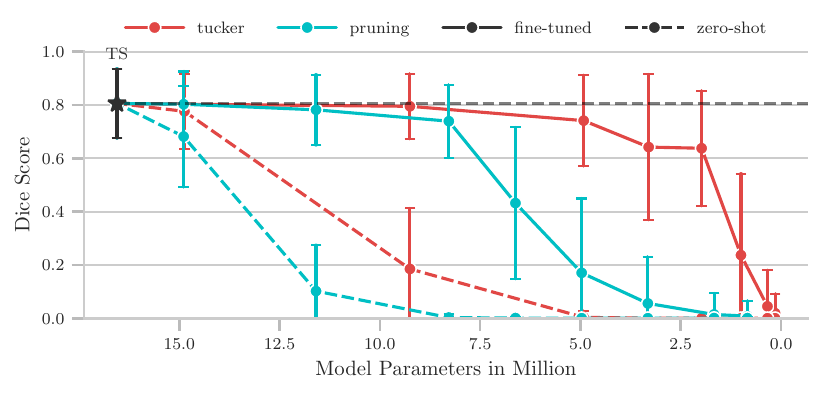}
        \caption{3mm model.}
        \label{fig:dice-overview-fast}
    \end{subfigure}
    \caption{
    Dice score aggregated over all classes for the TS test set using the 1.5mm (left) and 3mm (right) TS models.
    The performance of the original TS model is compared against the Tucker decomposition-based approach (red) and filter pruning (blue). Both compression methods are evaluated with (solid line) and without (dashed line) additional fine-tuning.
    Error bars represent the standard deviation across different classes.
    }
    \label{fig:dice-overview}
\end{figure}
\begin{figure}[t]
    \centering
    \begin{subfigure}[b]{0.48\textwidth}
        \includegraphics[width=\textwidth]{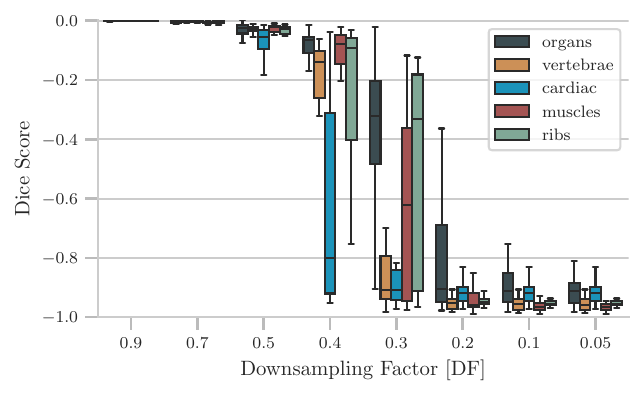}
        \caption{Zero-shot 1.5mm TS model.}
        \label{fig:dice-diff-zero}
    \end{subfigure}
    \hfill
    \begin{subfigure}[b]{0.48\textwidth}
        \includegraphics[width=\textwidth]{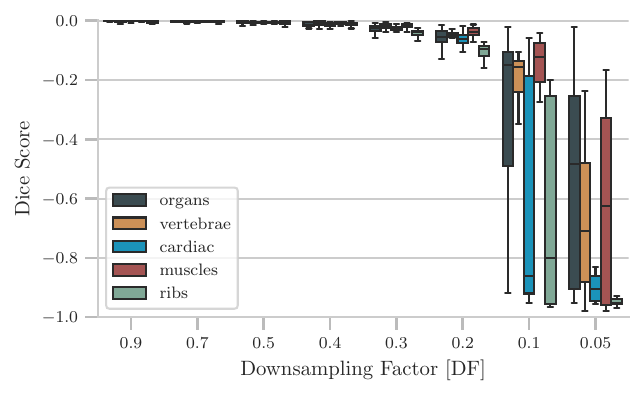}
        \caption{Fine-tuned 1.5mm TS model.}
        \label{fig:dice-diff-fine}
    \end{subfigure}
    \begin{subfigure}[b]{0.48\textwidth}
        \includegraphics[width=\textwidth]{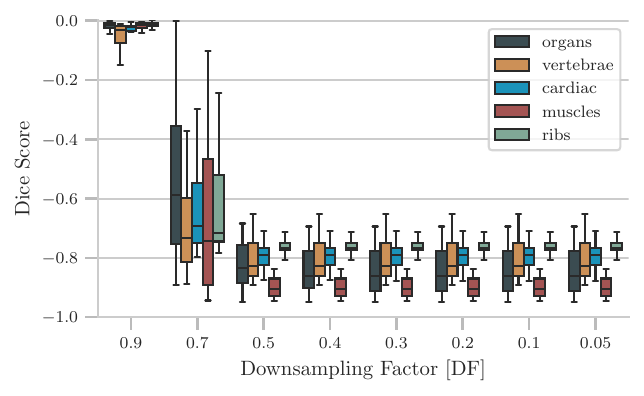}
        \caption{Zero-shot 3mm TS model.}
        \label{fig:dice-diff-zero-fast}
    \end{subfigure}
    \hfill
    \begin{subfigure}[b]{0.48\textwidth}
        \includegraphics[width=\textwidth]{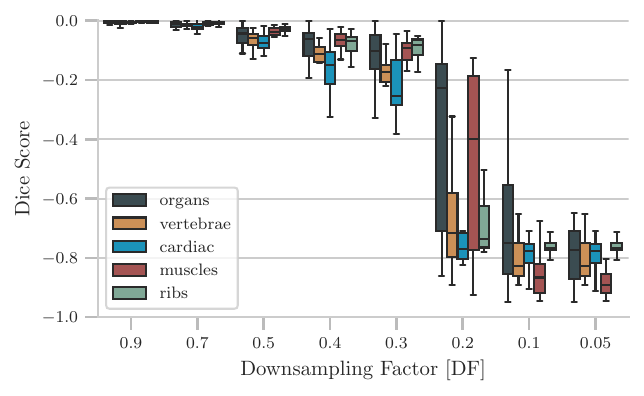}
        \caption{Fine-tuned 3mm TS model.}
        \label{fig:dice-diff-fine-fast}
    \end{subfigure}
    \caption{Difference in achieved Dice scores between the compressed and original model for each segmentation group (colors) across different downsampling factors (x-axis) and different models (subplots).}
    \label{fig:dice-diff}
\end{figure}
%

Figure~\ref{fig:dice-overview} illustrates the segmentation performance, measured as average Dice Score, of Tucker-decomposed convolution versus filter-based pruning.
Generally, the mean dice score declined with fewer model parameters, i.e., larger CRs, across all evaluated methods.
Fine-tuning the compressed model enhanced the performance of both compression techniques.
Tucker-decomposed models allowed higher CRs than pruning.
Notably, in the case of the 1.5mm model, the fine-tuned pruning model yielded lower dice scores against the zero-shot Tucker-decomposed model over all CRs.
Tucker decomposition with fine-tuning demonstrated substantial efficacy, maintaining metrics that matched the original TS model for a CR of $\approx 5$ with a Dice score of $0.92$ versus $0.93$ in the full TS.
A CR of $\approx 17$ with a total of $9.25$M parameters across the five-member ensemble still resulted in a Dice score of $0.84$.
However, the same compression potential was not observed in the fine-tuned 3mm variant, as parity to the TS occurred with a marginal CR of $\approx 2$, yielding a $9.25$M parameter model.
Figure~\ref{fig:dice-diff} displays the segmentation performance of Tucker-decomposed models, evaluated separately for each anatomical group.
The different groups exhibited different potential for compression.
For example, in the zero-shot 1.5mm model with a DF of $0.4$, the groups \textit{muscles} and \textit{organs} barely reached a maximal Dice discrepancy of $-0.2$, whereas the \textit{cardiac} group exhibited a partial collapse.
The fine-tuned 1.5mm model with a DF of $0.2$, demonstrated a slight decrease in performance across all groups, except for an outlier behavior in the \textit{ribs} group.
Similar, if not identical, results were observed for the NSD metric, available in Supplement~\ref{app:performance-nsd}.

To better understand the implications of model compression, we tested the hypothesis that the Tucker-decomposed models performed comparably to the full TS model for every individual class.
Tables containing the computed p-values are presented in Supplement~\ref{app:performance-test}.
The zero-shot 1.5mm model with a DF of $0.7$ enabled over $44\%$ less parameters with no significant difference in performance compared to TS.
Following fine-tuning, the 1.5mm model aligned with the TS in $96\%$ ($113/117$) of the classes, utilizing $88.17\%$ fewer parameters (DF of $0.3$), and remained competitive at a DF of $0.2$ for $82\%$ ($97/117$) of the classes, with a $94.06\%$ reduction in parameters.
However, specific classes such as \textit{lung\_upper\_lobe\_left}, \textit{rib\_right\_11}, \textit{rib\_right\_12} and \textit{subclavian\_artery\_right} exhibited instability in our compression experiments.
Significant degeneration with downsampling factors ranging from $0.4$ to $0.3$ was observed.

\paragraph{Practical Speedup.}
\label{par:pract-speedup}
In addition to the amount of parameter and FLOP reduction as well as the impact on segmentation performance, we evaluated the efficacy of our approach using different GPUs.
The average execution time in milliseconds for full- and half-precision is displayed in Figure~\ref{fig:gpu-bench} and reported in detail in Supplement~\ref{app:gpu-inference}.
The resulting speedup is shown in Table~\ref{tab:gpu-speedup}.
Consumer cards profited with speedups $> 2\times$, for example, the RTX 2070 with $2.57\times$ and the RTX 2080 Ti with $2.36\times$.
There was a substantial difference in performance gains between the groups of cost-intensive data-center (A6000, A100) and consumer cards.
For DFs like $0.05$, the A6000 and A100 had a speedup of $1.73\times$ and $1.43\times$.
It can be seen that employing the Tucker decomposition did not automatically result in a realized speedup, e.g. the DF of $0.9$ consistently had a speedup smaller than 1x, implying an actual deceleration compared to the original TS.
The minimal DF for achieving a speedup larger than 1x was highly dependent on the chosen GPU.
In the 32-bit regime, the GTX 1080 profited from the Tucker decomposition starting with a DF of $0.7$, whereas the A6000 required a DF of $0.1$.
Furthermore, the application of 16-bit precision for inference runtime was crucial.
While the general execution time shrinked across all evaluated models for GPUs other than the GTX series, the overall span of achieved speedup diminished as well.
The speedup factor stayed over 1x but did not breach the mark of 2x as demonstrated in the realms of full-precision, e.g. for the DF $0.2$ the speedup previously was $2.01\times$ but then was $1.33\times$ on the RTX 3060 with half-precision enabled.
\begin{figure}[t]
    \centering
    \includegraphics[width=\textwidth]{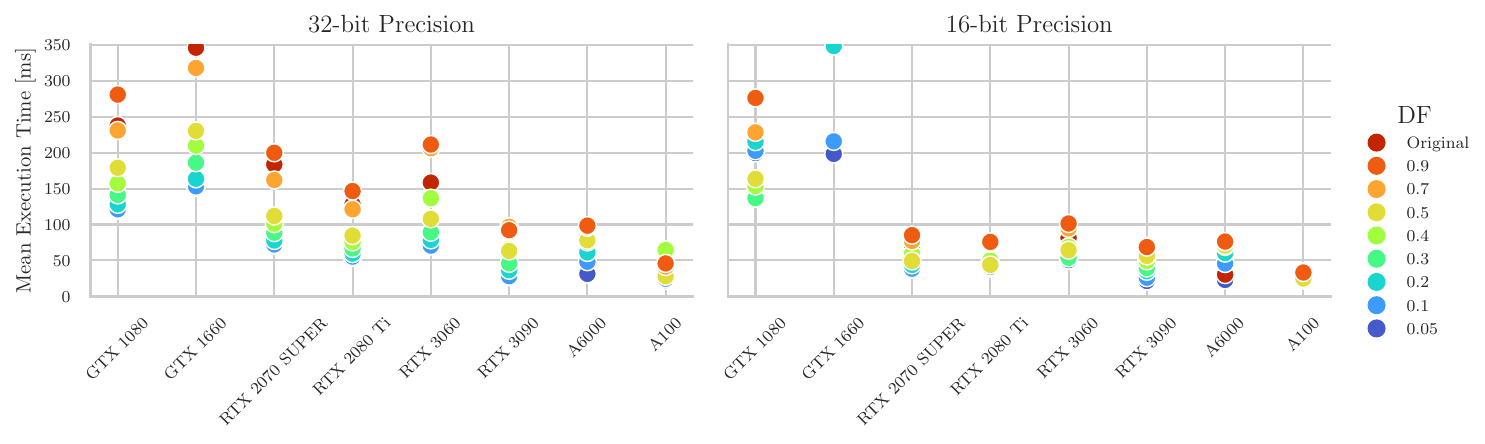}
    \caption{Mean execution times of 10 network forward passes in milliseconds (ms) for different GPUs and downsampling factors (DF) using the \textit{organ} 1.5mm model. The GPUs are ranked according to their computational capabilities. The {left} figure shows the performance using full 32-bit precision, while the {right} figure investigates mixed precision with 16-bit floats.}
    \label{fig:gpu-bench}
\end{figure}
\begin{table}[t]
\resizebox{\textwidth}{!}{
    \centering
\begin{tabular}{cc|cccccccc}
	\toprule
	 & \textbf{DF} & \textbf{GTX 1080} & \textbf{GTX 1660} & \textbf{RTX 2070} & \textbf{RTX 2080 Ti} & \textbf{RTX 3060} & \textbf{RTX 3090} & \textbf{A6000} & \textbf{A100} \\
\midrule
\multirow{9}{*}{\rotatebox{90}{\textbf{32-bit}}} & 0.9 & 0.85$\times$ & 0.90$\times$ & 0.92$\times$ & 0.87$\times$ & 0.75$\times$ & 0.65$\times$ & 0.55$\times$ & 0.70$\times$ \\
 & 0.7 & 1.03$\times$ & 1.09$\times$ & 1.13$\times$ & 1.05$\times$ & 0.77$\times$ & 0.62$\times$ & 0.56$\times$ & 0.77$\times$ \\
 & 0.5 & 1.33$\times$ & 1.50$\times$ & 1.64$\times$ & 1.51$\times$ & 1.47$\times$ & 0.95$\times$ & 0.70$\times$ & 1.15$\times$ \\
 & 0.4 & 1.51$\times$ & 1.65$\times$ & 1.82$\times$ & 1.69$\times$ & 1.16$\times$ & 0.92$\times$ & 0.64$\times$ & 0.50$\times$ \\
 & 0.3 & 1.68$\times$ & 1.86$\times$ & 2.07$\times$ & 1.92$\times$ & 1.77$\times$ & 1.31$\times$ & 0.72$\times$ & 0.50$\times$ \\
 & 0.2 & 1.86$\times$ & 2.11$\times$ & 2.36$\times$ & 2.15$\times$ & 2.01$\times$ & 1.65$\times$ & 0.89$\times$ & 0.79$\times$ \\
 & 0.1 & 1.96$\times$ & 2.26$\times$ & 2.54$\times$ & 2.31$\times$ & 2.24$\times$ & 2.14$\times$ & 1.12$\times$ & 1.31$\times$ \\
 & 0.05 & 1.98$\times$ & 2.29$\times$ & 2.57$\times$ & 2.36$\times$ & 2.26$\times$ & 2.21$\times$ & 1.73$\times$ & 1.43$\times$ \\
\midrule
\multirow{9}{*}{\rotatebox{90}{\textbf{16-bit}}} & 0.9 & 0.83$\times$ & 0.84$\times$ & 0.76$\times$ & 0.61$\times$ & 0.80$\times$ & 0.48$\times$ & 0.40$\times$ & 0.82$\times$ \\
 & 0.7 & 1.00$\times$ & 0.94$\times$ & 0.84$\times$ & 0.63$\times$ & 0.86$\times$ & 0.47$\times$ & 0.40$\times$ & 0.79$\times$ \\
 & 0.5 & 1.39$\times$ & 2.55$\times$ & 1.31$\times$ & 1.06$\times$ & 1.26$\times$ & 0.59$\times$ & 0.43$\times$ & 1.10$\times$ \\
 & 0.4 & 1.49$\times$ & 1.63$\times$ & 1.09$\times$ & 0.92$\times$ & 1.18$\times$ & 0.67$\times$ & 0.42$\times$ & 0.95$\times$ \\
 & 0.3 & 1.67$\times$ & 1.91$\times$ & 1.21$\times$ & 1.04$\times$ & 1.51$\times$ & 0.84$\times$ & 0.44$\times$ & 1.02$\times$ \\
 & 0.2 & 1.06$\times$ & 3.01$\times$ & 1.47$\times$ & 1.00$\times$ & 1.33$\times$ & 0.94$\times$ & 0.50$\times$ & 1.12$\times$ \\
 & 0.1 & 1.13$\times$ & 4.86$\times$ & 1.68$\times$ & 1.13$\times$ & 1.61$\times$ & 1.27$\times$ & 0.66$\times$ & 1.19$\times$ \\
 & 0.05 & 1.14$\times$ & 5.28$\times$ & 1.73$\times$ & 1.18$\times$ & 1.62$\times$ & 1.54$\times$ & 1.33$\times$ & 1.21$\times$ \\
\bottomrule
\end{tabular}}
\vspace{2mm}
\caption{Average speedup for the \textit{organ} 1.5mm model across different GPUs (columns) and downsampling factors (rows). The {upper} section shows the speedup using full 32-bit precision, while the {lower} section displays results for mixed precision with 16-bit floats .}
\label{tab:gpu-speedup}
\end{table}

\paragraph{Kernel Reconstruction Quality.}
As a general indicator of the Tucker approximation's quality, we reconstructed the weight tensor from the factorization and investigated how much variance is still being explained.
Supplement~\ref{app:kernel-rec} displays the amount of preserved variance for each layer in the 3mm and one 1.5mm model utilizing a grid of various core tensor dimensions.
We observed that the interaction of $T_I$ and $T_O$ is mostly symmetric, indicating an absence of bias towards a more substantial reduction in either input or output channels.
Noticeably, the transposed convolution layers for both models allowed extreme compression.
When comparing general compressibility, the 3mm model's layers generally showcased a larger lower bound on $T_I$ and $T_O$ than the 1.5mm model to preserve an adequate level of explained variance.

\section{Discussion}

In the present study, we investigated the feasibility of network compression for segmentation CNNs using Tucker-decomposed layers, using the widely used and publicly available TotalSegmentator model as a case example.
Our findings indicate that our approach allows for a compression of over $88$\% in parameters with no significant deterioration of segmentation performance in the majority of classes.
Further, the GPU architecture plays a crucial role in the realizable speed-up, where less-powerful hardware benefits more from our approach.

\paragraph{Related Work.} Another established method in weight factorization, CP-decomposition \cite{lebedev2015speeding}, factorizes each dimension of the target kernel.
In our scenario, this approach would necessitate the addition of five layers for each decomposed convolution.
We opted against this measure because, as previously discussed, extending the neural network sequence raises the minimum runtime lower bound.
Adding five layers rather than three reduces the performance gains on modern GPU hardware.
Furthermore, studies have reported instability issues with CP-decomposition \cite{de2008tensor, lebedev2015speeding, yong2016tucker}.
Other studies have explored the integration of weight quantization into CP-decomposition \cite{cherniuk2023quantization} and Tucker decomposition \cite{liu2024towards}.
Hierarchical Tucker decomposition was proposed to address the challenges in compressing recurrent neural networks \cite{yin2020compressing}.
Moreover, \cite{cheng2020novel} tackled the problem of optimal tensor rank selection using reinforcement learning, whereas \cite{idelbayev2020low} formulated this as a discrete-continuous optimization task.
\cite{bhalgaonkar2024model} highlighted the priority of accuracy over speed when using network compression in medical diagnostics.
Closely related to our domain, \cite{ashtari2021low} utilized a Tucker-inspired low-rank convolution kernel as a regularization strategy for the training of brain tumor segmentation models.
To the best of our knowledge, our study is the first to explore network compression with Tucker decomposition in medical image segmentation and its implications for clinical practice.

\paragraph{Practicability.} The results indicate that inference speedups did not match the theoretical FLOP reductions.
The GTX series cards, lacking tensor cores, do not benefit from performance gains using 16-bit precision due to the absence of specialized hardware for efficient accelerated computation.
For practitioners without memory constraints, we recommend avoiding using 16-bit precision on GTX series GPUs.

Another observation was that the execution time of the compressed model can be higher than the full original model.
While in theory, the number of FLOPs was reduced, the sequential nature of replacing one layer with a series of three layers came with additional overhead.
For most consumer cards, there was still a notable speedup with runtimes twice as fast as the original model.
High-performance data center GPUs, such as the NVIDIA A100, possess a sufficient number of CUDA cores to process the original models with a speed comparable to that of their compressed counterparts.

\paragraph{Segmentation Models.} We observed notable differences between the compressibility in terms of the DF of the single 3mm model with all labels and the five separate 1.5mm models, which was also evidenced by the rapidly decreasing explained variance in the respective layers. We argue that models with a larger task set, like the single 3mm model, require rich representations, i.e., a higher intrinsic rank, to handle the full range of segmentation classes.
This complexity makes it more difficult to compress models without sacrificing performance, which is visible in the lower viable DF.
In contrast, the specialized 1.5mm ``expert models'', focusing on smaller label sets, avoid the complex embedding required by the generalized 3mm model.
Due to its simplicity, this task-specific feature space enables more efficient compression.
However, comparing the absolute number of parameters per model might be misleading, as the 1.5mm consists of $5$ model weights, each having nearly twice the amount of parameters as the 3mm model.
On the other hand, dividing a problem into subtasks, using a specialist ensemble to obtain all predictions, and compressing the individual experts appears to be a viable strategy to achieve high performance at low cost.
This also allows for different compression ratios for different expert models, i.e., a rib model with DF 0.3 and an organ model with DF 0.4. 
Interestingly this effect seems to be reduced by finetuning the compressed models.
Here, the decline in model performance for both models started in the range between 2.5-5M parameters, which might indicate a lower bound of model parameters for the given task.

\paragraph{Limitations and Future Research.} Our study has several limitations.
With the nnU-Net Total Segmentator package, we focused on a single group of publicly available, open-source models.
While our analysis showed the advantages of decomposing 3D convolutions for this specific network architecture, further studies are required to examine the proposal's effects for other architectures or deriving new decompositions for, e.g., transformer-based networks.
In addition, our decomposition strategy handles every layer identically by using a relative DF.
Other layer-specific rank selection criteria like Variational Bayesian Matrix Factorization \cite{nakajima13a} could be further investigated to ensure optimal layer compression. 
Furthermore, while the selection of evaluated GPUs reflects the spectrum of currently available hardware resources, runtime speedups might further differ for other cards and series.

For some of the cards we tested, we were able to reduce the forward pass time by half without a significant performance decrease.
However, the model execution is only one building block in the full software stack of packages like TS.
Parts of the model's overall overhead boils down to I/O operations, i.e., loading and decoding image arrays, saving results, and further preprocessing, which is not addressed or affected by our proposed approach.
Making those operations more efficient could substantially accelerate the overall process, paving the way for more practicability.

\paragraph{Conclusion.} As the trend of scaling models continues, we advocate maintaining their accessibility to clinical practitioners without access to high-performance computing clusters.
In conclusion, we have presented a method for post-hoc compression of the TotalSegmentator tissue segmentation models.
Our results demonstrated that our compression method yields models with $88$\% fewer parameters while maintaining competitive segmentation performance.

\section*{Acknowledgments}

The authors gratefully acknowledge LMU Klinikum for providing computing resources on their Clinical Open Research Engine (CORE).
This work has been partially funded by the Deutsche Forschungsgemeinschaft (DFG, German Research Foundation) as part of BERD@NFDI - grant number 460037581.

\printbibliography
\clearpage

\clearpage

\setcounter{figure}{0}
\setcounter{table}{0} 

\appendix
\renewcommand\thefigure{\thesection.\arabic{figure}}
\renewcommand\thetable{\thesection.\arabic{table}}

\section*{Supplemental Material}

\section{Tucker Decomposition for 3D CNN Kernels}
\label{app:tucker-meth}
Given a volumetric input tensor $\mathcal{X} \in \mathbb{R}^{I \times H \times W \times D}$, where $I$ represents the number of input channels, and $H$, $W$, and $D$ are the height, width, and depth of the tensor, and a 3D convolution kernel $\mathcal{K} \in \mathbb{R}^{O \times I \times K_H \times K_W \times K_D}$, with $O$ denoting the number of output channels and $K_H$, $K_W$, and $K_D$ representing the kernel's dimensions, the convolution operation $\mathcal{X} * \mathcal{K}$ yields an output tensor $\mathcal{Y} \in \mathbb{R}^{O \times H' \times W' \times D'}$.

We can compute the size of the spatial output dimensions $H'$, $W'$, and $D'$ as follows:
\begin{equation}
    H' = \left\lfloor \frac{H - K_H + 2P_H}{S_H} + 1 \right\rfloor \eqendp
\end{equation}
$K_H$ is the spatial kernel size, $P_H$ is the padding, and $S_H$, is the kernel's stride.
The computation of $W'$ and $D'$ follows the above equation equivalently.

A single output voxel $\mathcal{Y}_{o,h',w',d'}$ of $\mathcal{Y}$ is computed as
\begin{equation} \label{eq:y_out}
    \mathcal{Y}_{o,h',w',d'} = \sum_{i=1}^{I} \sum_{j=1}^{K_H} \sum_{k=1}^{K_W} \sum_{m=1}^{K_D} \mathcal{K}_{o,i,j,k,m} \; \mathcal{X}_{i,h_j,w_k,d_m} \eqendc
\end{equation}
where $h_j = (h' - 1) S_h + j - P_h$ with $S_h$ being the stride and $P_h$ the padding along the $H$ axis.
The variables $w_k$ and $d_m$ can be characterized as analogous extensions of $h_j$ within the dimensions $W$ and $D$, respectively.

The Tucker decomposition \cite{tucker1966some} forms a generalized version of a singular value decomposition (SVD).
In this paper, the target of the decomposition is the convolutional kernel $\mathcal{K}$.
Following \cite{tucker1966some}, a $n$-order tensor can be decomposed into a core tensor and a set of $n$ factor matrices.
The original tensor $\mathcal{K}$ can be recovered via the $i$-mode product of the $i$-th factor matrix with the core tensor.
Instead of fully decomposing $\mathcal{K}$, \cite{yong2016tucker} proposed a partial decomposition of only the axes incorporating the channel dimensions.
As the kernel's spatial dimensions are rather small by nature, a full decomposition would not result in any substantial performance gain.
Thus, the partial Tucker decomposition results in a core tensor $\mathcal{C} \in \mathbb{R}^{R_o \times R_i \times K_H \times K_W \times K_D}$ and factor matrices $U^{O} \in \mathbb{R}^{O \times R_o}$ and $U^{I} \in \mathbb{R}^{I \times R_i}$, where
\begin{equation} \label{eq:tucker-product}
    \mathcal{K} \approx \mathcal{C} \times_1 U^{O} \times_2 U^{I} \eqendc
\end{equation}
with $R_o$ and $R_i$ being the internal ranks of the core tensor, which corresponds to the amount of compression for the output channels $O$ and input channels $I$, respectively.
A single kernel element in this decomposition is therefore given as 
\begin{equation}
    \mathcal{K}_{o,i,j,k,m} = \sum_{r_i=1}^{R_i} \sum_{r_o=1}^{R_o} \mathcal{C}_{r_o,r_i,j,k,m} \; U^{O}_{o, r_o} \; U^{I}_{i, r_i} \eqendp
\end{equation}

By substituting \eqref{eq:tucker-product} into the computation of the convolutions output $\mathcal{Y}$ in Equation~(\ref{eq:y_out}) and reordering terms, we get:

\begin{equation} \label{eq:tucker-full}
    \mathcal{Y}_{o,h',w',d'} = \color{cb-blue} \sum_{r_o=1}^{R_o} U^{O}_{o, r_o} \color{cb-green-sea}\sum_{r_i=1}^{R_i} \sum_{j=1}^{K_H} \sum_{k=1}^{K_W} \sum_{m=1}^{K_D} \mathcal{C}_{r_o,r_i,j,k,m} \color{cb-burgundy} \sum_{i=1}^{I} U^{I}_{i, r_i} \; \mathcal{X}_{i,h_j,w_k,d_m} \eqendp
\end{equation}

This practically translates into a sequence of separate convolutions:
\textcolor{cb-burgundy}{(i)} a pointwise convolution projecting the input channels on the rank dimension $R_i$, \textcolor{cb-green-sea}{(ii)} a conventional convolution operating with the original spatial kernel dimensions projecting the channels from $R_i$ to $R_o$, and \textcolor{cb-blue}{(iii)} a pointwise convolution utilized as projection from $R_o$ channels to the aspired output channel size $O$.
A schematic overview of Equation~\ref{eq:tucker-full} is presented in Figure~\ref{fig:tucker-high}.
Ultimately, for $R_o < O$ and $R_i < I$ this series of three convolutions has substantially fewer arithmetic operations than the original convolution.

\newpage
\section{Kernel Reconstruction Error}
\label{app:kernel-rec}

\begin{figure}[H]
    \centering
    \includegraphics[width=\textwidth]{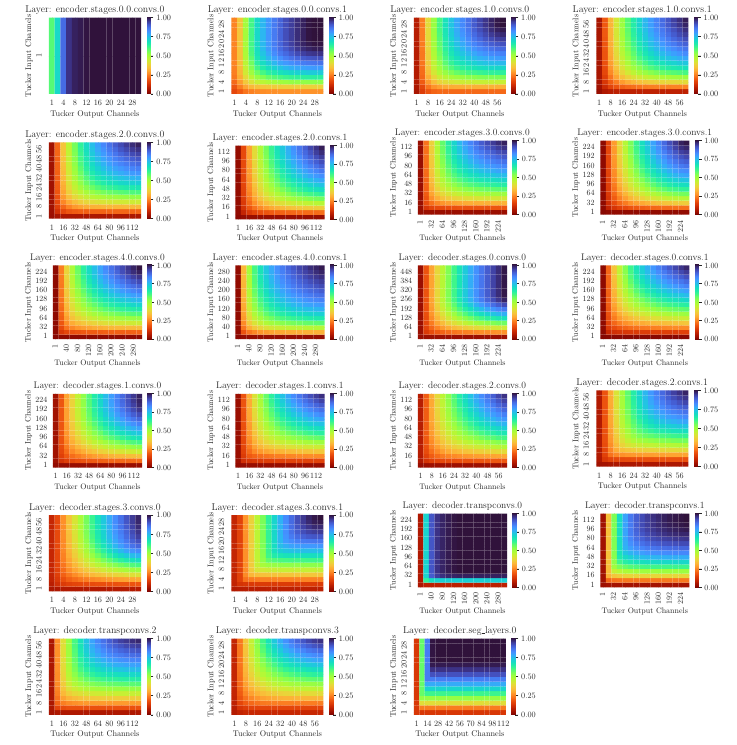}
    \caption{Explained variance over different Tucker decomposition ranks for the 3mm TS model.}
    \label{app:kernel3mm}
\end{figure}

\begin{figure}[H]
    \centering
    \includegraphics[width=\textwidth]{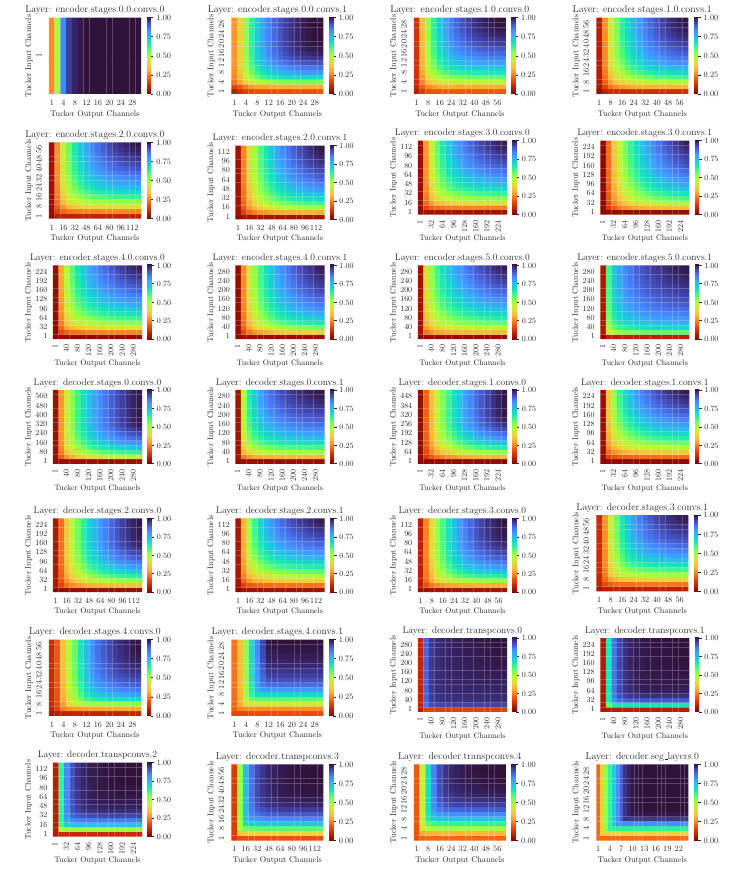}
    \caption{Explained variance over different Tucker decomposition ranks for the 1.5mm TS organ model.}
    \label{app:kernel15mm}
\end{figure}

In the following, we analyze the quality of the kernel approximation via Tucker decomposition.
For a kernel $\mathcal{K}$ and its reconstruction $\hat{\mathcal{K}}$ following Equation~\ref{eq:tucker-product}, we compute
\begin{equation}
    1 - \frac{||\mathcal{K} - \hat{\mathcal{K}}||^2}{||\mathcal{K}||^2}
\end{equation}
as a measure of the explained variance of the original kernel by the decomposed kernel.
In the following, every layer of the 1.5mm organ model \ref{app:kernel15mm} and the 3mm model \ref{app:kernel3mm} is decomposed with varying internal Tucker ranks, showcasing the complex information landscape contained in each kernel.
Each cell in the heatmap represents a unique setting for Tucker compression of the core tensor, with input and output channels being chosen independently.
The blue color denotes a high level of recovered variance, while red indicates a low amount.
\newpage
\section{Segmentation Performance: Normalized Surface Distance}
\label{app:performance-nsd}

Alongside the figures in the main document that display the model's segmentation performance using the Dice score, the accompanying plots demonstrate the behavior for the NSD metric.
\begin{figure}[H]
    \centering
    \begin{subfigure}[b]{0.48\textwidth}
        \includegraphics[width=\textwidth]{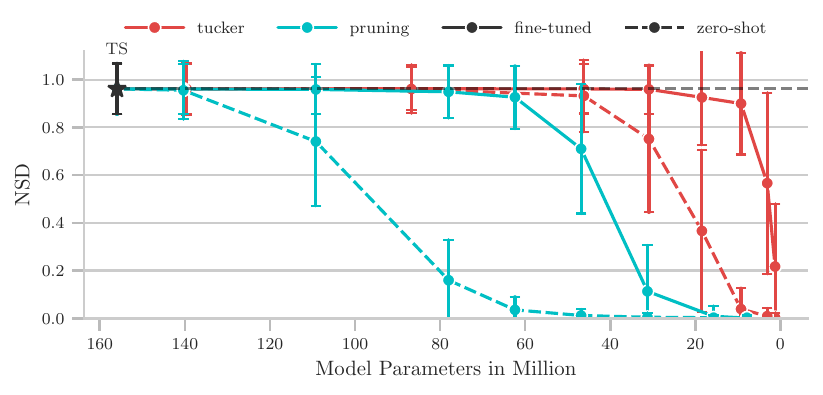}
        \caption{1.5mm model.}
        \label{fig:nsd-overview-slow}
    \end{subfigure}
    \begin{subfigure}[b]{0.48\textwidth}
        \includegraphics[width=\textwidth]{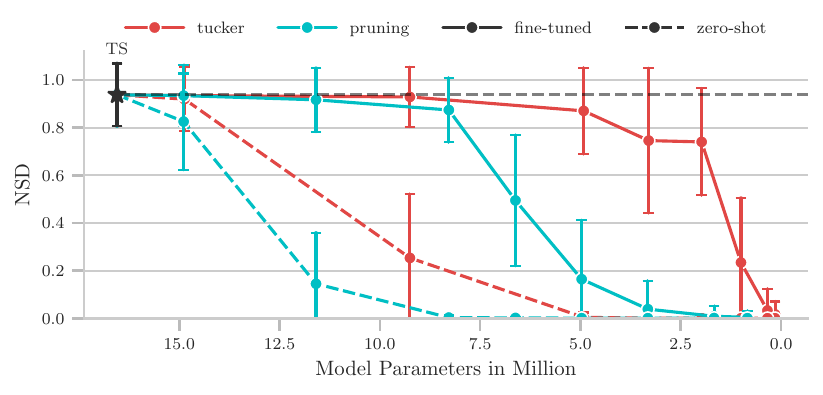}
        \caption{3mm model.}
        \label{fig:nsd-overview-fast}
    \end{subfigure}
    \caption{
    NSD aggregated over all classes for the TS test set using the 1.5mm (left) and 3mm (right) TS models.
    The performance of the original TS model is compared against the Tucker decomposition-based approach (red) and filter pruning (blue). Both compression methods are evaluated with (solid line) and without (dashed line) additional fine-tuning.
    Error bars represent the standard deviation across different classes.
    }
    \label{fig:nsd-overview}
\end{figure}
\begin{figure}[H]
    \centering
    \begin{subfigure}[b]{0.48\textwidth}
        \includegraphics[width=\textwidth]{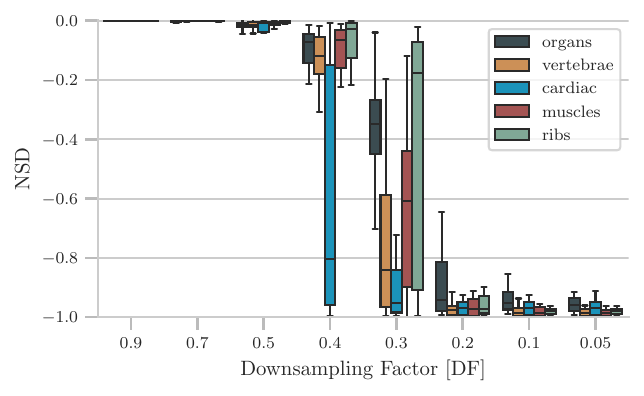}
        \caption{Zero-shot 1.5mm TS model.}
        \label{fig:nsd-diff-zero}
    \end{subfigure}
    \hfill
    \begin{subfigure}[b]{0.48\textwidth}
        \includegraphics[width=\textwidth]{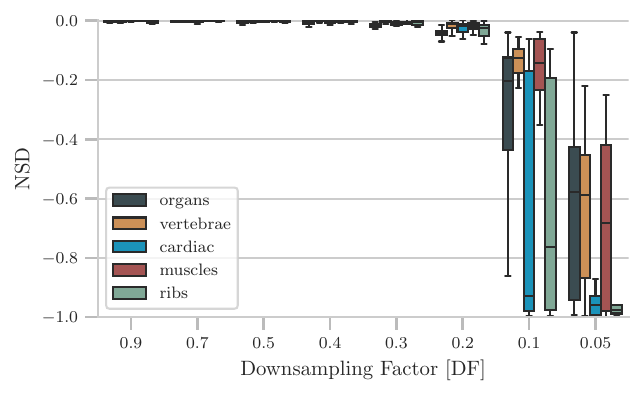}
        \caption{Fine-tuned 1.5mm TS model.}
        \label{fig:nsd-diff-fine}
    \end{subfigure}
    \begin{subfigure}[b]{0.48\textwidth}
        \includegraphics[width=\textwidth]{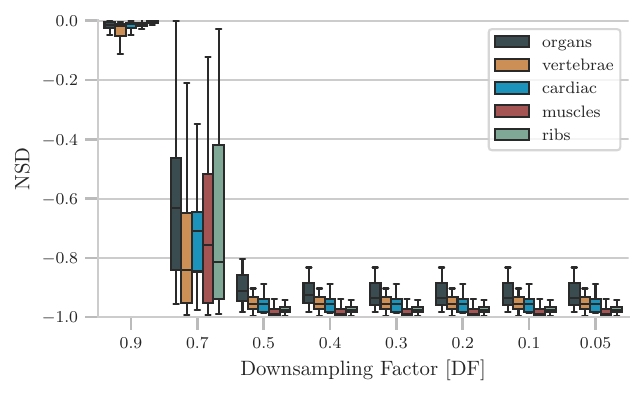}
        \caption{Zero-shot 3mm TS model.}
        \label{fig:nsd-diff-zero-fast}
    \end{subfigure}
    \hfill
    \begin{subfigure}[b]{0.48\textwidth}
        \includegraphics[width=\textwidth]{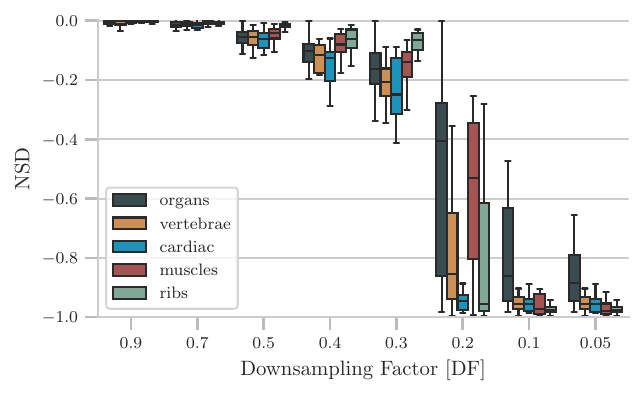}
        \caption{Fine-tuned 3mm TS model.}
        \label{fig:nsd-diff-fine-fast}
    \end{subfigure}
    \caption{Difference in achieved NSD between the compressed and original model for each segmentation group (colors) across different downsampling factors (x-axis) and different models (subplots).}
    \label{fig:nsd-diff}
\end{figure}

\newpage
\section{Segmentation Performance Testing}
\label{app:performance-test}

In the following, we analyze the segmentation performance across the evaluated models in comparison to the original full TotalSegmentator models.
The p-values were computed concerning the resulting NSD metrics.

\subsection{1.5mm Model - Zero-Shot}
\begin{table}[htbp]
    \centering
    \resizebox{\textwidth}{!}{
    \input{tables/pval_zero_1}
    }
    \vspace{2mm}
    \caption{p-values evaluating the hypothesis whether the Tucker-decomposed zero-shot 1.5mm model performs worse than the original TS model across different downsampling factors and all classes of the groups \textit{cardiac}, \textit{muscles}, and \textit{organs}. Red-colored cells indicate the significantly worse performance of the compressed model, whereas green-colored cells indicate that the null hypothesis of equal performance cannot be rejected.}
    \label{app:pval-zero-1}
\end{table}
\begin{table}[htbp]
    \centering
    \resizebox{\textwidth}{!}{
    \input{tables/pval_zero_2}
    }
    \vspace{2mm}
    \caption{p-values for evaluating whether the Tucker-decomposed zero-shot 1.5mm model performs worse than the original TS model across different downsampling factors and all classes of the groups \textit{ribs} and \textit{vertebrae}. Red-colored cells indicate the significantly worse performance of the compressed model, whereas green-colored cells indicate that the null hypothesis of equal performance cannot be rejected.}
    \label{app:pval-zero-2}
\end{table}

\clearpage

\subsection{1.5mm Model - Fine-Tuning}

\begin{table}[htbp]
    \centering
    \resizebox{\textwidth}{!}{
    \input{tables/pval_table_1}
    }
    \vspace{2mm}
    \caption{p-values evaluating the hypothesis whether the Tucker-decomposed fine-tuned 1.5mm model performs worse than the original TS model across different downsampling factors and all classes of the groups \textit{cardiac}, \textit{muscles}, and \textit{organs}. Red-colored cells indicate the significantly worse performance of the compressed model, whereas green-colored cells indicate that the null hypothesis of equal performance cannot be rejected.}
    \label{tab:pval-1}
\end{table}
\begin{table}[htbp]
    \centering
    \resizebox{\textwidth}{!}{
    \input{tables/pval_table_2}
    }
    \vspace{2mm}
    \caption{p-values for evaluating whether the Tucker-decomposed fine-tuned 1.5mm model performs worse than the original TS model across different downsampling factors and all classes of the groups \textit{ribs} and \textit{vertebrae}. Red-colored cells indicate the significantly worse performance of the compressed model, whereas green-colored cells indicate that the null hypothesis of equal performance cannot be rejected.}
    \label{tab:pval-2}
\end{table}

\clearpage

\subsection{3mm Model - Zero-Shot}
\begin{table}[htbp]
    \centering
    \resizebox{\textwidth}{!}{
    \input{tables/pval_zero_fast_1}
    }
    \vspace{2mm}
    \caption{p-values evaluating the hypothesis whether the Tucker-decomposed zero-shot 3mm model performs worse than the original TS model across different downsampling factors and all classes of the groups \textit{cardiac}, \textit{muscles}, and \textit{organs}. Red-colored cells indicate the significantly worse performance of the compressed model, whereas green-colored cells indicate that the null hypothesis of equal performance cannot be rejected.}
    \label{app:pval-zero-fast-1}
\end{table}
\begin{table}[htbp]
    \centering
    \resizebox{\textwidth}{!}{
    \input{tables/pval_zero_fast_2}
    }
    \vspace{2mm}
    \caption{p-values for evaluating whether the Tucker-decomposed zero-shot 1.5mm model performs worse than the original TS model across different downsampling factors and all classes of the groups \textit{ribs} and \textit{vertebrae}. Red-colored cells indicate the significantly worse performance of the compressed model, whereas green-colored cells indicate that the null hypothesis of equal performance cannot be rejected.}

    \label{app:pval-zero-fast-2}
\end{table}

\clearpage

\subsection{3mm Model - Fine-Tuning}
\begin{table}[htbp]
    \centering
    \resizebox{\textwidth}{!}{
    \input{tables/pval_fine_fast_1}
    }
    \vspace{2mm}
    \caption{p-values evaluating the hypothesis whether the Tucker-decomposed fine-tuned 3mm model performs worse than the original TS model across different downsampling factors and all classes of the groups \textit{cardiac}, \textit{muscles}, and \textit{organs}. Red-colored cells indicate the significantly worse performance of the compressed model, whereas green-colored cells indicate that the null hypothesis of equal performance cannot be rejected.}
    \label{app:pval-fine-fast-1}
\end{table}
\begin{table}[htbp]
    \centering
    \resizebox{\textwidth}{!}{
    \input{tables/pval_fine_fast_2}
    }
    \vspace{2mm}
    \caption{p-values for evaluating whether the Tucker-decomposed fine-tuned 3mm model performs worse than the original TS model across different downsampling factors and all classes of the groups \textit{ribs} and \textit{vertebrae}. Red-colored cells indicate the significantly worse performance of the compressed model, whereas green-colored cells indicate that the null hypothesis of equal performance cannot be rejected.}

    \label{app:pval-fine-fast-2}
\end{table}

\clearpage

\section{GPU Inference Speed}
\label{app:gpu-inference}

The below table supplements Paragraph~\ref{par:pract-speedup} in the results section and displays the full measurements times responsible for Figure~\ref{fig:gpu-bench}.
\begin{table}[htbp]
\resizebox{\textwidth}{!}{
    \centering
\begin{tabular}{cc|rrrrrrrr}
	\toprule
	 & \textbf{DF} & \textbf{GTX 1080} & \textbf{GTX 1660} & \textbf{RTX 2070} & \textbf{RTX 2080 Ti} & \textbf{RTX 3060} & \textbf{RTX 3090} & \textbf{A6000} & \textbf{A100} \\
	\midrule
 \multirow{9}{*}{\rotatebox{90}{\textbf{32-bit}}} & Original & 237.71$\pm$1.31 & 345.89$\pm$0.27 & 183.46$\pm$0.33 & 127.82$\pm$0.51 & 158.31$\pm$0.23 & 60.13$\pm$0.01 & 54.25$\pm$0.13 & 32.13$\pm$0.04 \\
 & 0.9 & 280.73$\pm$0.99 & 385.09$\pm$0.35 & 199.77$\pm$0.41 & 146.50$\pm$0.24 & 211.25$\pm$0.36 & 92.10$\pm$0.19 & 98.40$\pm$0.08 & 45.82$\pm$0.01 \\
 & 0.7 & 230.86$\pm$0.86 & 317.79$\pm$0.54 & 161.92$\pm$0.20 & 121.17$\pm$0.69 & 206.28$\pm$0.24 & 96.83$\pm$0.20 & 96.24$\pm$0.06 & 41.56$\pm$0.05 \\
 & 0.5 & 178.95$\pm$0.44 & 230.23$\pm$0.40 & 111.86$\pm$0.13 & 84.60$\pm$0.45 & 107.86$\pm$1.25 & 63.18$\pm$0.04 & 77.99$\pm$0.02 & 27.98$\pm$0.01 \\
 & 0.4 & 157.26$\pm$0.19 & 209.77$\pm$0.29 & 101.05$\pm$0.22 & 75.80$\pm$0.22 & 136.60$\pm$1.92 & 65.44$\pm$0.20 & 84.36$\pm$0.04 & 64.59$\pm$0.02 \\
 & 0.3 & 141.13$\pm$0.20 & 185.98$\pm$0.28 & 88.68$\pm$0.13 & 66.61$\pm$0.17 & 89.24$\pm$0.30 & 45.96$\pm$0.07 & 75.79$\pm$0.03 & 63.84$\pm$0.01 \\
 & 0.2 & 128.12$\pm$0.06 & 163.56$\pm$0.28 & 77.76$\pm$0.08 & 59.56$\pm$0.07 & 78.61$\pm$0.34 & 36.38$\pm$0.06 & 61.14$\pm$0.01 & 40.42$\pm$0.01 \\
 & 0.1 & 121.17$\pm$0.07 & 153.08$\pm$0.28 & 72.25$\pm$0.11 & 55.30$\pm$0.12 & 70.64$\pm$1.70 & 28.14$\pm$0.04 & 48.23$\pm$0.02 & 24.55$\pm$0.01 \\
 & 0.05 & 119.92$\pm$0.05 & 151.35$\pm$0.33 & 71.35$\pm$0.13 & 54.17$\pm$0.13 & 70.10$\pm$1.53 & 27.26$\pm$0.02 & 31.32$\pm$0.01 & 22.54$\pm$0.01 \\
	\midrule
 \multirow{9}{*}{\rotatebox{90}{\textbf{16-bit}}} & Original & 228.10$\pm$1.28 & 1047.08$\pm$0.60 & 64.51$\pm$0.12 & 46.31$\pm$0.09 & 81.30$\pm$1.74 & 32.98$\pm$0.04 & 30.35$\pm$0.08 & 27.26$\pm$1.43 \\
 & 0.9 & 276.06$\pm$0.71 & 1249.18$\pm$0.60 & 85.15$\pm$0.24 & 75.77$\pm$0.18 & 101.38$\pm$0.26 & 68.61$\pm$0.04 & 76.23$\pm$0.06 & 33.17$\pm$0.05 \\
 & 0.7 & 228.21$\pm$0.74 & 1114.11$\pm$0.59 & 77.16$\pm$0.15 & 73.91$\pm$0.13 & 94.59$\pm$0.15 & 69.64$\pm$0.06 & 75.93$\pm$0.14 & 34.52$\pm$0.05 \\
 & 0.5 & 163.54$\pm$0.36 & 411.42$\pm$0.32 & 49.26$\pm$0.12 & 43.75$\pm$0.12 & 64.30$\pm$1.94 & 56.31$\pm$0.06 & 70.69$\pm$0.06 & 24.80$\pm$0.02 \\
 & 0.4 & 153.53$\pm$0.14 & 643.12$\pm$0.18 & 59.24$\pm$0.13 & 50.13$\pm$0.15 & 68.97$\pm$3.07 & 49.42$\pm$0.08 & 71.65$\pm$0.04 & 28.56$\pm$0.07 \\
 & 0.3 & 136.67$\pm$0.09 & 547.10$\pm$0.18 & 53.38$\pm$0.08 & 44.52$\pm$0.14 & 53.69$\pm$1.96 & 39.40$\pm$0.05 & 69.50$\pm$0.03 & 26.60$\pm$0.02 \\
 & 0.2 & 215.26$\pm$0.21 & 348.28$\pm$0.32 & 43.85$\pm$0.09 & 46.38$\pm$0.08 & 61.33$\pm$0.69 & 34.95$\pm$0.12 & 60.22$\pm$0.04 & 24.24$\pm$0.01 \\
 & 0.1 & 202.55$\pm$0.17 & 215.65$\pm$0.34 & 38.43$\pm$0.04 & 41.01$\pm$0.08 & 50.44$\pm$0.07 & 26.00$\pm$0.06 & 45.70$\pm$0.04 & 22.93$\pm$0.01 \\
 & 0.05 & 199.55$\pm$0.17 & 198.22$\pm$0.34 & 37.35$\pm$0.08 & 39.36$\pm$0.07 & 50.25$\pm$0.31 & 21.45$\pm$0.07 & 22.89$\pm$0.03 & 22.60$\pm$0.01 \\
\bottomrule
\end{tabular}}
\vspace{2mm}
\caption{Mean execution of 10 network forward passes in milliseconds (ms) over a variety of GPUs and downsampling factors (DF) for one 1.5mm model. The \textbf{upper} section shows the performance using full 32-bit precision, while the \textbf{lower} section displays mixed precision with 16-bit floats enabled.}
\label{tab:gpu-bench}
\end{table}

\end{document}

%% file: tables/pval_zero_1.tex
\newcolumntype{C}[1]{>{\centering\arraybackslash}p{#1}}

\begin{tabular}{llC{15mm}C{15mm}C{15mm}C{15mm}C{15mm}C{15mm}C{15mm}C{15mm}}
\toprule
& & \multicolumn{8}{c}{\textbf{Downsampling Factor [DF]}} \\
\cmidrule{3-10}
 & \textbf{Class Label} & \textbf{0.9} & \textbf{0.7} & \textbf{0.5} & \textbf{0.4} & \textbf{0.3} & \textbf{0.2} & \textbf{0.1} & \textbf{0.05} \\
\midrule
cardiac & aorta & \cellcolor{custom-green!50} $1.0$ & \cellcolor{custom-green!50} $1.0$ & \cellcolor{custom-green!50} $1.0$ & \cellcolor{custom-red!80} $\num{2E-07}$ & \cellcolor{custom-red!80} $<\num{2E-16}$ & \cellcolor{custom-red!80} $<\num{2E-16}$ & \cellcolor{custom-red!80} $<\num{2E-16}$ & \cellcolor{custom-red!80} $<\num{2E-16}$ \\
cardiac & atrial\_appendage\_left & \cellcolor{custom-green!50} $1.0$ & \cellcolor{custom-green!50} $1.0$ & \cellcolor{custom-green!50} $1.0$ & \cellcolor{custom-red!80} $<\num{2E-16}$ & \cellcolor{custom-red!80} $<\num{2E-16}$ & \cellcolor{custom-red!80} $<\num{2E-16}$ & \cellcolor{custom-red!80} $<\num{2E-16}$ & \cellcolor{custom-red!80} $<\num{2E-16}$ \\
cardiac & brachiocephalic\_trunk & \cellcolor{custom-green!50} $1.0$ & \cellcolor{custom-green!50} $1.0$ & \cellcolor{custom-green!50} $1.0$ & \cellcolor{custom-green!50} $1.0$ & \cellcolor{custom-green!50} $1.0$ & \cellcolor{custom-red!80} $<\num{2E-16}$ & \cellcolor{custom-red!80} $<\num{2E-16}$ & \cellcolor{custom-red!80} $<\num{2E-16}$ \\
cardiac & brachiocephalic\_vein\_left & \cellcolor{custom-green!50} $1.0$ & \cellcolor{custom-green!50} $1.0$ & \cellcolor{custom-green!50} $1.0$ & \cellcolor{custom-green!50} $1.0$ & \cellcolor{custom-red!80} $<\num{2E-16}$ & \cellcolor{custom-red!80} $<\num{2E-16}$ & \cellcolor{custom-red!80} $<\num{2E-16}$ & \cellcolor{custom-red!80} $<\num{2E-16}$ \\
\bottomrule
\end{tabular}

%% file: tables/pval_zero_2.tex
\newcolumntype{C}[1]{>{\centering\arraybackslash}p{#1}}

\begin{tabular}{llC{15mm}C{15mm}C{15mm}C{15mm}C{15mm}C{15mm}C{15mm}C{15mm}}
\toprule
& & \multicolumn{8}{c}{\textbf{Downsampling Factor [DF]}} \\
\cmidrule{3-10}
 & \textbf{Class Label} & \textbf{0.9} & \textbf{0.7} & \textbf{0.5} & \textbf{0.4} & \textbf{0.3} & \textbf{0.2} & \textbf{0.1} & \textbf{0.05} \\
\midrule
ribs & costal\_cartilages & \cellcolor{custom-green!50} $1.0$ & \cellcolor{custom-green!50} $1.0$ & \cellcolor{custom-green!50} $1.0$ & \cellcolor{custom-green!50} $\num{4E-01}$ & \cellcolor{custom-red!80} $\num{3E-10}$ & \cellcolor{custom-red!80} $<\num{2E-16}$ & \cellcolor{custom-red!80} $<\num{2E-16}$ & \cellcolor{custom-red!80} $<\num{2E-16}$ \\
ribs & rib\_left\_1 & \cellcolor{custom-green!50} $1.0$ & \cellcolor{custom-green!50} $1.0$ & \cellcolor{custom-green!50} $1.0$ & \cellcolor{custom-green!50} $1.0$ & \cellcolor{custom-red!80} $<\num{2E-16}$ & \cellcolor{custom-red!80} $<\num{2E-16}$ & \cellcolor{custom-red!80} $<\num{2E-16}$ & \cellcolor{custom-red!80} $<\num{2E-16}$ \\
ribs & rib\_left\_10 & \cellcolor{custom-green!50} $1.0$ & \cellcolor{custom-green!50} $1.0$ & \cellcolor{custom-green!50} $1.0$ & \cellcolor{custom-green!50} $\num{4E-01}$ & \cellcolor{custom-red!80} $<\num{2E-16}$ & \cellcolor{custom-red!80} $<\num{2E-16}$ & \cellcolor{custom-red!80} $<\num{2E-16}$ & \cellcolor{custom-red!80} $<\num{2E-16}$ \\
ribs & rib\_left\_11 & \cellcolor{custom-green!50} $1.0$ & \cellcolor{custom-green!50} $1.0$ & \cellcolor{custom-green!50} $1.0$ & \cellcolor{custom-green!50} $1.0$ & \cellcolor{custom-green!50} $\num{8E-01}$ & \cellcolor{custom-red!80} $<\num{2E-16}$ & \cellcolor{custom-red!80} $<\num{2E-16}$ & \cellcolor{custom-red!80} $<\num{2E-16}$ \\
ribs & rib\_left\_12 & \cellcolor{custom-green!50} $1.0$ & \cellcolor{custom-green!50} $1.0$ & \cellcolor{custom-green!50} $1.0$ & \cellcolor{custom-red!80} $\num{1E-11}$ & \cellcolor{custom-red!80} $<\num{2E-16}$ & \cellcolor{custom-red!80} $<\num{2E-16}$ & \cellcolor{custom-red!80} $<\num{2E-16}$ & \cellcolor{custom-red!80} $<\num{2E-16}$ \\
ribs & rib\_left\_2 & \cellcolor{custom-green!50} $1.0$ & \cellcolor{custom-green!50} $1.0$ & \cellcolor{custom-green!50} $1.0$ & \cellcolor{custom-green!50} $1.0$ & \cellcolor{custom-red!80} $\num{2E-04}$ & \cellcolor{custom-red!80} $<\num{2E-16}$ & \cellcolor{custom-red!80} $<\num{2E-16}$ & \cellcolor{custom-red!80} $<\num{2E-16}$ \\
ribs & rib\_left\_3 & \cellcolor{custom-green!50} $1.0$ & \cellcolor{custom-green!50} $1.0$ & \cellcolor{custom-green!50} $1.0$ & \cellcolor{custom-green!50} $1.0$ & \cellcolor{custom-red!80} $\num{1E-05}$ & \cellcolor{custom-red!80} $<\num{2E-16}$ & \cellcolor{custom-red!80} $<\num{2E-16}$ & \cellcolor{custom-red!80} $<\num{2E-16}$ \\
ribs & rib\_left\_4 & \cellcolor{custom-green!50} $1.0$ & \cellcolor{custom-green!50} $1.0$ & \cellcolor{custom-green!50} $1.0$ & \cellcolor{custom-green!50} $1.0$ & \cellcolor{custom-red!80} $\num{2E-02}$ & \cellcolor{custom-red!80} $<\num{2E-16}$ & \cellcolor{custom-red!80} $<\num{2E-16}$ & \cellcolor{custom-red!80} $<\num{2E-16}$ \\
ribs & rib\_left\_5 & \cellcolor{custom-green!50} $1.0$ & \cellcolor{custom-green!50} $1.0$ & \cellcolor{custom-green!50} $1.0$ & \cellcolor{custom-green!50} $1.0$ & \cellcolor{custom-green!50} $\num{2E-01}$ & \cellcolor{custom-red!80} $<\num{2E-16}$ & \cellcolor{custom-red!80} $<\num{2E-16}$ & \cellcolor{custom-red!80} $<\num{2E-16}$ \\
ribs & rib\_left\_6 & \cellcolor{custom-green!50} $1.0$ & \cellcolor{custom-green!50} $1.0$ & \cellcolor{custom-green!50} $1.0$ & \cellcolor{custom-green!50} $1.0$ & \cellcolor{custom-red!80} $\num{2E-03}$ & \cellcolor{custom-red!80} $<\num{2E-16}$ & \cellcolor{custom-red!80} $<\num{2E-16}$ & \cellcolor{custom-red!80} $<\num{2E-16}$ \\
ribs & rib\_left\_7 & \cellcolor{custom-green!50} $1.0$ & \cellcolor{custom-green!50} $1.0$ & \cellcolor{custom-green!50} $1.0$ & \cellcolor{custom-green!50} $\num{8E-02}$ & \cellcolor{custom-red!80} $<\num{2E-16}$ & \cellcolor{custom-red!80} $<\num{2E-16}$ & \cellcolor{custom-red!80} $<\num{2E-16}$ & \cellcolor{custom-red!80} $<\num{2E-16}$ \\
ribs & rib\_left\_8 & \cellcolor{custom-green!50} $1.0$ & \cellcolor{custom-green!50} $1.0$ & \cellcolor{custom-green!50} $\num{1E-01}$ & \cellcolor{custom-red!80} $\num{3E-11}$ & \cellcolor{custom-red!80} $<\num{2E-16}$ & \cellcolor{custom-red!80} $<\num{2E-16}$ & \cellcolor{custom-red!80} $<\num{2E-16}$ & \cellcolor{custom-red!80} $<\num{2E-16}$ \\
ribs & rib\_left\_9 & \cellcolor{custom-green!50} $1.0$ & \cellcolor{custom-green!50} $1.0$ & \cellcolor{custom-green!50} $1.0$ & \cellcolor{custom-green!50} $1.0$ & \cellcolor{custom-red!80} $\num{1E-02}$ & \cellcolor{custom-red!80} $<\num{2E-16}$ & \cellcolor{custom-red!80} $<\num{2E-16}$ & \cellcolor{custom-red!80} $<\num{2E-16}$ \\
ribs & rib\_right\_1 & \cellcolor{custom-green!50} $1.0$ & \cellcolor{custom-green!50} $1.0$ & \cellcolor{custom-green!50} $1.0$ & \cellcolor{custom-green!50} $1.0$ & \cellcolor{custom-red!80} $<\num{2E-16}$ & \cellcolor{custom-red!80} $<\num{2E-16}$ & \cellcolor{custom-red!80} $<\num{2E-16}$ & \cellcolor{custom-red!80} $<\num{2E-16}$ \\
ribs & rib\_right\_10 & \cellcolor{custom-green!50} $1.0$ & \cellcolor{custom-green!50} $1.0$ & \cellcolor{custom-green!50} $1.0$ & \cellcolor{custom-green!50} $1.0$ & \cellcolor{custom-green!50} $\num{9E-02}$ & \cellcolor{custom-red!80} $<\num{2E-16}$ & \cellcolor{custom-red!80} $<\num{2E-16}$ & \cellcolor{custom-red!80} $<\num{2E-16}$ \\
ribs & rib\_right\_11 & \cellcolor{custom-green!50} $1.0$ & \cellcolor{custom-green!50} $1.0$ & \cellcolor{custom-green!50} $1.0$ & \cellcolor{custom-red!80} $<\num{2E-16}$ & \cellcolor{custom-red!80} $<\num{2E-16}$ & \cellcolor{custom-red!80} $<\num{2E-16}$ & \cellcolor{custom-red!80} $<\num{2E-16}$ & \cellcolor{custom-red!80} $<\num{2E-16}$ \\
ribs & rib\_right\_12 & \cellcolor{custom-green!50} $1.0$ & \cellcolor{custom-green!50} $1.0$ & \cellcolor{custom-red!80} $\num{9E-04}$ & \cellcolor{custom-red!80} $<\num{2E-16}$ & \cellcolor{custom-red!80} $<\num{2E-16}$ & \cellcolor{custom-red!80} $<\num{2E-16}$ & \cellcolor{custom-red!80} $<\num{2E-16}$ & \cellcolor{custom-red!80} $<\num{2E-16}$ \\
ribs & rib\_right\_2 & \cellcolor{custom-green!50} $1.0$ & \cellcolor{custom-green!50} $1.0$ & \cellcolor{custom-green!50} $1.0$ & \cellcolor{custom-green!50} $1.0$ & \cellcolor{custom-red!80} $<\num{2E-16}$ & \cellcolor{custom-red!80} $<\num{2E-16}$ & \cellcolor{custom-red!80} $<\num{2E-16}$ & \cellcolor{custom-red!80} $<\num{2E-16}$ \\
ribs & rib\_right\_3 & \cellcolor{custom-green!50} $1.0$ & \cellcolor{custom-green!50} $1.0$ & \cellcolor{custom-green!50} $1.0$ & \cellcolor{custom-green!50} $1.0$ & \cellcolor{custom-red!80} $\num{2E-07}$ & \cellcolor{custom-red!80} $<\num{2E-16}$ & \cellcolor{custom-red!80} $<\num{2E-16}$ & \cellcolor{custom-red!80} $<\num{2E-16}$ \\
ribs & rib\_right\_4 & \cellcolor{custom-green!50} $1.0$ & \cellcolor{custom-green!50} $1.0$ & \cellcolor{custom-green!50} $1.0$ & \cellcolor{custom-green!50} $1.0$ & \cellcolor{custom-red!80} $\num{2E-04}$ & \cellcolor{custom-red!80} $<\num{2E-16}$ & \cellcolor{custom-red!80} $<\num{2E-16}$ & \cellcolor{custom-red!80} $<\num{2E-16}$ \\
ribs & rib\_right\_5 & \cellcolor{custom-green!50} $1.0$ & \cellcolor{custom-green!50} $1.0$ & \cellcolor{custom-green!50} $1.0$ & \cellcolor{custom-green!50} $1.0$ & \cellcolor{custom-red!80} $\num{9E-04}$ & \cellcolor{custom-red!80} $<\num{2E-16}$ & \cellcolor{custom-red!80} $<\num{2E-16}$ & \cellcolor{custom-red!80} $<\num{2E-16}$ \\
ribs & rib\_right\_6 & \cellcolor{custom-green!50} $1.0$ & \cellcolor{custom-green!50} $1.0$ & \cellcolor{custom-green!50} $1.0$ & \cellcolor{custom-green!50} $1.0$ & \cellcolor{custom-green!50} $\num{4E-01}$ & \cellcolor{custom-red!80} $<\num{2E-16}$ & \cellcolor{custom-red!80} $<\num{2E-16}$ & \cellcolor{custom-red!80} $<\num{2E-16}$ \\
ribs & rib\_right\_7 & \cellcolor{custom-green!50} $1.0$ & \cellcolor{custom-green!50} $1.0$ & \cellcolor{custom-green!50} $1.0$ & \cellcolor{custom-green!50} $1.0$ & \cellcolor{custom-green!50} $\num{3E-01}$ & \cellcolor{custom-red!80} $<\num{2E-16}$ & \cellcolor{custom-red!80} $<\num{2E-16}$ & \cellcolor{custom-red!80} $<\num{2E-16}$ \\
ribs & rib\_right\_8 & \cellcolor{custom-green!50} $1.0$ & \cellcolor{custom-green!50} $1.0$ & \cellcolor{custom-green!50} $1.0$ & \cellcolor{custom-green!50} $1.0$ & \cellcolor{custom-red!80} $\num{2E-02}$ & \cellcolor{custom-red!80} $<\num{2E-16}$ & \cellcolor{custom-red!80} $<\num{2E-16}$ & \cellcolor{custom-red!80} $<\num{2E-16}$ \\
ribs & rib\_right\_9 & \cellcolor{custom-green!50} $1.0$ & \cellcolor{custom-green!50} $1.0$ & \cellcolor{custom-green!50} $1.0$ & \cellcolor{custom-green!50} $1.0$ & \cellcolor{custom-red!80} $\num{3E-07}$ & \cellcolor{custom-red!80} $<\num{2E-16}$ & \cellcolor{custom-red!80} $<\num{2E-16}$ & \cellcolor{custom-red!80} $<\num{2E-16}$ \\
ribs & sternum & \cellcolor{custom-green!50} $1.0$ & \cellcolor{custom-green!50} $1.0$ & \cellcolor{custom-green!50} $\num{1E-01}$ & \cellcolor{custom-red!80} $\num{9E-15}$ & \cellcolor{custom-red!80} $<\num{2E-16}$ & \cellcolor{custom-red!80} $<\num{2E-16}$ & \cellcolor{custom-red!80} $<\num{2E-16}$ & \cellcolor{custom-red!80} $<\num{2E-16}$ \\
\midrule
vertebrae & sacrum & \cellcolor{custom-green!50} $1.0$ & \cellcolor{custom-green!50} $1.0$ & \cellcolor{custom-green!50} $1.0$ & \cellcolor{custom-red!80} $\num{2E-06}$ & \cellcolor{custom-red!80} $<\num{2E-16}$ & \cellcolor{custom-red!80} $<\num{2E-16}$ & \cellcolor{custom-red!80} $<\num{2E-16}$ & \cellcolor{custom-red!80} $<\num{2E-16}$ \\
vertebrae & vertebrae\_C1 & \cellcolor{custom-green!50} $1.0$ & \cellcolor{custom-green!50} $1.0$ & \cellcolor{custom-green!50} $1.0$ & \cellcolor{custom-green!50} $\num{2E-01}$ & \cellcolor{custom-red!80} $<\num{2E-16}$ & \cellcolor{custom-red!80} $<\num{2E-16}$ & \cellcolor{custom-red!80} $<\num{2E-16}$ & \cellcolor{custom-red!80} $<\num{2E-16}$ \\
vertebrae & vertebrae\_C2 & \cellcolor{custom-green!50} $1.0$ & \cellcolor{custom-green!50} $1.0$ & \cellcolor{custom-green!50} $1.0$ & \cellcolor{custom-green!50} $1.0$ & \cellcolor{custom-red!80} $\num{1E-07}$ & \cellcolor{custom-red!80} $<\num{2E-16}$ & \cellcolor{custom-red!80} $<\num{2E-16}$ & \cellcolor{custom-red!80} $<\num{2E-16}$ \\
vertebrae & vertebrae\_C3 & \cellcolor{custom-green!50} $1.0$ & \cellcolor{custom-green!50} $1.0$ & \cellcolor{custom-green!50} $1.0$ & \cellcolor{custom-green!50} $1.0$ & \cellcolor{custom-red!80} $<\num{2E-16}$ & \cellcolor{custom-red!80} $<\num{2E-16}$ & \cellcolor{custom-red!80} $<\num{2E-16}$ & \cellcolor{custom-red!80} $<\num{2E-16}$ \\
vertebrae & vertebrae\_C4 & \cellcolor{custom-green!50} $1.0$ & \cellcolor{custom-green!50} $1.0$ & \cellcolor{custom-green!50} $1.0$ & \cellcolor{custom-green!50} $1.0$ & \cellcolor{custom-red!80} $\num{5E-09}$ & \cellcolor{custom-red!80} $<\num{2E-16}$ & \cellcolor{custom-red!80} $<\num{2E-16}$ & \cellcolor{custom-red!80} $<\num{2E-16}$ \\
vertebrae & vertebrae\_C5 & \cellcolor{custom-green!50} $1.0$ & \cellcolor{custom-green!50} $1.0$ & \cellcolor{custom-green!50} $1.0$ & \cellcolor{custom-red!80} $\num{3E-14}$ & \cellcolor{custom-red!80} $<\num{2E-16}$ & \cellcolor{custom-red!80} $<\num{2E-16}$ & \cellcolor{custom-red!80} $<\num{2E-16}$ & \cellcolor{custom-red!80} $<\num{2E-16}$ \\
vertebrae & vertebrae\_C6 & \cellcolor{custom-green!50} $1.0$ & \cellcolor{custom-green!50} $1.0$ & \cellcolor{custom-green!50} $\num{3E-01}$ & \cellcolor{custom-red!80} $<\num{2E-16}$ & \cellcolor{custom-red!80} $<\num{2E-16}$ & \cellcolor{custom-red!80} $<\num{2E-16}$ & \cellcolor{custom-red!80} $<\num{2E-16}$ & \cellcolor{custom-red!80} $<\num{2E-16}$ \\
vertebrae & vertebrae\_C7 & \cellcolor{custom-green!50} $1.0$ & \cellcolor{custom-green!50} $1.0$ & \cellcolor{custom-green!50} $1.0$ & \cellcolor{custom-red!80} $\num{6E-03}$ & \cellcolor{custom-red!80} $<\num{2E-16}$ & \cellcolor{custom-red!80} $<\num{2E-16}$ & \cellcolor{custom-red!80} $<\num{2E-16}$ & \cellcolor{custom-red!80} $<\num{2E-16}$ \\
vertebrae & vertebrae\_L1 & \cellcolor{custom-green!50} $1.0$ & \cellcolor{custom-green!50} $1.0$ & \cellcolor{custom-green!50} $1.0$ & \cellcolor{custom-red!80} $\num{8E-03}$ & \cellcolor{custom-red!80} $<\num{2E-16}$ & \cellcolor{custom-red!80} $<\num{2E-16}$ & \cellcolor{custom-red!80} $<\num{2E-16}$ & \cellcolor{custom-red!80} $<\num{2E-16}$ \\
vertebrae & vertebrae\_L2 & \cellcolor{custom-green!50} $1.0$ & \cellcolor{custom-green!50} $1.0$ & \cellcolor{custom-green!50} $1.0$ & \cellcolor{custom-green!50} $1.0$ & \cellcolor{custom-red!80} $<\num{2E-16}$ & \cellcolor{custom-red!80} $<\num{2E-16}$ & \cellcolor{custom-red!80} $<\num{2E-16}$ & \cellcolor{custom-red!80} $<\num{2E-16}$ \\
vertebrae & vertebrae\_L3 & \cellcolor{custom-green!50} $1.0$ & \cellcolor{custom-green!50} $1.0$ & \cellcolor{custom-green!50} $1.0$ & \cellcolor{custom-green!50} $\num{1E-01}$ & \cellcolor{custom-red!80} $<\num{2E-16}$ & \cellcolor{custom-red!80} $<\num{2E-16}$ & \cellcolor{custom-red!80} $<\num{2E-16}$ & \cellcolor{custom-red!80} $<\num{2E-16}$ \\
vertebrae & vertebrae\_L4 & \cellcolor{custom-green!50} $1.0$ & \cellcolor{custom-green!50} $1.0$ & \cellcolor{custom-green!50} $1.0$ & \cellcolor{custom-red!80} $\num{2E-02}$ & \cellcolor{custom-red!80} $<\num{2E-16}$ & \cellcolor{custom-red!80} $<\num{2E-16}$ & \cellcolor{custom-red!80} $<\num{2E-16}$ & \cellcolor{custom-red!80} $<\num{2E-16}$ \\
vertebrae & vertebrae\_L5 & \cellcolor{custom-green!50} $1.0$ & \cellcolor{custom-green!50} $1.0$ & \cellcolor{custom-green!50} $1.0$ & \cellcolor{custom-red!80} $\num{6E-04}$ & \cellcolor{custom-red!80} $<\num{2E-16}$ & \cellcolor{custom-red!80} $<\num{2E-16}$ & \cellcolor{custom-red!80} $<\num{2E-16}$ & \cellcolor{custom-red!80} $<\num{2E-16}$ \\
vertebrae & vertebrae\_S1 & \cellcolor{custom-green!50} $1.0$ & \cellcolor{custom-green!50} $1.0$ & \cellcolor{custom-green!50} $1.0$ & \cellcolor{custom-green!50} $\num{1E-01}$ & \cellcolor{custom-red!80} $<\num{2E-16}$ & \cellcolor{custom-red!80} $<\num{2E-16}$ & \cellcolor{custom-red!80} $<\num{2E-16}$ & \cellcolor{custom-red!80} $<\num{2E-16}$ \\
vertebrae & vertebrae\_T1 & \cellcolor{custom-green!50} $1.0$ & \cellcolor{custom-green!50} $1.0$ & \cellcolor{custom-green!50} $1.0$ & \cellcolor{custom-red!80} $\num{2E-02}$ & \cellcolor{custom-red!80} $<\num{2E-16}$ & \cellcolor{custom-red!80} $<\num{2E-16}$ & \cellcolor{custom-red!80} $<\num{2E-16}$ & \cellcolor{custom-red!80} $<\num{2E-16}$ \\
vertebrae & vertebrae\_T10 & \cellcolor{custom-green!50} $1.0$ & \cellcolor{custom-green!50} $1.0$ & \cellcolor{custom-green!50} $1.0$ & \cellcolor{custom-green!50} $1.0$ & \cellcolor{custom-red!80} $\num{7E-09}$ & \cellcolor{custom-red!80} $<\num{2E-16}$ & \cellcolor{custom-red!80} $<\num{2E-16}$ & \cellcolor{custom-red!80} $<\num{2E-16}$ \\
vertebrae & vertebrae\_T11 & \cellcolor{custom-green!50} $1.0$ & \cellcolor{custom-green!50} $1.0$ & \cellcolor{custom-green!50} $1.0$ & \cellcolor{custom-green!50} $\num{6E-02}$ & \cellcolor{custom-red!80} $<\num{2E-16}$ & \cellcolor{custom-red!80} $<\num{2E-16}$ & \cellcolor{custom-red!80} $<\num{2E-16}$ & \cellcolor{custom-red!80} $<\num{2E-16}$ \\
vertebrae & vertebrae\_T12 & \cellcolor{custom-green!50} $1.0$ & \cellcolor{custom-green!50} $1.0$ & \cellcolor{custom-green!50} $1.0$ & \cellcolor{custom-red!80} $\num{4E-02}$ & \cellcolor{custom-red!80} $<\num{2E-16}$ & \cellcolor{custom-red!80} $<\num{2E-16}$ & \cellcolor{custom-red!80} $<\num{2E-16}$ & \cellcolor{custom-red!80} $<\num{2E-16}$ \\
vertebrae & vertebrae\_T2 & \cellcolor{custom-green!50} $1.0$ & \cellcolor{custom-green!50} $1.0$ & \cellcolor{custom-green!50} $1.0$ & \cellcolor{custom-green!50} $1.0$ & \cellcolor{custom-red!80} $<\num{2E-16}$ & \cellcolor{custom-red!80} $<\num{2E-16}$ & \cellcolor{custom-red!80} $<\num{2E-16}$ & \cellcolor{custom-red!80} $<\num{2E-16}$ \\
vertebrae & vertebrae\_T3 & \cellcolor{custom-green!50} $1.0$ & \cellcolor{custom-green!50} $1.0$ & \cellcolor{custom-green!50} $1.0$ & \cellcolor{custom-green!50} $1.0$ & \cellcolor{custom-red!80} $<\num{2E-16}$ & \cellcolor{custom-red!80} $<\num{2E-16}$ & \cellcolor{custom-red!80} $<\num{2E-16}$ & \cellcolor{custom-red!80} $<\num{2E-16}$ \\
vertebrae & vertebrae\_T4 & \cellcolor{custom-green!50} $1.0$ & \cellcolor{custom-green!50} $1.0$ & \cellcolor{custom-green!50} $1.0$ & \cellcolor{custom-green!50} $\num{2E-01}$ & \cellcolor{custom-red!80} $<\num{2E-16}$ & \cellcolor{custom-red!80} $<\num{2E-16}$ & \cellcolor{custom-red!80} $<\num{2E-16}$ & \cellcolor{custom-red!80} $<\num{2E-16}$ \\
vertebrae & vertebrae\_T5 & \cellcolor{custom-green!50} $1.0$ & \cellcolor{custom-green!50} $1.0$ & \cellcolor{custom-green!50} $1.0$ & \cellcolor{custom-green!50} $\num{8E-02}$ & \cellcolor{custom-red!80} $<\num{2E-16}$ & \cellcolor{custom-red!80} $<\num{2E-16}$ & \cellcolor{custom-red!80} $<\num{2E-16}$ & \cellcolor{custom-red!80} $<\num{2E-16}$ \\
vertebrae & vertebrae\_T6 & \cellcolor{custom-green!50} $1.0$ & \cellcolor{custom-green!50} $1.0$ & \cellcolor{custom-green!50} $1.0$ & \cellcolor{custom-red!80} $\num{5E-05}$ & \cellcolor{custom-red!80} $<\num{2E-16}$ & \cellcolor{custom-red!80} $<\num{2E-16}$ & \cellcolor{custom-red!80} $<\num{2E-16}$ & \cellcolor{custom-red!80} $<\num{2E-16}$ \\
vertebrae & vertebrae\_T7 & \cellcolor{custom-green!50} $1.0$ & \cellcolor{custom-green!50} $1.0$ & \cellcolor{custom-green!50} $1.0$ & \cellcolor{custom-green!50} $\num{1E-01}$ & \cellcolor{custom-red!80} $<\num{2E-16}$ & \cellcolor{custom-red!80} $<\num{2E-16}$ & \cellcolor{custom-red!80} $<\num{2E-16}$ & \cellcolor{custom-red!80} $<\num{2E-16}$ \\
vertebrae & vertebrae\_T8 & \cellcolor{custom-green!50} $1.0$ & \cellcolor{custom-green!50} $1.0$ & \cellcolor{custom-green!50} $1.0$ & \cellcolor{custom-green!50} $1.0$ & \cellcolor{custom-red!80} $\num{5E-14}$ & \cellcolor{custom-red!80} $<\num{2E-16}$ & \cellcolor{custom-red!80} $<\num{2E-16}$ & \cellcolor{custom-red!80} $<\num{2E-16}$ \\
vertebrae & vertebrae\_T9 & \cellcolor{custom-green!50} $1.0$ & \cellcolor{custom-green!50} $1.0$ & \cellcolor{custom-green!50} $1.0$ & \cellcolor{custom-green!50} $1.0$ & \cellcolor{custom-red!80} $<\num{2E-16}$ & \cellcolor{custom-red!80} $<\num{2E-16}$ & \cellcolor{custom-red!80} $<\num{2E-16}$ & \cellcolor{custom-red!80} $<\num{2E-16}$ \\
\bottomrule
\end{tabular}

%% file: tables/pval_table_1.tex
\newcolumntype{C}[1]{>{\centering\arraybackslash}p{#1}}

\begin{tabular}{llC{15mm}C{15mm}C{15mm}C{15mm}C{15mm}C{15mm}C{15mm}C{15mm}}
\toprule
& & \multicolumn{8}{c}{\textbf{Downsampling Factor [DF]}} \\
\cmidrule{3-10}
 & \textbf{Class Label} & \textbf{0.9} & \textbf{0.7} & \textbf{0.5} & \textbf{0.4} & \textbf{0.3} & \textbf{0.2} & \textbf{0.1} & \textbf{0.05} \\
\midrule
cardiac & aorta & \cellcolor{custom-green!50} $1.0$ & \cellcolor{custom-green!50} $1.0$ & \cellcolor{custom-green!50} $1.0$ & \cellcolor{custom-green!50} $1.0$ & \cellcolor{custom-green!50} $1.0$ & \cellcolor{custom-green!50} $1.0$ & \cellcolor{custom-red!80} $\num{2E-05}$ & \cellcolor{custom-red!80} $\num{7E-06}$ \\
 cardiac & atrial\_appendage\_left & \cellcolor{custom-green!50} $1.0$ & \cellcolor{custom-green!50} $1.0$ & \cellcolor{custom-green!50} $1.0$ & \cellcolor{custom-green!50} $1.0$ & \cellcolor{custom-green!50} $1.0$ & \cellcolor{custom-red!80} $<\num{2E-16}$ & \cellcolor{custom-red!80} $<\num{2E-16}$ & \cellcolor{custom-red!80} $<\num{2E-16}$ \\
 cardiac & brachiocephalic\_trunk & \cellcolor{custom-green!50} $1.0$ & \cellcolor{custom-green!50} $1.0$ & \cellcolor{custom-green!50} $1.0$ & \cellcolor{custom-green!50} $1.0$ & \cellcolor{custom-green!50} $1.0$ & \cellcolor{custom-green!50} $1.0$ & \cellcolor{custom-red!80} $\num{1E-02}$ & \cellcolor{custom-red!80} $<\num{2E-16}$ \\
 cardiac & brachiocephalic\_vein\_left & \cellcolor{custom-green!50} $1.0$ & \cellcolor{custom-green!50} $1.0$ & \cellcolor{custom-green!50} $1.0$ & \cellcolor{custom-green!50} $1.0$ & \cellcolor{custom-green!50} $1.0$ & \cellcolor{custom-green!50} $1.0$ & \cellcolor{custom-red!80} $\num{4E-02}$ & \cellcolor{custom-red!80} $<\num{2E-16}$ \\
 cardiac & brachiocephalic\_vein\_right & \cellcolor{custom-green!50} $1.0$ & \cellcolor{custom-green!50} $1.0$ & \cellcolor{custom-green!50} $1.0$ & \cellcolor{custom-green!50} $1.0$ & \cellcolor{custom-green!50} $1.0$ & \cellcolor{custom-green!50} $1.0$ & \cellcolor{custom-red!80} $<\num{2E-16}$ & \cellcolor{custom-red!80} $<\num{2E-16}$ \\
\bottomrule
\end{tabular}

%% file: tables/pval_table_2.tex
\newcolumntype{C}[1]{>{\centering\arraybackslash}p{#1}}

\begin{tabular}{llC{15mm}C{15mm}C{15mm}C{15mm}C{15mm}C{15mm}C{15mm}C{15mm}}
\toprule
& & \multicolumn{8}{c}{\textbf{Downsampling Factor [DF]}} \\
\cmidrule{3-10}
 & \textbf{Class Label} & \textbf{0.9} & \textbf{0.7} & \textbf{0.5} & \textbf{0.4} & \textbf{0.3} & \textbf{0.2} & \textbf{0.1} & \textbf{0.05} \\
\midrule
ribs & costal\_cartilages & \cellcolor{custom-green!50} $1.0$ & \cellcolor{custom-green!50} $1.0$ & \cellcolor{custom-green!50} $1.0$ & \cellcolor{custom-green!50} $1.0$ & \cellcolor{custom-green!50} $1.0$ & \cellcolor{custom-red!80} $\num{3E-02}$ & \cellcolor{custom-red!80} $\num{2E-11}$ & \cellcolor{custom-red!80} $\num{6E-11}$ \\
 ribs & rib\_left\_1 & \cellcolor{custom-green!50} $1.0$ & \cellcolor{custom-green!50} $1.0$ & \cellcolor{custom-green!50} $1.0$ & \cellcolor{custom-green!50} $1.0$ & \cellcolor{custom-green!50} $1.0$ & \cellcolor{custom-green!50} $1.0$ & \cellcolor{custom-red!80} $<\num{2E-16}$ & \cellcolor{custom-red!80} $<\num{2E-16}$ \\
 ribs & rib\_left\_10 & \cellcolor{custom-green!50} $1.0$ & \cellcolor{custom-green!50} $1.0$ & \cellcolor{custom-green!50} $1.0$ & \cellcolor{custom-green!50} $1.0$ & \cellcolor{custom-green!50} $1.0$ & \cellcolor{custom-green!50} $\num{1E-01}$ & \cellcolor{custom-red!80} $<\num{2E-16}$ & \cellcolor{custom-red!80} $<\num{2E-16}$ \\
 ribs & rib\_left\_11 & \cellcolor{custom-green!50} $1.0$ & \cellcolor{custom-green!50} $1.0$ & \cellcolor{custom-green!50} $1.0$ & \cellcolor{custom-green!50} $1.0$ & \cellcolor{custom-green!50} $1.0$ & \cellcolor{custom-green!50} $1.0$ & \cellcolor{custom-red!80} $<\num{2E-16}$ & \cellcolor{custom-red!80} $<\num{2E-16}$ \\
 ribs & rib\_left\_12 & \cellcolor{custom-green!50} $1.0$ & \cellcolor{custom-green!50} $1.0$ & \cellcolor{custom-green!50} $1.0$ & \cellcolor{custom-green!50} $1.0$ & \cellcolor{custom-green!50} $\num{7E-01}$ & \cellcolor{custom-red!80} $<\num{2E-16}$ & \cellcolor{custom-red!80} $<\num{2E-16}$ & \cellcolor{custom-red!80} $<\num{2E-16}$ \\
 ribs & rib\_left\_2 & \cellcolor{custom-green!50} $1.0$ & \cellcolor{custom-green!50} $1.0$ & \cellcolor{custom-green!50} $1.0$ & \cellcolor{custom-green!50} $1.0$ & \cellcolor{custom-green!50} $1.0$ & \cellcolor{custom-green!50} $1.0$ & \cellcolor{custom-red!80} $\num{4E-09}$ & \cellcolor{custom-red!80} $<\num{2E-16}$ \\
 ribs & rib\_left\_3 & \cellcolor{custom-green!50} $1.0$ & \cellcolor{custom-green!50} $1.0$ & \cellcolor{custom-green!50} $1.0$ & \cellcolor{custom-green!50} $1.0$ & \cellcolor{custom-green!50} $1.0$ & \cellcolor{custom-green!50} $1.0$ & \cellcolor{custom-red!80} $<\num{2E-16}$ & \cellcolor{custom-red!80} $<\num{2E-16}$ \\
 ribs & rib\_left\_4 & \cellcolor{custom-green!50} $1.0$ & \cellcolor{custom-green!50} $1.0$ & \cellcolor{custom-green!50} $1.0$ & \cellcolor{custom-green!50} $1.0$ & \cellcolor{custom-green!50} $1.0$ & \cellcolor{custom-green!50} $\num{8E-01}$ & \cellcolor{custom-red!80} $<\num{2E-16}$ & \cellcolor{custom-red!80} $<\num{2E-16}$ \\
 ribs & rib\_left\_5 & \cellcolor{custom-green!50} $1.0$ & \cellcolor{custom-green!50} $1.0$ & \cellcolor{custom-green!50} $1.0$ & \cellcolor{custom-green!50} $1.0$ & \cellcolor{custom-green!50} $1.0$ & \cellcolor{custom-red!80} $\num{2E-04}$ & \cellcolor{custom-red!80} $<\num{2E-16}$ & \cellcolor{custom-red!80} $<\num{2E-16}$ \\
 ribs & rib\_left\_6 & \cellcolor{custom-green!50} $1.0$ & \cellcolor{custom-green!50} $1.0$ & \cellcolor{custom-green!50} $1.0$ & \cellcolor{custom-green!50} $1.0$ & \cellcolor{custom-green!50} $1.0$ & \cellcolor{custom-red!80} $\num{7E-05}$ & \cellcolor{custom-red!80} $<\num{2E-16}$ & \cellcolor{custom-red!80} $<\num{2E-16}$ \\
 ribs & rib\_left\_7 & \cellcolor{custom-green!50} $1.0$ & \cellcolor{custom-green!50} $1.0$ & \cellcolor{custom-green!50} $1.0$ & \cellcolor{custom-green!50} $1.0$ & \cellcolor{custom-green!50} $1.0$ & \cellcolor{custom-green!50} $\num{6E-02}$ & \cellcolor{custom-red!80} $<\num{2E-16}$ & \cellcolor{custom-red!80} $<\num{2E-16}$ \\
 ribs & rib\_left\_8 & \cellcolor{custom-green!50} $1.0$ & \cellcolor{custom-green!50} $1.0$ & \cellcolor{custom-green!50} $1.0$ & \cellcolor{custom-green!50} $1.0$ & \cellcolor{custom-green!50} $1.0$ & \cellcolor{custom-green!50} $\num{1E-01}$ & \cellcolor{custom-red!80} $<\num{2E-16}$ & \cellcolor{custom-red!80} $<\num{2E-16}$ \\
 ribs & rib\_left\_9 & \cellcolor{custom-green!50} $1.0$ & \cellcolor{custom-green!50} $1.0$ & \cellcolor{custom-green!50} $1.0$ & \cellcolor{custom-green!50} $1.0$ & \cellcolor{custom-green!50} $1.0$ & \cellcolor{custom-green!50} $\num{4E-01}$ & \cellcolor{custom-red!80} $<\num{2E-16}$ & \cellcolor{custom-red!80} $<\num{2E-16}$ \\
 ribs & rib\_right\_1 & \cellcolor{custom-green!50} $1.0$ & \cellcolor{custom-green!50} $1.0$ & \cellcolor{custom-green!50} $1.0$ & \cellcolor{custom-green!50} $1.0$ & \cellcolor{custom-green!50} $1.0$ & \cellcolor{custom-green!50} $1.0$ & \cellcolor{custom-red!80} $\num{3E-05}$ & \cellcolor{custom-red!80} $<\num{2E-16}$ \\
 ribs & rib\_right\_10 & \cellcolor{custom-green!50} $1.0$ & \cellcolor{custom-green!50} $1.0$ & \cellcolor{custom-green!50} $1.0$ & \cellcolor{custom-green!50} $1.0$ & \cellcolor{custom-green!50} $1.0$ & \cellcolor{custom-green!50} $1.0$ & \cellcolor{custom-red!80} $<\num{2E-16}$ & \cellcolor{custom-red!80} $<\num{2E-16}$ \\
 ribs & rib\_right\_11 & \cellcolor{custom-green!50} $1.0$ & \cellcolor{custom-green!50} $1.0$ & \cellcolor{custom-green!50} $\num{4E-01}$ & \cellcolor{custom-red!80} $\num{3E-05}$ & \cellcolor{custom-red!80} $<\num{2E-16}$ & \cellcolor{custom-red!80} $<\num{2E-16}$ & \cellcolor{custom-red!80} $<\num{2E-16}$ & \cellcolor{custom-red!80} $<\num{2E-16}$ \\
 ribs & rib\_right\_12 & \cellcolor{custom-green!50} $1.0$ & \cellcolor{custom-green!50} $1.0$ & \cellcolor{custom-green!50} $1.0$ & \cellcolor{custom-green!50} $1.0$ & \cellcolor{custom-red!80} $<\num{2E-16}$ & \cellcolor{custom-red!80} $<\num{2E-16}$ & \cellcolor{custom-red!80} $<\num{2E-16}$ & \cellcolor{custom-red!80} $<\num{2E-16}$ \\
 ribs & rib\_right\_2 & \cellcolor{custom-green!50} $1.0$ & \cellcolor{custom-green!50} $1.0$ & \cellcolor{custom-green!50} $1.0$ & \cellcolor{custom-green!50} $1.0$ & \cellcolor{custom-green!50} $1.0$ & \cellcolor{custom-green!50} $1.0$ & \cellcolor{custom-green!50} $1.0$ & \cellcolor{custom-red!80} $<\num{2E-16}$ \\
 ribs & rib\_right\_3 & \cellcolor{custom-green!50} $1.0$ & \cellcolor{custom-green!50} $1.0$ & \cellcolor{custom-green!50} $1.0$ & \cellcolor{custom-green!50} $1.0$ & \cellcolor{custom-green!50} $1.0$ & \cellcolor{custom-green!50} $1.0$ & \cellcolor{custom-green!50} $\num{5E-01}$ & \cellcolor{custom-red!80} $<\num{2E-16}$ \\
 ribs & rib\_right\_4 & \cellcolor{custom-green!50} $1.0$ & \cellcolor{custom-green!50} $1.0$ & \cellcolor{custom-green!50} $1.0$ & \cellcolor{custom-green!50} $1.0$ & \cellcolor{custom-green!50} $1.0$ & \cellcolor{custom-green!50} $1.0$ & \cellcolor{custom-red!80} $\num{3E-03}$ & \cellcolor{custom-red!80} $<\num{2E-16}$ \\
 ribs & rib\_right\_5 & \cellcolor{custom-green!50} $1.0$ & \cellcolor{custom-green!50} $1.0$ & \cellcolor{custom-green!50} $1.0$ & \cellcolor{custom-green!50} $1.0$ & \cellcolor{custom-green!50} $1.0$ & \cellcolor{custom-green!50} $1.0$ & \cellcolor{custom-red!80} $\num{2E-16}$ & \cellcolor{custom-red!80} $<\num{2E-16}$ \\
 ribs & rib\_right\_6 & \cellcolor{custom-green!50} $1.0$ & \cellcolor{custom-green!50} $1.0$ & \cellcolor{custom-green!50} $1.0$ & \cellcolor{custom-green!50} $1.0$ & \cellcolor{custom-green!50} $1.0$ & \cellcolor{custom-green!50} $\num{5E-01}$ & \cellcolor{custom-red!80} $<\num{2E-16}$ & \cellcolor{custom-red!80} $<\num{2E-16}$ \\
 ribs & rib\_right\_7 & \cellcolor{custom-green!50} $1.0$ & \cellcolor{custom-green!50} $1.0$ & \cellcolor{custom-green!50} $1.0$ & \cellcolor{custom-green!50} $1.0$ & \cellcolor{custom-green!50} $1.0$ & \cellcolor{custom-green!50} $\num{5E-02}$ & \cellcolor{custom-red!80} $<\num{2E-16}$ & \cellcolor{custom-red!80} $<\num{2E-16}$ \\
 ribs & rib\_right\_8 & \cellcolor{custom-green!50} $1.0$ & \cellcolor{custom-green!50} $1.0$ & \cellcolor{custom-green!50} $1.0$ & \cellcolor{custom-green!50} $1.0$ & \cellcolor{custom-green!50} $1.0$ & \cellcolor{custom-green!50} $1.0$ & \cellcolor{custom-red!80} $<\num{2E-16}$ & \cellcolor{custom-red!80} $<\num{2E-16}$ \\
 ribs & rib\_right\_9 & \cellcolor{custom-green!50} $1.0$ & \cellcolor{custom-green!50} $1.0$ & \cellcolor{custom-green!50} $1.0$ & \cellcolor{custom-green!50} $1.0$ & \cellcolor{custom-green!50} $1.0$ & \cellcolor{custom-green!50} $\num{6E-01}$ & \cellcolor{custom-red!80} $<\num{2E-16}$ & \cellcolor{custom-red!80} $<\num{2E-16}$ \\
 ribs & sternum & \cellcolor{custom-green!50} $1.0$ & \cellcolor{custom-green!50} $1.0$ & \cellcolor{custom-green!50} $1.0$ & \cellcolor{custom-green!50} $1.0$ & \cellcolor{custom-green!50} $1.0$ & \cellcolor{custom-green!50} $\num{7E-02}$ & \cellcolor{custom-red!80} $\num{1E-14}$ & \cellcolor{custom-red!80} $\num{2E-16}$ \\
\midrule
vertebrae & sacrum & \cellcolor{custom-green!50} $1.0$ & \cellcolor{custom-green!50} $1.0$ & \cellcolor{custom-green!50} $1.0$ & \cellcolor{custom-green!50} $1.0$ & \cellcolor{custom-green!50} $\num{3E-01}$ & \cellcolor{custom-red!80} $\num{6E-06}$ & \cellcolor{custom-red!80} $\num{5E-09}$ & \cellcolor{custom-red!80} $\num{1E-07}$ \\
 vertebrae & vertebrae\_C1 & \cellcolor{custom-green!50} $1.0$ & \cellcolor{custom-green!50} $1.0$ & \cellcolor{custom-green!50} $1.0$ & \cellcolor{custom-green!50} $1.0$ & \cellcolor{custom-green!50} $1.0$ & \cellcolor{custom-green!50} $1.0$ & \cellcolor{custom-red!80} $<\num{2E-16}$ & \cellcolor{custom-red!80} $<\num{2E-16}$ \\
 vertebrae & vertebrae\_C2 & \cellcolor{custom-green!50} $1.0$ & \cellcolor{custom-green!50} $1.0$ & \cellcolor{custom-green!50} $1.0$ & \cellcolor{custom-green!50} $1.0$ & \cellcolor{custom-green!50} $1.0$ & \cellcolor{custom-green!50} $1.0$ & \cellcolor{custom-red!80} $\num{8E-06}$ & \cellcolor{custom-red!80} $<\num{2E-16}$ \\
 vertebrae & vertebrae\_C3 & \cellcolor{custom-green!50} $1.0$ & \cellcolor{custom-green!50} $1.0$ & \cellcolor{custom-green!50} $1.0$ & \cellcolor{custom-green!50} $1.0$ & \cellcolor{custom-green!50} $1.0$ & \cellcolor{custom-green!50} $1.0$ & \cellcolor{custom-green!50} $1.0$ & \cellcolor{custom-red!80} $<\num{2E-16}$ \\
 vertebrae & vertebrae\_C4 & \cellcolor{custom-green!50} $1.0$ & \cellcolor{custom-green!50} $1.0$ & \cellcolor{custom-green!50} $1.0$ & \cellcolor{custom-green!50} $1.0$ & \cellcolor{custom-green!50} $1.0$ & \cellcolor{custom-green!50} $1.0$ & \cellcolor{custom-green!50} $1.0$ & \cellcolor{custom-red!80} $\num{9E-16}$ \\
 vertebrae & vertebrae\_C5 & \cellcolor{custom-green!50} $1.0$ & \cellcolor{custom-green!50} $1.0$ & \cellcolor{custom-green!50} $1.0$ & \cellcolor{custom-green!50} $1.0$ & \cellcolor{custom-green!50} $1.0$ & \cellcolor{custom-green!50} $1.0$ & \cellcolor{custom-red!80} $\num{8E-09}$ & \cellcolor{custom-red!80} $<\num{2E-16}$ \\
 vertebrae & vertebrae\_C6 & \cellcolor{custom-green!50} $1.0$ & \cellcolor{custom-green!50} $1.0$ & \cellcolor{custom-green!50} $1.0$ & \cellcolor{custom-green!50} $1.0$ & \cellcolor{custom-green!50} $1.0$ & \cellcolor{custom-red!80} $\num{1E-02}$ & \cellcolor{custom-red!80} $<\num{2E-16}$ & \cellcolor{custom-red!80} $<\num{2E-16}$ \\
 vertebrae & vertebrae\_C7 & \cellcolor{custom-green!50} $1.0$ & \cellcolor{custom-green!50} $1.0$ & \cellcolor{custom-green!50} $1.0$ & \cellcolor{custom-green!50} $1.0$ & \cellcolor{custom-green!50} $1.0$ & \cellcolor{custom-green!50} $1.0$ & \cellcolor{custom-red!80} $\num{1E-07}$ & \cellcolor{custom-red!80} $<\num{2E-16}$ \\
 vertebrae & vertebrae\_L1 & \cellcolor{custom-green!50} $1.0$ & \cellcolor{custom-green!50} $1.0$ & \cellcolor{custom-green!50} $1.0$ & \cellcolor{custom-green!50} $1.0$ & \cellcolor{custom-green!50} $1.0$ & \cellcolor{custom-green!50} $1.0$ & \cellcolor{custom-red!80} $\num{8E-11}$ & \cellcolor{custom-red!80} $\num{3E-13}$ \\
 vertebrae & vertebrae\_L2 & \cellcolor{custom-green!50} $1.0$ & \cellcolor{custom-green!50} $1.0$ & \cellcolor{custom-green!50} $1.0$ & \cellcolor{custom-green!50} $1.0$ & \cellcolor{custom-green!50} $1.0$ & \cellcolor{custom-green!50} $1.0$ & \cellcolor{custom-red!80} $\num{3E-04}$ & \cellcolor{custom-red!80} $\num{6E-11}$ \\
 vertebrae & vertebrae\_L3 & \cellcolor{custom-green!50} $1.0$ & \cellcolor{custom-green!50} $1.0$ & \cellcolor{custom-green!50} $1.0$ & \cellcolor{custom-green!50} $1.0$ & \cellcolor{custom-green!50} $1.0$ & \cellcolor{custom-green!50} $1.0$ & \cellcolor{custom-red!80} $\num{8E-07}$ & \cellcolor{custom-red!80} $\num{5E-14}$ \\
 vertebrae & vertebrae\_L4 & \cellcolor{custom-green!50} $1.0$ & \cellcolor{custom-green!50} $1.0$ & \cellcolor{custom-green!50} $1.0$ & \cellcolor{custom-green!50} $1.0$ & \cellcolor{custom-green!50} $1.0$ & \cellcolor{custom-green!50} $1.0$ & \cellcolor{custom-red!80} $\num{2E-09}$ & \cellcolor{custom-red!80} $\num{7E-13}$ \\
 vertebrae & vertebrae\_L5 & \cellcolor{custom-green!50} $1.0$ & \cellcolor{custom-green!50} $1.0$ & \cellcolor{custom-green!50} $1.0$ & \cellcolor{custom-green!50} $1.0$ & \cellcolor{custom-green!50} $1.0$ & \cellcolor{custom-green!50} $1.0$ & \cellcolor{custom-red!80} $\num{2E-03}$ & \cellcolor{custom-red!80} $\num{3E-10}$ \\
 vertebrae & vertebrae\_S1 & \cellcolor{custom-green!50} $1.0$ & \cellcolor{custom-green!50} $1.0$ & \cellcolor{custom-green!50} $1.0$ & \cellcolor{custom-green!50} $1.0$ & \cellcolor{custom-green!50} $1.0$ & \cellcolor{custom-green!50} $1.0$ & \cellcolor{custom-red!80} $\num{3E-06}$ & \cellcolor{custom-red!80} $<\num{2E-16}$ \\
 vertebrae & vertebrae\_T1 & \cellcolor{custom-green!50} $1.0$ & \cellcolor{custom-green!50} $1.0$ & \cellcolor{custom-green!50} $1.0$ & \cellcolor{custom-green!50} $1.0$ & \cellcolor{custom-green!50} $1.0$ & \cellcolor{custom-green!50} $1.0$ & \cellcolor{custom-red!80} $\num{3E-02}$ & \cellcolor{custom-red!80} $<\num{2E-16}$ \\
 vertebrae & vertebrae\_T10 & \cellcolor{custom-green!50} $1.0$ & \cellcolor{custom-green!50} $1.0$ & \cellcolor{custom-green!50} $1.0$ & \cellcolor{custom-green!50} $1.0$ & \cellcolor{custom-green!50} $1.0$ & \cellcolor{custom-green!50} $1.0$ & \cellcolor{custom-red!80} $\num{2E-03}$ & \cellcolor{custom-red!80} $\num{2E-03}$ \\
 vertebrae & vertebrae\_T11 & \cellcolor{custom-green!50} $1.0$ & \cellcolor{custom-green!50} $1.0$ & \cellcolor{custom-green!50} $1.0$ & \cellcolor{custom-green!50} $1.0$ & \cellcolor{custom-green!50} $1.0$ & \cellcolor{custom-green!50} $1.0$ & \cellcolor{custom-red!80} $\num{2E-03}$ & \cellcolor{custom-red!80} $\num{2E-02}$ \\
 vertebrae & vertebrae\_T12 & \cellcolor{custom-green!50} $1.0$ & \cellcolor{custom-green!50} $1.0$ & \cellcolor{custom-green!50} $1.0$ & \cellcolor{custom-green!50} $1.0$ & \cellcolor{custom-green!50} $1.0$ & \cellcolor{custom-red!80} $\num{2E-03}$ & \cellcolor{custom-red!80} $\num{5E-06}$ & \cellcolor{custom-red!80} $\num{2E-04}$ \\
 vertebrae & vertebrae\_T2 & \cellcolor{custom-green!50} $1.0$ & \cellcolor{custom-green!50} $1.0$ & \cellcolor{custom-green!50} $1.0$ & \cellcolor{custom-green!50} $1.0$ & \cellcolor{custom-green!50} $1.0$ & \cellcolor{custom-green!50} $1.0$ & \cellcolor{custom-green!50} $1.0$ & \cellcolor{custom-red!80} $\num{2E-14}$ \\
 vertebrae & vertebrae\_T3 & \cellcolor{custom-green!50} $1.0$ & \cellcolor{custom-green!50} $1.0$ & \cellcolor{custom-green!50} $1.0$ & \cellcolor{custom-green!50} $1.0$ & \cellcolor{custom-green!50} $1.0$ & \cellcolor{custom-green!50} $1.0$ & \cellcolor{custom-green!50} $\num{7E-01}$ & \cellcolor{custom-red!80} $\num{2E-07}$ \\
 vertebrae & vertebrae\_T4 & \cellcolor{custom-green!50} $1.0$ & \cellcolor{custom-green!50} $1.0$ & \cellcolor{custom-green!50} $1.0$ & \cellcolor{custom-green!50} $1.0$ & \cellcolor{custom-green!50} $1.0$ & \cellcolor{custom-green!50} $1.0$ & \cellcolor{custom-green!50} $\num{4E-01}$ & \cellcolor{custom-red!80} $\num{1E-06}$ \\
 vertebrae & vertebrae\_T5 & \cellcolor{custom-green!50} $1.0$ & \cellcolor{custom-green!50} $1.0$ & \cellcolor{custom-green!50} $1.0$ & \cellcolor{custom-green!50} $1.0$ & \cellcolor{custom-green!50} $1.0$ & \cellcolor{custom-green!50} $1.0$ & \cellcolor{custom-red!80} $\num{6E-03}$ & \cellcolor{custom-red!80} $\num{4E-09}$ \\
 vertebrae & vertebrae\_T6 & \cellcolor{custom-green!50} $1.0$ & \cellcolor{custom-green!50} $1.0$ & \cellcolor{custom-green!50} $1.0$ & \cellcolor{custom-green!50} $1.0$ & \cellcolor{custom-green!50} $1.0$ & \cellcolor{custom-green!50} $1.0$ & \cellcolor{custom-red!80} $\num{2E-03}$ & \cellcolor{custom-red!80} $\num{6E-16}$ \\
 vertebrae & vertebrae\_T7 & \cellcolor{custom-green!50} $1.0$ & \cellcolor{custom-green!50} $1.0$ & \cellcolor{custom-green!50} $1.0$ & \cellcolor{custom-green!50} $1.0$ & \cellcolor{custom-green!50} $1.0$ & \cellcolor{custom-green!50} $1.0$ & \cellcolor{custom-red!80} $\num{8E-05}$ & \cellcolor{custom-red!80} $\num{3E-08}$ \\
 vertebrae & vertebrae\_T8 & \cellcolor{custom-green!50} $1.0$ & \cellcolor{custom-green!50} $1.0$ & \cellcolor{custom-green!50} $1.0$ & \cellcolor{custom-green!50} $1.0$ & \cellcolor{custom-green!50} $1.0$ & \cellcolor{custom-green!50} $1.0$ & \cellcolor{custom-red!80} $\num{7E-08}$ & \cellcolor{custom-red!80} $\num{3E-08}$ \\
 vertebrae & vertebrae\_T9 & \cellcolor{custom-green!50} $1.0$ & \cellcolor{custom-green!50} $1.0$ & \cellcolor{custom-green!50} $1.0$ & \cellcolor{custom-green!50} $1.0$ & \cellcolor{custom-green!50} $1.0$ & \cellcolor{custom-green!50} $1.0$ & \cellcolor{custom-red!80} $\num{6E-03}$ & \cellcolor{custom-red!80} $\num{3E-06}$ \\
\bottomrule
\end{tabular}

%% file: tables/pval_zero_fast_1.tex
\newcolumntype{C}[1]{>{\centering\arraybackslash}p{#1}}

\begin{tabular}{llC{15mm}C{15mm}C{15mm}C{15mm}C{15mm}C{15mm}C{15mm}C{15mm}}
\toprule
& & \multicolumn{8}{c}{\textbf{Downsampling Factor [DF]}} \\
\cmidrule{3-10}
 & \textbf{Class Label} & \textbf{0.9} & \textbf{0.7} & \textbf{0.5} & \textbf{0.4} & \textbf{0.3} & \textbf{0.2} & \textbf{0.1} & \textbf{0.05} \\
\midrule
cardiac & aorta & \cellcolor{custom-green!50} $\num{2E-01}$ & \cellcolor{custom-red!80} $<\num{2E-16}$ & \cellcolor{custom-red!80} $<\num{2E-16}$ & \cellcolor{custom-red!80} $<\num{2E-16}$ & \cellcolor{custom-red!80} $<\num{2E-16}$ & \cellcolor{custom-red!80} $<\num{2E-16}$ & \cellcolor{custom-red!80} $<\num{2E-16}$ & \cellcolor{custom-red!80} $<\num{2E-16}$ \\
cardiac & atrial\_appendage\_left & \cellcolor{custom-green!50} $\num{9E-02}$ & \cellcolor{custom-red!80} $<\num{2E-16}$ & \cellcolor{custom-red!80} $<\num{2E-16}$ & \cellcolor{custom-red!80} $<\num{2E-16}$ & \cellcolor{custom-red!80} $<\num{2E-16}$ & \cellcolor{custom-red!80} $<\num{2E-16}$ & \cellcolor{custom-red!80} $<\num{2E-16}$ & \cellcolor{custom-red!80} $<\num{2E-16}$ \\
cardiac & brachiocephalic\_trunk & \cellcolor{custom-red!80} $\num{2E-02}$ & \cellcolor{custom-red!80} $<\num{2E-16}$ & \cellcolor{custom-red!80} $<\num{2E-16}$ & \cellcolor{custom-red!80} $<\num{2E-16}$ & \cellcolor{custom-red!80} $<\num{2E-16}$ & \cellcolor{custom-red!80} $<\num{2E-16}$ & \cellcolor{custom-red!80} $<\num{2E-16}$ & \cellcolor{custom-red!80} $<\num{2E-16}$ \\
cardiac & brachiocephalic\_vein\_left & \cellcolor{custom-green!50} $1.0$ & \cellcolor{custom-red!80} $\num{7E-15}$ & \cellcolor{custom-red!80} $<\num{2E-16}$ & \cellcolor{custom-red!80} $<\num{2E-16}$ & \cellcolor{custom-red!80} $<\num{2E-16}$ & \cellcolor{custom-red!80} $<\num{2E-16}$ & \cellcolor{custom-red!80} $<\num{2E-16}$ & \cellcolor{custom-red!80} $<\num{2E-16}$ \\
\bottomrule
\end{tabular}

%% file: tables/pval_zero_fast_2.tex
\newcolumntype{C}[1]{>{\centering\arraybackslash}p{#1}}

\begin{tabular}{llC{15mm}C{15mm}C{15mm}C{15mm}C{15mm}C{15mm}C{15mm}C{15mm}}
\toprule
& & \multicolumn{8}{c}{\textbf{Downsampling Factor [DF]}} \\
\cmidrule{3-10}
 & \textbf{Class Label} & \textbf{0.9} & \textbf{0.7} & \textbf{0.5} & \textbf{0.4} & \textbf{0.3} & \textbf{0.2} & \textbf{0.1} & \textbf{0.05} \\
\midrule
ribs & costal\_cartilages & \cellcolor{custom-green!50} $\num{7E-01}$ & \cellcolor{custom-red!80} $\num{2E-09}$ & \cellcolor{custom-red!80} $<\num{2E-16}$ & \cellcolor{custom-red!80} $<\num{2E-16}$ & \cellcolor{custom-red!80} $<\num{2E-16}$ & \cellcolor{custom-red!80} $<\num{2E-16}$ & \cellcolor{custom-red!80} $<\num{2E-16}$ & \cellcolor{custom-red!80} $<\num{2E-16}$ \\
ribs & rib\_left\_1 & \cellcolor{custom-green!50} $1.0$ & \cellcolor{custom-red!80} $\num{8E-04}$ & \cellcolor{custom-red!80} $<\num{2E-16}$ & \cellcolor{custom-red!80} $<\num{2E-16}$ & \cellcolor{custom-red!80} $<\num{2E-16}$ & \cellcolor{custom-red!80} $<\num{2E-16}$ & \cellcolor{custom-red!80} $<\num{2E-16}$ & \cellcolor{custom-red!80} $<\num{2E-16}$ \\
ribs & rib\_left\_10 & \cellcolor{custom-green!50} $1.0$ & \cellcolor{custom-red!80} $\num{1E-02}$ & \cellcolor{custom-red!80} $<\num{2E-16}$ & \cellcolor{custom-red!80} $<\num{2E-16}$ & \cellcolor{custom-red!80} $<\num{2E-16}$ & \cellcolor{custom-red!80} $<\num{2E-16}$ & \cellcolor{custom-red!80} $<\num{2E-16}$ & \cellcolor{custom-red!80} $<\num{2E-16}$ \\
ribs & rib\_left\_11 & \cellcolor{custom-green!50} $\num{4E-01}$ & \cellcolor{custom-red!80} $<\num{2E-16}$ & \cellcolor{custom-red!80} $<\num{2E-16}$ & \cellcolor{custom-red!80} $<\num{2E-16}$ & \cellcolor{custom-red!80} $<\num{2E-16}$ & \cellcolor{custom-red!80} $<\num{2E-16}$ & \cellcolor{custom-red!80} $<\num{2E-16}$ & \cellcolor{custom-red!80} $<\num{2E-16}$ \\
ribs & rib\_left\_12 & \cellcolor{custom-red!80} $\num{3E-02}$ & \cellcolor{custom-red!80} $<\num{2E-16}$ & \cellcolor{custom-red!80} $<\num{2E-16}$ & \cellcolor{custom-red!80} $<\num{2E-16}$ & \cellcolor{custom-red!80} $<\num{2E-16}$ & \cellcolor{custom-red!80} $<\num{2E-16}$ & \cellcolor{custom-red!80} $<\num{2E-16}$ & \cellcolor{custom-red!80} $<\num{2E-16}$ \\
ribs & rib\_left\_2 & \cellcolor{custom-red!80} $\num{5E-04}$ & \cellcolor{custom-red!80} $<\num{2E-16}$ & \cellcolor{custom-red!80} $<\num{2E-16}$ & \cellcolor{custom-red!80} $<\num{2E-16}$ & \cellcolor{custom-red!80} $<\num{2E-16}$ & \cellcolor{custom-red!80} $<\num{2E-16}$ & \cellcolor{custom-red!80} $<\num{2E-16}$ & \cellcolor{custom-red!80} $<\num{2E-16}$ \\
ribs & rib\_left\_3 & \cellcolor{custom-green!50} $1.0$ & \cellcolor{custom-red!80} $\num{1E-02}$ & \cellcolor{custom-red!80} $<\num{2E-16}$ & \cellcolor{custom-red!80} $<\num{2E-16}$ & \cellcolor{custom-red!80} $<\num{2E-16}$ & \cellcolor{custom-red!80} $<\num{2E-16}$ & \cellcolor{custom-red!80} $<\num{2E-16}$ & \cellcolor{custom-red!80} $<\num{2E-16}$ \\
ribs & rib\_left\_4 & \cellcolor{custom-red!80} $\num{2E-02}$ & \cellcolor{custom-red!80} $<\num{2E-16}$ & \cellcolor{custom-red!80} $<\num{2E-16}$ & \cellcolor{custom-red!80} $<\num{2E-16}$ & \cellcolor{custom-red!80} $<\num{2E-16}$ & \cellcolor{custom-red!80} $<\num{2E-16}$ & \cellcolor{custom-red!80} $<\num{2E-16}$ & \cellcolor{custom-red!80} $<\num{2E-16}$ \\
ribs & rib\_left\_5 & \cellcolor{custom-red!80} $\num{5E-04}$ & \cellcolor{custom-red!80} $<\num{2E-16}$ & \cellcolor{custom-red!80} $<\num{2E-16}$ & \cellcolor{custom-red!80} $<\num{2E-16}$ & \cellcolor{custom-red!80} $<\num{2E-16}$ & \cellcolor{custom-red!80} $<\num{2E-16}$ & \cellcolor{custom-red!80} $<\num{2E-16}$ & \cellcolor{custom-red!80} $<\num{2E-16}$ \\
ribs & rib\_left\_6 & \cellcolor{custom-red!80} $\num{5E-04}$ & \cellcolor{custom-red!80} $<\num{2E-16}$ & \cellcolor{custom-red!80} $<\num{2E-16}$ & \cellcolor{custom-red!80} $<\num{2E-16}$ & \cellcolor{custom-red!80} $<\num{2E-16}$ & \cellcolor{custom-red!80} $<\num{2E-16}$ & \cellcolor{custom-red!80} $<\num{2E-16}$ & \cellcolor{custom-red!80} $<\num{2E-16}$ \\
ribs & rib\_left\_7 & \cellcolor{custom-red!80} $\num{2E-05}$ & \cellcolor{custom-red!80} $<\num{2E-16}$ & \cellcolor{custom-red!80} $<\num{2E-16}$ & \cellcolor{custom-red!80} $<\num{2E-16}$ & \cellcolor{custom-red!80} $<\num{2E-16}$ & \cellcolor{custom-red!80} $<\num{2E-16}$ & \cellcolor{custom-red!80} $<\num{2E-16}$ & \cellcolor{custom-red!80} $<\num{2E-16}$ \\
ribs & rib\_left\_8 & \cellcolor{custom-red!80} $\num{4E-05}$ & \cellcolor{custom-red!80} $<\num{2E-16}$ & \cellcolor{custom-red!80} $<\num{2E-16}$ & \cellcolor{custom-red!80} $<\num{2E-16}$ & \cellcolor{custom-red!80} $<\num{2E-16}$ & \cellcolor{custom-red!80} $<\num{2E-16}$ & \cellcolor{custom-red!80} $<\num{2E-16}$ & \cellcolor{custom-red!80} $<\num{2E-16}$ \\
ribs & rib\_left\_9 & \cellcolor{custom-red!80} $\num{3E-03}$ & \cellcolor{custom-red!80} $<\num{2E-16}$ & \cellcolor{custom-red!80} $<\num{2E-16}$ & \cellcolor{custom-red!80} $<\num{2E-16}$ & \cellcolor{custom-red!80} $<\num{2E-16}$ & \cellcolor{custom-red!80} $<\num{2E-16}$ & \cellcolor{custom-red!80} $<\num{2E-16}$ & \cellcolor{custom-red!80} $<\num{2E-16}$ \\
ribs & rib\_right\_1 & \cellcolor{custom-green!50} $1.0$ & \cellcolor{custom-red!80} $<\num{2E-16}$ & \cellcolor{custom-red!80} $<\num{2E-16}$ & \cellcolor{custom-red!80} $<\num{2E-16}$ & \cellcolor{custom-red!80} $<\num{2E-16}$ & \cellcolor{custom-red!80} $<\num{2E-16}$ & \cellcolor{custom-red!80} $<\num{2E-16}$ & \cellcolor{custom-red!80} $<\num{2E-16}$ \\
ribs & rib\_right\_10 & \cellcolor{custom-green!50} $\num{8E-01}$ & \cellcolor{custom-red!80} $<\num{2E-16}$ & \cellcolor{custom-red!80} $<\num{2E-16}$ & \cellcolor{custom-red!80} $<\num{2E-16}$ & \cellcolor{custom-red!80} $<\num{2E-16}$ & \cellcolor{custom-red!80} $<\num{2E-16}$ & \cellcolor{custom-red!80} $<\num{2E-16}$ & \cellcolor{custom-red!80} $<\num{2E-16}$ \\
ribs & rib\_right\_11 & \cellcolor{custom-green!50} $1.0$ & \cellcolor{custom-red!80} $<\num{2E-16}$ & \cellcolor{custom-red!80} $<\num{2E-16}$ & \cellcolor{custom-red!80} $<\num{2E-16}$ & \cellcolor{custom-red!80} $<\num{2E-16}$ & \cellcolor{custom-red!80} $<\num{2E-16}$ & \cellcolor{custom-red!80} $<\num{2E-16}$ & \cellcolor{custom-red!80} $<\num{2E-16}$ \\
ribs & rib\_right\_12 & \cellcolor{custom-green!50} $1.0$ & \cellcolor{custom-red!80} $\num{1E-02}$ & \cellcolor{custom-red!80} $<\num{2E-16}$ & \cellcolor{custom-red!80} $<\num{2E-16}$ & \cellcolor{custom-red!80} $<\num{2E-16}$ & \cellcolor{custom-red!80} $<\num{2E-16}$ & \cellcolor{custom-red!80} $<\num{2E-16}$ & \cellcolor{custom-red!80} $<\num{2E-16}$ \\
ribs & rib\_right\_2 & \cellcolor{custom-green!50} $1.0$ & \cellcolor{custom-red!80} $\num{8E-10}$ & \cellcolor{custom-red!80} $<\num{2E-16}$ & \cellcolor{custom-red!80} $<\num{2E-16}$ & \cellcolor{custom-red!80} $<\num{2E-16}$ & \cellcolor{custom-red!80} $<\num{2E-16}$ & \cellcolor{custom-red!80} $<\num{2E-16}$ & \cellcolor{custom-red!80} $<\num{2E-16}$ \\
ribs & rib\_right\_3 & \cellcolor{custom-green!50} $1.0$ & \cellcolor{custom-red!80} $\num{5E-03}$ & \cellcolor{custom-red!80} $<\num{2E-16}$ & \cellcolor{custom-red!80} $<\num{2E-16}$ & \cellcolor{custom-red!80} $<\num{2E-16}$ & \cellcolor{custom-red!80} $<\num{2E-16}$ & \cellcolor{custom-red!80} $<\num{2E-16}$ & \cellcolor{custom-red!80} $<\num{2E-16}$ \\
ribs & rib\_right\_4 & \cellcolor{custom-red!80} $\num{1E-03}$ & \cellcolor{custom-red!80} $<\num{2E-16}$ & \cellcolor{custom-red!80} $<\num{2E-16}$ & \cellcolor{custom-red!80} $<\num{2E-16}$ & \cellcolor{custom-red!80} $<\num{2E-16}$ & \cellcolor{custom-red!80} $<\num{2E-16}$ & \cellcolor{custom-red!80} $<\num{2E-16}$ & \cellcolor{custom-red!80} $<\num{2E-16}$ \\
ribs & rib\_right\_5 & \cellcolor{custom-red!80} $\num{2E-02}$ & \cellcolor{custom-red!80} $<\num{2E-16}$ & \cellcolor{custom-red!80} $<\num{2E-16}$ & \cellcolor{custom-red!80} $<\num{2E-16}$ & \cellcolor{custom-red!80} $<\num{2E-16}$ & \cellcolor{custom-red!80} $<\num{2E-16}$ & \cellcolor{custom-red!80} $<\num{2E-16}$ & \cellcolor{custom-red!80} $<\num{2E-16}$ \\
ribs & rib\_right\_6 & \cellcolor{custom-green!50} $\num{6E-02}$ & \cellcolor{custom-red!80} $<\num{2E-16}$ & \cellcolor{custom-red!80} $<\num{2E-16}$ & \cellcolor{custom-red!80} $<\num{2E-16}$ & \cellcolor{custom-red!80} $<\num{2E-16}$ & \cellcolor{custom-red!80} $<\num{2E-16}$ & \cellcolor{custom-red!80} $<\num{2E-16}$ & \cellcolor{custom-red!80} $<\num{2E-16}$ \\
ribs & rib\_right\_7 & \cellcolor{custom-green!50} $\num{4E-01}$ & \cellcolor{custom-red!80} $\num{2E-12}$ & \cellcolor{custom-red!80} $<\num{2E-16}$ & \cellcolor{custom-red!80} $<\num{2E-16}$ & \cellcolor{custom-red!80} $<\num{2E-16}$ & \cellcolor{custom-red!80} $<\num{2E-16}$ & \cellcolor{custom-red!80} $<\num{2E-16}$ & \cellcolor{custom-red!80} $<\num{2E-16}$ \\
ribs & rib\_right\_8 & \cellcolor{custom-red!80} $\num{1E-03}$ & \cellcolor{custom-red!80} $<\num{2E-16}$ & \cellcolor{custom-red!80} $<\num{2E-16}$ & \cellcolor{custom-red!80} $<\num{2E-16}$ & \cellcolor{custom-red!80} $<\num{2E-16}$ & \cellcolor{custom-red!80} $<\num{2E-16}$ & \cellcolor{custom-red!80} $<\num{2E-16}$ & \cellcolor{custom-red!80} $<\num{2E-16}$ \\
ribs & rib\_right\_9 & \cellcolor{custom-red!80} $\num{2E-04}$ & \cellcolor{custom-red!80} $<\num{2E-16}$ & \cellcolor{custom-red!80} $<\num{2E-16}$ & \cellcolor{custom-red!80} $<\num{2E-16}$ & \cellcolor{custom-red!80} $<\num{2E-16}$ & \cellcolor{custom-red!80} $<\num{2E-16}$ & \cellcolor{custom-red!80} $<\num{2E-16}$ & \cellcolor{custom-red!80} $<\num{2E-16}$ \\
ribs & sternum & \cellcolor{custom-red!80} $\num{2E-05}$ & \cellcolor{custom-red!80} $<\num{2E-16}$ & \cellcolor{custom-red!80} $<\num{2E-16}$ & \cellcolor{custom-red!80} $<\num{2E-16}$ & \cellcolor{custom-red!80} $<\num{2E-16}$ & \cellcolor{custom-red!80} $<\num{2E-16}$ & \cellcolor{custom-red!80} $<\num{2E-16}$ & \cellcolor{custom-red!80} $<\num{2E-16}$ \\
\midrule
vertebrae & sacrum & \cellcolor{custom-red!80} $\num{9E-03}$ & \cellcolor{custom-red!80} $<\num{2E-16}$ & \cellcolor{custom-red!80} $<\num{2E-16}$ & \cellcolor{custom-red!80} $<\num{2E-16}$ & \cellcolor{custom-red!80} $<\num{2E-16}$ & \cellcolor{custom-red!80} $<\num{2E-16}$ & \cellcolor{custom-red!80} $<\num{2E-16}$ & \cellcolor{custom-red!80} $<\num{2E-16}$ \\
vertebrae & vertebrae\_C1 & \cellcolor{custom-red!80} $\num{1E-02}$ & \cellcolor{custom-red!80} $\num{2E-15}$ & \cellcolor{custom-red!80} $<\num{2E-16}$ & \cellcolor{custom-red!80} $<\num{2E-16}$ & \cellcolor{custom-red!80} $<\num{2E-16}$ & \cellcolor{custom-red!80} $<\num{2E-16}$ & \cellcolor{custom-red!80} $<\num{2E-16}$ & \cellcolor{custom-red!80} $<\num{2E-16}$ \\
vertebrae & vertebrae\_C2 & \cellcolor{custom-red!80} $\num{3E-02}$ & \cellcolor{custom-red!80} $<\num{2E-16}$ & \cellcolor{custom-red!80} $<\num{2E-16}$ & \cellcolor{custom-red!80} $<\num{2E-16}$ & \cellcolor{custom-red!80} $<\num{2E-16}$ & \cellcolor{custom-red!80} $<\num{2E-16}$ & \cellcolor{custom-red!80} $<\num{2E-16}$ & \cellcolor{custom-red!80} $<\num{2E-16}$ \\
vertebrae & vertebrae\_C3 & \cellcolor{custom-green!50} $1.0$ & \cellcolor{custom-red!80} $\num{8E-15}$ & \cellcolor{custom-red!80} $<\num{2E-16}$ & \cellcolor{custom-red!80} $<\num{2E-16}$ & \cellcolor{custom-red!80} $<\num{2E-16}$ & \cellcolor{custom-red!80} $<\num{2E-16}$ & \cellcolor{custom-red!80} $<\num{2E-16}$ & \cellcolor{custom-red!80} $<\num{2E-16}$ \\
vertebrae & vertebrae\_C4 & \cellcolor{custom-green!50} $\num{1E-01}$ & \cellcolor{custom-red!80} $\num{1E-10}$ & \cellcolor{custom-red!80} $<\num{2E-16}$ & \cellcolor{custom-red!80} $<\num{2E-16}$ & \cellcolor{custom-red!80} $<\num{2E-16}$ & \cellcolor{custom-red!80} $<\num{2E-16}$ & \cellcolor{custom-red!80} $<\num{2E-16}$ & \cellcolor{custom-red!80} $<\num{2E-16}$ \\
vertebrae & vertebrae\_C5 & \cellcolor{custom-red!80} $\num{3E-03}$ & \cellcolor{custom-red!80} $\num{9E-15}$ & \cellcolor{custom-red!80} $<\num{2E-16}$ & \cellcolor{custom-red!80} $<\num{2E-16}$ & \cellcolor{custom-red!80} $<\num{2E-16}$ & \cellcolor{custom-red!80} $<\num{2E-16}$ & \cellcolor{custom-red!80} $<\num{2E-16}$ & \cellcolor{custom-red!80} $<\num{2E-16}$ \\
vertebrae & vertebrae\_C6 & \cellcolor{custom-red!80} $\num{2E-05}$ & \cellcolor{custom-red!80} $\num{3E-16}$ & \cellcolor{custom-red!80} $<\num{2E-16}$ & \cellcolor{custom-red!80} $<\num{2E-16}$ & \cellcolor{custom-red!80} $<\num{2E-16}$ & \cellcolor{custom-red!80} $<\num{2E-16}$ & \cellcolor{custom-red!80} $<\num{2E-16}$ & \cellcolor{custom-red!80} $<\num{2E-16}$ \\
vertebrae & vertebrae\_C7 & \cellcolor{custom-green!50} $\num{2E-01}$ & \cellcolor{custom-red!80} $<\num{2E-16}$ & \cellcolor{custom-red!80} $<\num{2E-16}$ & \cellcolor{custom-red!80} $<\num{2E-16}$ & \cellcolor{custom-red!80} $<\num{2E-16}$ & \cellcolor{custom-red!80} $<\num{2E-16}$ & \cellcolor{custom-red!80} $<\num{2E-16}$ & \cellcolor{custom-red!80} $<\num{2E-16}$ \\
vertebrae & vertebrae\_L1 & \cellcolor{custom-red!80} $\num{8E-05}$ & \cellcolor{custom-red!80} $<\num{2E-16}$ & \cellcolor{custom-red!80} $<\num{2E-16}$ & \cellcolor{custom-red!80} $<\num{2E-16}$ & \cellcolor{custom-red!80} $<\num{2E-16}$ & \cellcolor{custom-red!80} $<\num{2E-16}$ & \cellcolor{custom-red!80} $<\num{2E-16}$ & \cellcolor{custom-red!80} $<\num{2E-16}$ \\
vertebrae & vertebrae\_L2 & \cellcolor{custom-red!80} $\num{2E-02}$ & \cellcolor{custom-red!80} $<\num{2E-16}$ & \cellcolor{custom-red!80} $<\num{2E-16}$ & \cellcolor{custom-red!80} $<\num{2E-16}$ & \cellcolor{custom-red!80} $<\num{2E-16}$ & \cellcolor{custom-red!80} $<\num{2E-16}$ & \cellcolor{custom-red!80} $<\num{2E-16}$ & \cellcolor{custom-red!80} $<\num{2E-16}$ \\
vertebrae & vertebrae\_L3 & \cellcolor{custom-green!50} $\num{9E-01}$ & \cellcolor{custom-red!80} $\num{1E-07}$ & \cellcolor{custom-red!80} $<\num{2E-16}$ & \cellcolor{custom-red!80} $<\num{2E-16}$ & \cellcolor{custom-red!80} $<\num{2E-16}$ & \cellcolor{custom-red!80} $<\num{2E-16}$ & \cellcolor{custom-red!80} $<\num{2E-16}$ & \cellcolor{custom-red!80} $<\num{2E-16}$ \\
vertebrae & vertebrae\_L4 & \cellcolor{custom-green!50} $\num{9E-01}$ & \cellcolor{custom-red!80} $\num{2E-07}$ & \cellcolor{custom-red!80} $<\num{2E-16}$ & \cellcolor{custom-red!80} $<\num{2E-16}$ & \cellcolor{custom-red!80} $<\num{2E-16}$ & \cellcolor{custom-red!80} $<\num{2E-16}$ & \cellcolor{custom-red!80} $<\num{2E-16}$ & \cellcolor{custom-red!80} $<\num{2E-16}$ \\
vertebrae & vertebrae\_L5 & \cellcolor{custom-green!50} $\num{3E-01}$ & \cellcolor{custom-red!80} $<\num{2E-16}$ & \cellcolor{custom-red!80} $<\num{2E-16}$ & \cellcolor{custom-red!80} $<\num{2E-16}$ & \cellcolor{custom-red!80} $<\num{2E-16}$ & \cellcolor{custom-red!80} $<\num{2E-16}$ & \cellcolor{custom-red!80} $<\num{2E-16}$ & \cellcolor{custom-red!80} $<\num{2E-16}$ \\
vertebrae & vertebrae\_S1 & \cellcolor{custom-red!80} $\num{8E-04}$ & \cellcolor{custom-red!80} $<\num{2E-16}$ & \cellcolor{custom-red!80} $<\num{2E-16}$ & \cellcolor{custom-red!80} $<\num{2E-16}$ & \cellcolor{custom-red!80} $<\num{2E-16}$ & \cellcolor{custom-red!80} $<\num{2E-16}$ & \cellcolor{custom-red!80} $<\num{2E-16}$ & \cellcolor{custom-red!80} $<\num{2E-16}$ \\
vertebrae & vertebrae\_T1 & \cellcolor{custom-green!50} $1.0$ & \cellcolor{custom-red!80} $<\num{2E-16}$ & \cellcolor{custom-red!80} $<\num{2E-16}$ & \cellcolor{custom-red!80} $<\num{2E-16}$ & \cellcolor{custom-red!80} $<\num{2E-16}$ & \cellcolor{custom-red!80} $<\num{2E-16}$ & \cellcolor{custom-red!80} $<\num{2E-16}$ & \cellcolor{custom-red!80} $<\num{2E-16}$ \\
vertebrae & vertebrae\_T10 & \cellcolor{custom-red!80} $\num{7E-06}$ & \cellcolor{custom-red!80} $<\num{2E-16}$ & \cellcolor{custom-red!80} $<\num{2E-16}$ & \cellcolor{custom-red!80} $<\num{2E-16}$ & \cellcolor{custom-red!80} $<\num{2E-16}$ & \cellcolor{custom-red!80} $<\num{2E-16}$ & \cellcolor{custom-red!80} $<\num{2E-16}$ & \cellcolor{custom-red!80} $<\num{2E-16}$ \\
vertebrae & vertebrae\_T11 & \cellcolor{custom-green!50} $\num{2E-01}$ & \cellcolor{custom-red!80} $\num{3E-12}$ & \cellcolor{custom-red!80} $<\num{2E-16}$ & \cellcolor{custom-red!80} $<\num{2E-16}$ & \cellcolor{custom-red!80} $<\num{2E-16}$ & \cellcolor{custom-red!80} $<\num{2E-16}$ & \cellcolor{custom-red!80} $<\num{2E-16}$ & \cellcolor{custom-red!80} $<\num{2E-16}$ \\
vertebrae & vertebrae\_T12 & \cellcolor{custom-red!80} $\num{2E-06}$ & \cellcolor{custom-red!80} $<\num{2E-16}$ & \cellcolor{custom-red!80} $<\num{2E-16}$ & \cellcolor{custom-red!80} $<\num{2E-16}$ & \cellcolor{custom-red!80} $<\num{2E-16}$ & \cellcolor{custom-red!80} $<\num{2E-16}$ & \cellcolor{custom-red!80} $<\num{2E-16}$ & \cellcolor{custom-red!80} $<\num{2E-16}$ \\
vertebrae & vertebrae\_T2 & \cellcolor{custom-green!50} $1.0$ & \cellcolor{custom-red!80} $<\num{2E-16}$ & \cellcolor{custom-red!80} $<\num{2E-16}$ & \cellcolor{custom-red!80} $<\num{2E-16}$ & \cellcolor{custom-red!80} $<\num{2E-16}$ & \cellcolor{custom-red!80} $<\num{2E-16}$ & \cellcolor{custom-red!80} $<\num{2E-16}$ & \cellcolor{custom-red!80} $<\num{2E-16}$ \\
vertebrae & vertebrae\_T3 & \cellcolor{custom-green!50} $\num{8E-01}$ & \cellcolor{custom-red!80} $\num{9E-10}$ & \cellcolor{custom-red!80} $<\num{2E-16}$ & \cellcolor{custom-red!80} $<\num{2E-16}$ & \cellcolor{custom-red!80} $<\num{2E-16}$ & \cellcolor{custom-red!80} $<\num{2E-16}$ & \cellcolor{custom-red!80} $<\num{2E-16}$ & \cellcolor{custom-red!80} $<\num{2E-16}$ \\
vertebrae & vertebrae\_T4 & \cellcolor{custom-red!80} $\num{5E-04}$ & \cellcolor{custom-red!80} $<\num{2E-16}$ & \cellcolor{custom-red!80} $<\num{2E-16}$ & \cellcolor{custom-red!80} $<\num{2E-16}$ & \cellcolor{custom-red!80} $<\num{2E-16}$ & \cellcolor{custom-red!80} $<\num{2E-16}$ & \cellcolor{custom-red!80} $<\num{2E-16}$ & \cellcolor{custom-red!80} $<\num{2E-16}$ \\
vertebrae & vertebrae\_T5 & \cellcolor{custom-red!80} $\num{3E-02}$ & \cellcolor{custom-red!80} $<\num{2E-16}$ & \cellcolor{custom-red!80} $<\num{2E-16}$ & \cellcolor{custom-red!80} $<\num{2E-16}$ & \cellcolor{custom-red!80} $<\num{2E-16}$ & \cellcolor{custom-red!80} $<\num{2E-16}$ & \cellcolor{custom-red!80} $<\num{2E-16}$ & \cellcolor{custom-red!80} $<\num{2E-16}$ \\
vertebrae & vertebrae\_T6 & \cellcolor{custom-red!80} $\num{8E-05}$ & \cellcolor{custom-red!80} $<\num{2E-16}$ & \cellcolor{custom-red!80} $<\num{2E-16}$ & \cellcolor{custom-red!80} $<\num{2E-16}$ & \cellcolor{custom-red!80} $<\num{2E-16}$ & \cellcolor{custom-red!80} $<\num{2E-16}$ & \cellcolor{custom-red!80} $<\num{2E-16}$ & \cellcolor{custom-red!80} $<\num{2E-16}$ \\
vertebrae & vertebrae\_T7 & \cellcolor{custom-red!80} $\num{1E-02}$ & \cellcolor{custom-red!80} $\num{4E-11}$ & \cellcolor{custom-red!80} $<\num{2E-16}$ & \cellcolor{custom-red!80} $<\num{2E-16}$ & \cellcolor{custom-red!80} $<\num{2E-16}$ & \cellcolor{custom-red!80} $<\num{2E-16}$ & \cellcolor{custom-red!80} $<\num{2E-16}$ & \cellcolor{custom-red!80} $<\num{2E-16}$ \\
vertebrae & vertebrae\_T8 & \cellcolor{custom-red!80} $\num{2E-07}$ & \cellcolor{custom-red!80} $<\num{2E-16}$ & \cellcolor{custom-red!80} $<\num{2E-16}$ & \cellcolor{custom-red!80} $<\num{2E-16}$ & \cellcolor{custom-red!80} $<\num{2E-16}$ & \cellcolor{custom-red!80} $<\num{2E-16}$ & \cellcolor{custom-red!80} $<\num{2E-16}$ & \cellcolor{custom-red!80} $<\num{2E-16}$ \\
vertebrae & vertebrae\_T9 & \cellcolor{custom-red!80} $\num{1E-05}$ & \cellcolor{custom-red!80} $<\num{2E-16}$ & \cellcolor{custom-red!80} $<\num{2E-16}$ & \cellcolor{custom-red!80} $<\num{2E-16}$ & \cellcolor{custom-red!80} $<\num{2E-16}$ & \cellcolor{custom-red!80} $<\num{2E-16}$ & \cellcolor{custom-red!80} $<\num{2E-16}$ & \cellcolor{custom-red!80} $<\num{2E-16}$ \\
\bottomrule
\end{tabular}

%% file: tables/pval_fine_fast_1.tex
\newcolumntype{C}[1]{>{\centering\arraybackslash}p{#1}}

\begin{tabular}{llC{15mm}C{15mm}C{15mm}C{15mm}C{15mm}C{15mm}C{15mm}C{15mm}}
\toprule
& & \multicolumn{8}{c}{\textbf{Downsampling Factor [DF]}} \\
\cmidrule{3-10}
 & \textbf{Class Label} & \textbf{0.9} & \textbf{0.7} & \textbf{0.5} & \textbf{0.4} & \textbf{0.3} & \textbf{0.2} & \textbf{0.1} & \textbf{0.05} \\
\midrule
cardiac & aorta & \cellcolor{custom-green!50} $1.0$ & \cellcolor{custom-green!50} $1.0$ & \cellcolor{custom-green!50} $1.0$ & \cellcolor{custom-green!50} $1.0$ & \cellcolor{custom-red!80} $\num{9E-06}$ & \cellcolor{custom-red!80} $<\num{2E-16}$ & \cellcolor{custom-red!80} $<\num{2E-16}$ & \cellcolor{custom-red!80} $<\num{2E-16}$ \\
cardiac & atrial\_appendage\_left & \cellcolor{custom-green!50} $1.0$ & \cellcolor{custom-green!50} $1.0$ & \cellcolor{custom-green!50} $1.0$ & \cellcolor{custom-green!50} $\num{6E-01}$ & \cellcolor{custom-red!80} $\num{4E-13}$ & \cellcolor{custom-red!80} $<\num{2E-16}$ & \cellcolor{custom-red!80} $<\num{2E-16}$ & \cellcolor{custom-red!80} $<\num{2E-16}$ \\
cardiac & brachiocephalic\_trunk & \cellcolor{custom-green!50} $1.0$ & \cellcolor{custom-green!50} $1.0$ & \cellcolor{custom-green!50} $1.0$ & \cellcolor{custom-green!50} $1.0$ & \cellcolor{custom-red!80} $\num{3E-06}$ & \cellcolor{custom-red!80} $<\num{2E-16}$ & \cellcolor{custom-red!80} $<\num{2E-16}$ & \cellcolor{custom-red!80} $<\num{2E-16}$ \\
cardiac & brachiocephalic\_vein\_left & \cellcolor{custom-green!50} $1.0$ & \cellcolor{custom-green!50} $1.0$ & \cellcolor{custom-green!50} $1.0$ & \cellcolor{custom-green!50} $\num{6E-02}$ & \cellcolor{custom-red!80} $<\num{2E-16}$ & \cellcolor{custom-red!80} $<\num{2E-16}$ & \cellcolor{custom-red!80} $<\num{2E-16}$ & \cellcolor{custom-red!80} $<\num{2E-16}$ \\
\bottomrule
\end{tabular}

%% file: tables/pval_fine_fast_2.tex
\newcolumntype{C}[1]{>{\centering\arraybackslash}p{#1}}

\begin{tabular}{llC{15mm}C{15mm}C{15mm}C{15mm}C{15mm}C{15mm}C{15mm}C{15mm}}
\toprule
& & \multicolumn{8}{c}{\textbf{Downsampling Factor [DF]}} \\
\cmidrule{3-10}
 & \textbf{Class Label} & \textbf{0.9} & \textbf{0.7} & \textbf{0.5} & \textbf{0.4} & \textbf{0.3} & \textbf{0.2} & \textbf{0.1} & \textbf{0.05} \\
\midrule
ribs & costal\_cartilages & \cellcolor{custom-green!50} $1.0$ & \cellcolor{custom-green!50} $1.0$ & \cellcolor{custom-green!50} $1.0$ & \cellcolor{custom-green!50} $\num{6E-01}$ & \cellcolor{custom-red!80} $\num{9E-07}$ & \cellcolor{custom-red!80} $<\num{2E-16}$ & \cellcolor{custom-red!80} $<\num{2E-16}$ & \cellcolor{custom-red!80} $<\num{2E-16}$ \\
ribs & rib\_left\_1 & \cellcolor{custom-green!50} $1.0$ & \cellcolor{custom-green!50} $1.0$ & \cellcolor{custom-green!50} $1.0$ & \cellcolor{custom-green!50} $1.0$ & \cellcolor{custom-red!80} $\num{1E-03}$ & \cellcolor{custom-red!80} $<\num{2E-16}$ & \cellcolor{custom-red!80} $<\num{2E-16}$ & \cellcolor{custom-red!80} $<\num{2E-16}$ \\
ribs & rib\_left\_10 & \cellcolor{custom-green!50} $1.0$ & \cellcolor{custom-green!50} $1.0$ & \cellcolor{custom-green!50} $1.0$ & \cellcolor{custom-green!50} $1.0$ & \cellcolor{custom-red!80} $\num{3E-03}$ & \cellcolor{custom-red!80} $<\num{2E-16}$ & \cellcolor{custom-red!80} $<\num{2E-16}$ & \cellcolor{custom-red!80} $<\num{2E-16}$ \\
ribs & rib\_left\_11 & \cellcolor{custom-green!50} $1.0$ & \cellcolor{custom-green!50} $1.0$ & \cellcolor{custom-red!80} $\num{5E-07}$ & \cellcolor{custom-red!80} $<\num{2E-16}$ & \cellcolor{custom-red!80} $<\num{2E-16}$ & \cellcolor{custom-red!80} $<\num{2E-16}$ & \cellcolor{custom-red!80} $<\num{2E-16}$ & \cellcolor{custom-red!80} $<\num{2E-16}$ \\
ribs & rib\_left\_12 & \cellcolor{custom-green!50} $1.0$ & \cellcolor{custom-green!50} $1.0$ & \cellcolor{custom-red!80} $\num{3E-06}$ & \cellcolor{custom-red!80} $<\num{2E-16}$ & \cellcolor{custom-red!80} $<\num{2E-16}$ & \cellcolor{custom-red!80} $<\num{2E-16}$ & \cellcolor{custom-red!80} $<\num{2E-16}$ & \cellcolor{custom-red!80} $<\num{2E-16}$ \\
ribs & rib\_left\_2 & \cellcolor{custom-green!50} $1.0$ & \cellcolor{custom-green!50} $1.0$ & \cellcolor{custom-green!50} $1.0$ & \cellcolor{custom-green!50} $1.0$ & \cellcolor{custom-red!80} $\num{1E-04}$ & \cellcolor{custom-red!80} $<\num{2E-16}$ & \cellcolor{custom-red!80} $<\num{2E-16}$ & \cellcolor{custom-red!80} $<\num{2E-16}$ \\
ribs & rib\_left\_3 & \cellcolor{custom-green!50} $1.0$ & \cellcolor{custom-green!50} $1.0$ & \cellcolor{custom-green!50} $1.0$ & \cellcolor{custom-green!50} $1.0$ & \cellcolor{custom-red!80} $\num{2E-06}$ & \cellcolor{custom-red!80} $<\num{2E-16}$ & \cellcolor{custom-red!80} $<\num{2E-16}$ & \cellcolor{custom-red!80} $<\num{2E-16}$ \\
ribs & rib\_left\_4 & \cellcolor{custom-green!50} $1.0$ & \cellcolor{custom-green!50} $1.0$ & \cellcolor{custom-green!50} $1.0$ & \cellcolor{custom-green!50} $1.0$ & \cellcolor{custom-red!80} $\num{4E-09}$ & \cellcolor{custom-red!80} $<\num{2E-16}$ & \cellcolor{custom-red!80} $<\num{2E-16}$ & \cellcolor{custom-red!80} $<\num{2E-16}$ \\
ribs & rib\_left\_5 & \cellcolor{custom-green!50} $1.0$ & \cellcolor{custom-green!50} $1.0$ & \cellcolor{custom-green!50} $1.0$ & \cellcolor{custom-green!50} $1.0$ & \cellcolor{custom-red!80} $\num{3E-09}$ & \cellcolor{custom-red!80} $<\num{2E-16}$ & \cellcolor{custom-red!80} $<\num{2E-16}$ & \cellcolor{custom-red!80} $<\num{2E-16}$ \\
ribs & rib\_left\_6 & \cellcolor{custom-green!50} $1.0$ & \cellcolor{custom-green!50} $1.0$ & \cellcolor{custom-green!50} $1.0$ & \cellcolor{custom-green!50} $\num{8E-01}$ & \cellcolor{custom-red!80} $\num{6E-12}$ & \cellcolor{custom-red!80} $<\num{2E-16}$ & \cellcolor{custom-red!80} $<\num{2E-16}$ & \cellcolor{custom-red!80} $<\num{2E-16}$ \\
ribs & rib\_left\_7 & \cellcolor{custom-green!50} $1.0$ & \cellcolor{custom-green!50} $1.0$ & \cellcolor{custom-green!50} $1.0$ & \cellcolor{custom-green!50} $1.0$ & \cellcolor{custom-red!80} $\num{1E-06}$ & \cellcolor{custom-red!80} $<\num{2E-16}$ & \cellcolor{custom-red!80} $<\num{2E-16}$ & \cellcolor{custom-red!80} $<\num{2E-16}$ \\
ribs & rib\_left\_8 & \cellcolor{custom-green!50} $1.0$ & \cellcolor{custom-green!50} $1.0$ & \cellcolor{custom-green!50} $1.0$ & \cellcolor{custom-green!50} $1.0$ & \cellcolor{custom-red!80} $\num{1E-04}$ & \cellcolor{custom-red!80} $<\num{2E-16}$ & \cellcolor{custom-red!80} $<\num{2E-16}$ & \cellcolor{custom-red!80} $<\num{2E-16}$ \\
ribs & rib\_left\_9 & \cellcolor{custom-green!50} $1.0$ & \cellcolor{custom-green!50} $1.0$ & \cellcolor{custom-green!50} $1.0$ & \cellcolor{custom-green!50} $1.0$ & \cellcolor{custom-red!80} $\num{4E-04}$ & \cellcolor{custom-red!80} $<\num{2E-16}$ & \cellcolor{custom-red!80} $<\num{2E-16}$ & \cellcolor{custom-red!80} $<\num{2E-16}$ \\
ribs & rib\_right\_1 & \cellcolor{custom-green!50} $1.0$ & \cellcolor{custom-green!50} $1.0$ & \cellcolor{custom-green!50} $1.0$ & \cellcolor{custom-green!50} $1.0$ & \cellcolor{custom-green!50} $\num{6E-02}$ & \cellcolor{custom-red!80} $<\num{2E-16}$ & \cellcolor{custom-red!80} $<\num{2E-16}$ & \cellcolor{custom-red!80} $<\num{2E-16}$ \\
ribs & rib\_right\_10 & \cellcolor{custom-green!50} $1.0$ & \cellcolor{custom-green!50} $1.0$ & \cellcolor{custom-green!50} $1.0$ & \cellcolor{custom-green!50} $1.0$ & \cellcolor{custom-red!80} $\num{7E-04}$ & \cellcolor{custom-red!80} $<\num{2E-16}$ & \cellcolor{custom-red!80} $<\num{2E-16}$ & \cellcolor{custom-red!80} $<\num{2E-16}$ \\
ribs & rib\_right\_11 & \cellcolor{custom-green!50} $1.0$ & \cellcolor{custom-green!50} $1.0$ & \cellcolor{custom-green!50} $1.0$ & \cellcolor{custom-green!50} $1.0$ & \cellcolor{custom-green!50} $1.0$ & \cellcolor{custom-red!80} $<\num{2E-16}$ & \cellcolor{custom-red!80} $<\num{2E-16}$ & \cellcolor{custom-red!80} $<\num{2E-16}$ \\
ribs & rib\_right\_12 & \cellcolor{custom-green!50} $1.0$ & \cellcolor{custom-green!50} $1.0$ & \cellcolor{custom-red!80} $\num{2E-03}$ & \cellcolor{custom-red!80} $<\num{2E-16}$ & \cellcolor{custom-red!80} $<\num{2E-16}$ & \cellcolor{custom-red!80} $<\num{2E-16}$ & \cellcolor{custom-red!80} $<\num{2E-16}$ & \cellcolor{custom-red!80} $<\num{2E-16}$ \\
ribs & rib\_right\_2 & \cellcolor{custom-green!50} $1.0$ & \cellcolor{custom-green!50} $1.0$ & \cellcolor{custom-green!50} $1.0$ & \cellcolor{custom-green!50} $1.0$ & \cellcolor{custom-red!80} $\num{5E-05}$ & \cellcolor{custom-red!80} $<\num{2E-16}$ & \cellcolor{custom-red!80} $<\num{2E-16}$ & \cellcolor{custom-red!80} $<\num{2E-16}$ \\
ribs & rib\_right\_3 & \cellcolor{custom-green!50} $1.0$ & \cellcolor{custom-green!50} $1.0$ & \cellcolor{custom-green!50} $1.0$ & \cellcolor{custom-green!50} $1.0$ & \cellcolor{custom-red!80} $\num{2E-09}$ & \cellcolor{custom-red!80} $<\num{2E-16}$ & \cellcolor{custom-red!80} $<\num{2E-16}$ & \cellcolor{custom-red!80} $<\num{2E-16}$ \\
ribs & rib\_right\_4 & \cellcolor{custom-green!50} $1.0$ & \cellcolor{custom-green!50} $1.0$ & \cellcolor{custom-green!50} $1.0$ & \cellcolor{custom-green!50} $1.0$ & \cellcolor{custom-red!80} $\num{6E-06}$ & \cellcolor{custom-red!80} $<\num{2E-16}$ & \cellcolor{custom-red!80} $<\num{2E-16}$ & \cellcolor{custom-red!80} $<\num{2E-16}$ \\
ribs & rib\_right\_5 & \cellcolor{custom-green!50} $1.0$ & \cellcolor{custom-green!50} $1.0$ & \cellcolor{custom-green!50} $1.0$ & \cellcolor{custom-green!50} $\num{7E-01}$ & \cellcolor{custom-red!80} $\num{1E-10}$ & \cellcolor{custom-red!80} $<\num{2E-16}$ & \cellcolor{custom-red!80} $<\num{2E-16}$ & \cellcolor{custom-red!80} $<\num{2E-16}$ \\
ribs & rib\_right\_6 & \cellcolor{custom-green!50} $1.0$ & \cellcolor{custom-green!50} $1.0$ & \cellcolor{custom-green!50} $1.0$ & \cellcolor{custom-green!50} $1.0$ & \cellcolor{custom-red!80} $\num{1E-09}$ & \cellcolor{custom-red!80} $<\num{2E-16}$ & \cellcolor{custom-red!80} $<\num{2E-16}$ & \cellcolor{custom-red!80} $<\num{2E-16}$ \\
ribs & rib\_right\_7 & \cellcolor{custom-green!50} $1.0$ & \cellcolor{custom-green!50} $1.0$ & \cellcolor{custom-green!50} $1.0$ & \cellcolor{custom-green!50} $1.0$ & \cellcolor{custom-green!50} $\num{9E-02}$ & \cellcolor{custom-red!80} $<\num{2E-16}$ & \cellcolor{custom-red!80} $<\num{2E-16}$ & \cellcolor{custom-red!80} $<\num{2E-16}$ \\
ribs & rib\_right\_8 & \cellcolor{custom-green!50} $1.0$ & \cellcolor{custom-green!50} $1.0$ & \cellcolor{custom-green!50} $1.0$ & \cellcolor{custom-green!50} $1.0$ & \cellcolor{custom-green!50} $\num{2E-01}$ & \cellcolor{custom-red!80} $<\num{2E-16}$ & \cellcolor{custom-red!80} $<\num{2E-16}$ & \cellcolor{custom-red!80} $<\num{2E-16}$ \\
ribs & rib\_right\_9 & \cellcolor{custom-green!50} $1.0$ & \cellcolor{custom-green!50} $1.0$ & \cellcolor{custom-green!50} $1.0$ & \cellcolor{custom-green!50} $1.0$ & \cellcolor{custom-red!80} $\num{4E-02}$ & \cellcolor{custom-red!80} $<\num{2E-16}$ & \cellcolor{custom-red!80} $<\num{2E-16}$ & \cellcolor{custom-red!80} $<\num{2E-16}$ \\
ribs & sternum & \cellcolor{custom-green!50} $1.0$ & \cellcolor{custom-green!50} $1.0$ & \cellcolor{custom-green!50} $1.0$ & \cellcolor{custom-green!50} $\num{1E-01}$ & \cellcolor{custom-red!80} $\num{4E-10}$ & \cellcolor{custom-red!80} $<\num{2E-16}$ & \cellcolor{custom-red!80} $<\num{2E-16}$ & \cellcolor{custom-red!80} $<\num{2E-16}$ \\
vertebrae & sacrum & \cellcolor{custom-green!50} $1.0$ & \cellcolor{custom-green!50} $1.0$ & \cellcolor{custom-green!50} $1.0$ & \cellcolor{custom-green!50} $\num{7E-01}$ & \cellcolor{custom-red!80} $\num{2E-06}$ & \cellcolor{custom-red!80} $<\num{2E-16}$ & \cellcolor{custom-red!80} $<\num{2E-16}$ & \cellcolor{custom-red!80} $<\num{2E-16}$ \\
vertebrae & vertebrae\_C1 & \cellcolor{custom-green!50} $1.0$ & \cellcolor{custom-green!50} $1.0$ & \cellcolor{custom-red!80} $\num{1E-06}$ & \cellcolor{custom-red!80} $<\num{2E-16}$ & \cellcolor{custom-red!80} $<\num{2E-16}$ & \cellcolor{custom-red!80} $<\num{2E-16}$ & \cellcolor{custom-red!80} $<\num{2E-16}$ & \cellcolor{custom-red!80} $<\num{2E-16}$ \\
vertebrae & vertebrae\_C2 & \cellcolor{custom-green!50} $1.0$ & \cellcolor{custom-green!50} $1.0$ & \cellcolor{custom-red!80} $\num{7E-07}$ & \cellcolor{custom-red!80} $<\num{2E-16}$ & \cellcolor{custom-red!80} $<\num{2E-16}$ & \cellcolor{custom-red!80} $<\num{2E-16}$ & \cellcolor{custom-red!80} $<\num{2E-16}$ & \cellcolor{custom-red!80} $<\num{2E-16}$ \\
vertebrae & vertebrae\_C3 & \cellcolor{custom-green!50} $1.0$ & \cellcolor{custom-green!50} $1.0$ & \cellcolor{custom-red!80} $\num{2E-02}$ & \cellcolor{custom-red!80} $<\num{2E-16}$ & \cellcolor{custom-red!80} $<\num{2E-16}$ & \cellcolor{custom-red!80} $<\num{2E-16}$ & \cellcolor{custom-red!80} $<\num{2E-16}$ & \cellcolor{custom-red!80} $<\num{2E-16}$ \\
vertebrae & vertebrae\_C4 & \cellcolor{custom-green!50} $1.0$ & \cellcolor{custom-green!50} $1.0$ & \cellcolor{custom-green!50} $\num{5E-02}$ & \cellcolor{custom-red!80} $\num{1E-12}$ & \cellcolor{custom-red!80} $<\num{2E-16}$ & \cellcolor{custom-red!80} $<\num{2E-16}$ & \cellcolor{custom-red!80} $<\num{2E-16}$ & \cellcolor{custom-red!80} $<\num{2E-16}$ \\
vertebrae & vertebrae\_C5 & \cellcolor{custom-green!50} $1.0$ & \cellcolor{custom-green!50} $1.0$ & \cellcolor{custom-red!80} $\num{2E-05}$ & \cellcolor{custom-red!80} $\num{3E-15}$ & \cellcolor{custom-red!80} $<\num{2E-16}$ & \cellcolor{custom-red!80} $<\num{2E-16}$ & \cellcolor{custom-red!80} $<\num{2E-16}$ & \cellcolor{custom-red!80} $<\num{2E-16}$ \\
vertebrae & vertebrae\_C6 & \cellcolor{custom-green!50} $1.0$ & \cellcolor{custom-green!50} $1.0$ & \cellcolor{custom-red!80} $\num{2E-05}$ & \cellcolor{custom-red!80} $\num{9E-14}$ & \cellcolor{custom-red!80} $<\num{2E-16}$ & \cellcolor{custom-red!80} $<\num{2E-16}$ & \cellcolor{custom-red!80} $<\num{2E-16}$ & \cellcolor{custom-red!80} $<\num{2E-16}$ \\
vertebrae & vertebrae\_C7 & \cellcolor{custom-green!50} $1.0$ & \cellcolor{custom-green!50} $1.0$ & \cellcolor{custom-green!50} $1.0$ & \cellcolor{custom-green!50} $1.0$ & \cellcolor{custom-red!80} $\num{2E-08}$ & \cellcolor{custom-red!80} $<\num{2E-16}$ & \cellcolor{custom-red!80} $<\num{2E-16}$ & \cellcolor{custom-red!80} $<\num{2E-16}$ \\
vertebrae & vertebrae\_L1 & \cellcolor{custom-green!50} $1.0$ & \cellcolor{custom-green!50} $1.0$ & \cellcolor{custom-green!50} $1.0$ & \cellcolor{custom-green!50} $1.0$ & \cellcolor{custom-red!80} $\num{1E-05}$ & \cellcolor{custom-red!80} $<\num{2E-16}$ & \cellcolor{custom-red!80} $<\num{2E-16}$ & \cellcolor{custom-red!80} $<\num{2E-16}$ \\
vertebrae & vertebrae\_L2 & \cellcolor{custom-green!50} $1.0$ & \cellcolor{custom-green!50} $1.0$ & \cellcolor{custom-green!50} $1.0$ & \cellcolor{custom-green!50} $\num{2E-01}$ & \cellcolor{custom-red!80} $\num{1E-10}$ & \cellcolor{custom-red!80} $<\num{2E-16}$ & \cellcolor{custom-red!80} $<\num{2E-16}$ & \cellcolor{custom-red!80} $<\num{2E-16}$ \\
vertebrae & vertebrae\_L3 & \cellcolor{custom-green!50} $1.0$ & \cellcolor{custom-green!50} $1.0$ & \cellcolor{custom-green!50} $\num{8E-01}$ & \cellcolor{custom-red!80} $\num{1E-04}$ & \cellcolor{custom-red!80} $<\num{2E-16}$ & \cellcolor{custom-red!80} $<\num{2E-16}$ & \cellcolor{custom-red!80} $<\num{2E-16}$ & \cellcolor{custom-red!80} $<\num{2E-16}$ \\
vertebrae & vertebrae\_L4 & \cellcolor{custom-green!50} $1.0$ & \cellcolor{custom-green!50} $1.0$ & \cellcolor{custom-green!50} $1.0$ & \cellcolor{custom-red!80} $\num{5E-03}$ & \cellcolor{custom-red!80} $<\num{2E-16}$ & \cellcolor{custom-red!80} $<\num{2E-16}$ & \cellcolor{custom-red!80} $<\num{2E-16}$ & \cellcolor{custom-red!80} $<\num{2E-16}$ \\
vertebrae & vertebrae\_L5 & \cellcolor{custom-green!50} $1.0$ & \cellcolor{custom-green!50} $1.0$ & \cellcolor{custom-green!50} $1.0$ & \cellcolor{custom-green!50} $\num{9E-02}$ & \cellcolor{custom-red!80} $<\num{2E-16}$ & \cellcolor{custom-red!80} $<\num{2E-16}$ & \cellcolor{custom-red!80} $<\num{2E-16}$ & \cellcolor{custom-red!80} $<\num{2E-16}$ \\
vertebrae & vertebrae\_S1 & \cellcolor{custom-green!50} $1.0$ & \cellcolor{custom-green!50} $1.0$ & \cellcolor{custom-green!50} $1.0$ & \cellcolor{custom-green!50} $\num{6E-02}$ & \cellcolor{custom-red!80} $<\num{2E-16}$ & \cellcolor{custom-red!80} $<\num{2E-16}$ & \cellcolor{custom-red!80} $<\num{2E-16}$ & \cellcolor{custom-red!80} $<\num{2E-16}$ \\
vertebrae & vertebrae\_T1 & \cellcolor{custom-green!50} $1.0$ & \cellcolor{custom-green!50} $1.0$ & \cellcolor{custom-green!50} $1.0$ & \cellcolor{custom-green!50} $1.0$ & \cellcolor{custom-red!80} $\num{5E-04}$ & \cellcolor{custom-red!80} $<\num{2E-16}$ & \cellcolor{custom-red!80} $<\num{2E-16}$ & \cellcolor{custom-red!80} $<\num{2E-16}$ \\
vertebrae & vertebrae\_T10 & \cellcolor{custom-green!50} $1.0$ & \cellcolor{custom-green!50} $1.0$ & \cellcolor{custom-green!50} $1.0$ & \cellcolor{custom-green!50} $\num{5E-01}$ & \cellcolor{custom-red!80} $\num{8E-10}$ & \cellcolor{custom-red!80} $<\num{2E-16}$ & \cellcolor{custom-red!80} $<\num{2E-16}$ & \cellcolor{custom-red!80} $<\num{2E-16}$ \\
vertebrae & vertebrae\_T11 & \cellcolor{custom-green!50} $1.0$ & \cellcolor{custom-green!50} $1.0$ & \cellcolor{custom-green!50} $1.0$ & \cellcolor{custom-green!50} $1.0$ & \cellcolor{custom-red!80} $\num{2E-05}$ & \cellcolor{custom-red!80} $<\num{2E-16}$ & \cellcolor{custom-red!80} $<\num{2E-16}$ & \cellcolor{custom-red!80} $<\num{2E-16}$ \\
vertebrae & vertebrae\_T12 & \cellcolor{custom-green!50} $1.0$ & \cellcolor{custom-green!50} $1.0$ & \cellcolor{custom-green!50} $1.0$ & \cellcolor{custom-green!50} $1.0$ & \cellcolor{custom-red!80} $\num{2E-04}$ & \cellcolor{custom-red!80} $<\num{2E-16}$ & \cellcolor{custom-red!80} $<\num{2E-16}$ & \cellcolor{custom-red!80} $<\num{2E-16}$ \\
vertebrae & vertebrae\_T2 & \cellcolor{custom-green!50} $1.0$ & \cellcolor{custom-green!50} $1.0$ & \cellcolor{custom-green!50} $1.0$ & \cellcolor{custom-green!50} $1.0$ & \cellcolor{custom-red!80} $\num{4E-06}$ & \cellcolor{custom-red!80} $<\num{2E-16}$ & \cellcolor{custom-red!80} $<\num{2E-16}$ & \cellcolor{custom-red!80} $<\num{2E-16}$ \\
vertebrae & vertebrae\_T3 & \cellcolor{custom-green!50} $1.0$ & \cellcolor{custom-green!50} $1.0$ & \cellcolor{custom-green!50} $1.0$ & \cellcolor{custom-green!50} $1.0$ & \cellcolor{custom-red!80} $\num{2E-07}$ & \cellcolor{custom-red!80} $<\num{2E-16}$ & \cellcolor{custom-red!80} $<\num{2E-16}$ & \cellcolor{custom-red!80} $<\num{2E-16}$ \\
vertebrae & vertebrae\_T4 & \cellcolor{custom-green!50} $1.0$ & \cellcolor{custom-green!50} $1.0$ & \cellcolor{custom-green!50} $1.0$ & \cellcolor{custom-green!50} $\num{8E-01}$ & \cellcolor{custom-red!80} $\num{2E-09}$ & \cellcolor{custom-red!80} $<\num{2E-16}$ & \cellcolor{custom-red!80} $<\num{2E-16}$ & \cellcolor{custom-red!80} $<\num{2E-16}$ \\
vertebrae & vertebrae\_T5 & \cellcolor{custom-green!50} $1.0$ & \cellcolor{custom-green!50} $1.0$ & \cellcolor{custom-green!50} $1.0$ & \cellcolor{custom-red!80} $\num{5E-05}$ & \cellcolor{custom-red!80} $<\num{2E-16}$ & \cellcolor{custom-red!80} $<\num{2E-16}$ & \cellcolor{custom-red!80} $<\num{2E-16}$ & \cellcolor{custom-red!80} $<\num{2E-16}$ \\
vertebrae & vertebrae\_T6 & \cellcolor{custom-green!50} $1.0$ & \cellcolor{custom-green!50} $1.0$ & \cellcolor{custom-green!50} $\num{2E-01}$ & \cellcolor{custom-red!80} $\num{3E-07}$ & \cellcolor{custom-red!80} $<\num{2E-16}$ & \cellcolor{custom-red!80} $<\num{2E-16}$ & \cellcolor{custom-red!80} $<\num{2E-16}$ & \cellcolor{custom-red!80} $<\num{2E-16}$ \\
vertebrae & vertebrae\_T7 & \cellcolor{custom-green!50} $1.0$ & \cellcolor{custom-green!50} $1.0$ & \cellcolor{custom-green!50} $\num{8E-02}$ & \cellcolor{custom-red!80} $\num{6E-07}$ & \cellcolor{custom-red!80} $<\num{2E-16}$ & \cellcolor{custom-red!80} $<\num{2E-16}$ & \cellcolor{custom-red!80} $<\num{2E-16}$ & \cellcolor{custom-red!80} $<\num{2E-16}$ \\
vertebrae & vertebrae\_T8 & \cellcolor{custom-green!50} $1.0$ & \cellcolor{custom-green!50} $1.0$ & \cellcolor{custom-green!50} $\num{5E-01}$ & \cellcolor{custom-red!80} $\num{1E-04}$ & \cellcolor{custom-red!80} $<\num{2E-16}$ & \cellcolor{custom-red!80} $<\num{2E-16}$ & \cellcolor{custom-red!80} $<\num{2E-16}$ & \cellcolor{custom-red!80} $<\num{2E-16}$ \\
vertebrae & vertebrae\_T9 & \cellcolor{custom-green!50} $1.0$ & \cellcolor{custom-green!50} $1.0$ & \cellcolor{custom-green!50} $1.0$ & \cellcolor{custom-red!80} $\num{3E-02}$ & \cellcolor{custom-red!80} $\num{4E-14}$ & \cellcolor{custom-red!80} $<\num{2E-16}$ & \cellcolor{custom-red!80} $<\num{2E-16}$ & \cellcolor{custom-red!80} $<\num{2E-16}$ \\
\bottomrule
\end{tabular}